\documentclass[trackchanges,twocolumn,twocolappendix]{aastex7}
\usepackage{amsmath}
\usepackage{subcaption} 

\newcommand{\comment}[1]{}
\usepackage{bm}

\newcommand{\Erase}[1]{}

\newcommand{\M}{{\ $\mid$}}

\newcommand{\W}{{$\lambda$}}

\newcommand{\q}{$q_{\rm ion}$}

\newcommand{\K}{{\rm K}}

\begin{document}
\title{
  Gas-Phase Metallicity and Nitrogen Abundances in Low-Mass Galaxies Down to $M_\star\simeq10^{6.6}\,M_\odot$ at $z\simeq4.5$--$9.8$ from JWST Lensing Cluster Surveys
}

\correspondingauthor{Hiroya Umeda}

\author[0009-0008-0167-5129]{Hiroya Umeda}
\affiliation{Center for Computational Sciences, University of Tsukuba, Ten-nodai, 1-1-1 Tsukuba, Ibaraki 305-8577, Japan}
\affiliation{Institute for Cosmic Ray Research,
The University of Tokyo,
5-1-5 Kashiwanoha, Kashiwa,
Chiba 277-8582, Japan}
\affiliation{Department of Physics, Graduate School of Science, The University of Tokyo, 7-3-1 Hongo, Bunkyo, Tokyo 113-0033, Japan}
\email[show]{ume@icrr.u-tokyo.ac.jp}

\author[0000-0002-1049-6658]{Masami Ouchi}
\affiliation{National Astronomical Observatory of Japan, 2-21-1 Osawa, Mitaka, Tokyo 181-8588, Japan}
\affiliation{Institute for Cosmic Ray Research,
The University of Tokyo,
5-1-5 Kashiwanoha, Kashiwa,
Chiba 277-8582, Japan}
\affiliation{Kavli Institute for the Physics and Mathematics of the Universe (WPI),
University of Tokyo, Kashiwa, Chiba 277-8583, Japan}
\email{ouchims@icrr.u-tokyo.ac.jp}

\author[0000-0003-2965-5070]{Kimihiko Nakajima}
\affiliation{Kanazawa University, Kakumamachi, Kanazawa, Ishikawa 920-1192, Japan}
\email{knakajima@staff.kanazawa-u.ac.jp}

\author[0000-0002-2740-3403]{Kuria Watanabe}
\affiliation{Department of Astronomical Science, SOKENDAI (The Graduate University for Advanced Studies),
2-21-1 Osawa, Mitaka, Tokyo, 181-8588, Japan}
\affiliation{National Astronomical Observatory of Japan, 2-21-1 Osawa, Mitaka, Tokyo 181-8588, Japan}
\email{kuria.watanabe@grad.nao.ac.jp}

\author[0000-0002-6047-430X]{Yuichi Harikane}
\affiliation{Institute for Cosmic Ray Research, The University of Tokyo, 5-1-5 Kashiwanoha, Kashiwa, Chiba 277-8582, Japan}
\email{hari@icrr.u-tokyo.ac.jp}

\author[0000-0001-7730-8634]{Yuki Isobe}
\affiliation{Kavli Institute for Cosmology, University of Cambridge, Madingley Road, Cambridge, CB3 0HA, UK}
\affiliation{Cavendish Laboratory, University of Cambridge, 19 JJ Thomson Avenue, Cambridge, CB3 0HE, UK}
\affiliation{Waseda Research Institute for Science and Engineering, Faculty of Science and Engineering, Waseda University, 3-4-1, Okubo, Shinjuku, Tokyo 169-8555, Japan}
\email{yi264@cam.ac.uk}

\author[0000-0003-4321-0975]{Moka Nishigaki}
\affiliation{Department of Astronomical Science, SOKENDAI (The Graduate University for Advanced Studies),
2-21-1 Osawa, Mitaka, Tokyo, 181-8588, Japan}
\affiliation{National Astronomical Observatory of Japan, 2-21-1 Osawa, Mitaka, Tokyo 181-8588, Japan}
\email{moka.nishigaki@grad.nao.ac.jp}

\author[0000-0001-9011-7605]{Yoshiaki Ono}
\affiliation{Institute for Cosmic Ray Research, The University of
Tokyo, 5-1-5 Kashiwanoha, Kashiwa, Chiba 277-8582, Japan}
\email{ono@icrr.u-tokyo.ac.jp}

\author[0000-0002-1319-3433]{Hidenobu Yajima}
\affiliation{Center for Computational Sciences, University of Tsukuba, Ten-nodai, 1-1-1 Tsukuba, Ibaraki 305-8577, Japan}
\email{yajima@ccs.tsukuba.ac.jp}

\author[0000-0002-5768-8235]{Yi Xu}
\affiliation{Cosmic Dawn Center (DAWN), Niels Bohr Institute, University of Copenhagen, Jagtvej 128, Copenhagen N, DK-2200, Denmark}
\email{yi.xu@nbi.ku.dk}

\begin{abstract}
		We analyze 405 deep JWST/NIRSpec spectra of star-forming galaxies at $z=4.5$--$9.8$ from multiple JWST lensing-cluster surveys, including DREAMS, SPURS, GLIMPSE-D, UNCOVER, CANUCS, GLASS, and ERO, to study chemical enrichment in intrinsically faint, low-mass galaxies. The sample covers $M_{\rm UV}\simeq-12$ to $-22$ and reaches $M_\star\simeq10^{5.7}\,M_\odot$, with 50\% of the sources at $M_{\rm UV}>-17.5$ magnified by $\mu>3$. From individual spectra and mass-binned stacks, we derive the gas-phase metallicity together with nitrogen and carbon abundances using the rest-frame UV and optical lines. We derive new empirical strong-line metallicity calibrations using direct-method measurements from stellar-mass-binned stacks, reaching a representative stellar mass of $M_\star\simeq10^{6.6}\,M_\odot$. Applying these calibrations, we trace the $z\sim6$ mass--metallicity relation down to $M_\star\simeq10^{6.6}\,M_\odot$, where it reaches $12+\log({\rm O/H})\simeq7.2$. A two-parameter linear fit gives a slope of $\alpha=0.318\pm0.037$, only slightly steeper than the local low-mass slope and broadly consistent with hydrodynamical simulations. In the $M_\star\simeq10^{7.7}\,M_\odot$ stack, N/O from N\,{\sc iv}]\,$\lambda\lambda\,1483,1486$ exceeds that from [N\,{\sc ii}]\,$\lambda6583$ by $\simeq1.4$ dex. The UV--optical difference could indicate a localized, highly ionized N-rich component whose high N/O and subsolar C/O resemble nitrogen-rich globular-cluster populations with $M_\star\lesssim10^6\,M_\odot$. The combination of these abundance patterns and a He\,{\sc ii}\,\W4686/H$\beta$ ratio of $\simeq0.03$, well above BPASS predictions, suggests that WR stars may contribute both prompt CNO-cycle enrichment and hard ionizing radiation, with the inferred WR population potentially supplying enough nitrogen to account for the excess on globular-cluster scales.
\end{abstract}



\section{Introduction} \label{intro}
The James Webb Space Telescope (JWST) has dramatically advanced the study of galaxy evolution in the early universe, enabling direct spectroscopic measurements of the physical and chemical properties of galaxies at high redshift \citep[e.g.,][]{isobe_jwst_2023,cameron_nitrogen_2023,schaerer_nitrogen_2026}. Early JWST observations have already suggested a number of possible anomalies in nebular abundance patterns and in the mass--metallicity relation (MZR), hinting that chemically young, low-mass galaxies in the distant universe may not follow the same relations established at low redshift \citep[e.g.,][]{nishigaki_dreamsii_2025,curti_jades_2024,nakajima_jwst_2023,asada_glimpse-ddt_2026,chemerynska_extreme_2024,stanton_jwst_2025}. Yet these findings are still based on limited samples and are often restricted by insufficient spectroscopic depth, which hampers the detection of faint diagnostic emission lines and leaves the low-mass end of the MZR poorly constrained.

Moreover, the chemical abundance patterns of some high-redshift galaxies show nitrogen enhancement relative to oxygen, which may indicate a significant contribution from enrichment channels such as Wolf-Rayet (WR) stars, supermassive stars, or tidal disruption events \citep[e.g.,][]{watanabe_chemical_2026}. However, the current evidence for nitrogen enhancement is still limited to a small number of bright galaxies. Nitrogen emission lines in both UV and optical are relatively weak compared to hydrogen or helium lines, and thus the detection of nitrogen emission lines could be strongly biased to the nitrogen-enhanced galaxies \citep[e.g.,][]{zhu_nature_2026}. Thus, it remains unclear whether the nitrogen enhancement is common among the broader population of star-forming galaxies at high redshift. A more comprehensive investigation of the chemical abundance patterns in a larger sample of faint galaxies is needed to understand the prevalence and implications of nitrogen enhancement in the early universe.

Recent developments in JWST lensing-cluster surveys provide a powerful way forward. Gravitational lensing makes it possible to access intrinsically faint galaxies, and the combination of multiple lensing-cluster datasets now enables deep spectroscopic studies over a broader dynamic range in luminosity and stellar mass \citep[e.g.,][]{nakajima_ultra-faint_2025,fujimoto_glimpse-d_2025,atek_jwsts_2025,bezanson_jwst_2024,willott_near-infrared_2022,treu_glass-jwst_2022}. Lensing surveys have now enabled detections of very faint (i.e., $M_{\rm UV}\simeq -12$) galaxies \citep{nakajima_ultra-faint_2025,atek_glimpse_2026}. Many of the Pop III candidates have also been discovered with the help of magnification by gravitational lensing \citep[e.g.,]{nakajima_ultra-faint_2025,vanzella_pristine_2026,cai_metal-free_2025,morishita_pristine_2025}. As such, the highly magnified objects at high redshift provide a unique pathway to probe the parameter space of the faint-end galaxy population.\par

Another strategy is to create stacked spectra of the galaxies. Multiple works including \cite{isobe_jades_2026-1} have stacked the individual spectra to maximize the signal to noise ratio of faint emission lines and successfully detect multiple chemical tracers. Moreover, because the stacking analysis can probe the characteristic spectral properties of the galaxy sample, the stacking analysis is a very useful tools to test whether the chemical anomalies seen in some of the bright galaxies can be universally seen in low-mass galaxy. Even with the stacking analysis, the detection of faint emission lines can suffer from the detection limit \citep{isobe_jades_2025}.\par

Motivated by these recent advances in the field, in this work we compile deep JWST spectra from multiple lensing-cluster surveys centered around the Deep Reconnaissance of Early Assemblies of Metal-poor Star formation (DREAMS) survey \citep{nakajima_ultra-faint_2025,nishigaki_dreamsii_2025} and  investigate the chemical properties of faint galaxies at $z\gtrsim4.5$ via the stacked spectra analysis. Using faint emission-line diagnostics, we study the mass-metallicity relation and the behavior of nitrogen and carbon abundances in the low-mass galaxy population.

This paper is organized as follows. Section~\ref{sec:data} describes the JWST datasets, data reduction, emission-line measurements, and sample selection. Section~\ref{sec:data:stack} presents the construction and analysis of the stacked spectra. Section~\ref{sec:res} reports the physical properties, gas-phase metallicities, mass--metallicity relation, and elemental abundance ratios of our galaxy sample. Section~\ref{sec:disc} discusses the implications for chemical enrichment in low-mass galaxies, including the discrepancy between UV- and optical-based nitrogen abundances. Section~6 summarizes our main findings. 

Throughout this paper, we assume a flat $\Lambda$ cold dark matter (CDM) cosmology consistent with the constraints from Planck \citep{planck_collaboration_planck_2020}: $h=0.6766$, $\Omega_{\rm m}=0.3103$, $\Omega_\Lambda=0.6897$, and  $\Omega_{\rm b} h^2=0.02234$. All magnitudes are in the AB system \citep{oke_secondary_1983}. We adopt the solar abundance composition measured by \cite{asplund_chemical_2009} with $12+\log({\rm O/H})_\odot=8.69$.

\section{Data \& Reduction}\label{sec:data}

\subsection{Lensing Cluster Observations and Data Reduction}
In this study, we use spectroscopic datasets from multiple JWST programs targeting lensing clusters, including DREAMS \citep[GO-4750; PI: K. Nakajima;][]{nakajima_ultra-faint_2025,nishigaki_dreamsii_2025,takechi_dreams_2026}, GLIMPSE-D \citep[GO-9223; PI: S. Fujimoto;][]{fujimoto_glimpse-d_2025}, GLASS \citep[ERS-1324; PI: T. Treu;][]{treu_glass-jwst_2022}, ERO-2736 \citep[PI: K. Pontoppidan;][]{pontoppidan_jwst_2022}, SPURS \citep[GO-9214; PIs: C. Mason \& Dan Stark;][]{tang_spurs_2026,chen_spurs_2026}, UNCOVER \citep[GO-2561; PIs: I. Labbe \& R. Bezanson;][]{bezanson_jwst_2024}, and CANUCS \citep[GTO-1208; PI: C. Willott;][]{willott_near-infrared_2022}. Collectively, these programs provide high-, medium-, and low-resolution NIRSpec spectra covering key rest-frame UV to optical emission lines, along with complementary NIRCam photometry. This combination enables us to construct a large, diverse sample of galaxies with high signal-to-noise ratio (S/N) spectra across a wide redshift range.

Table~\ref{tab:survey_summary} summarizes the number of galaxies contributed by each survey and the corresponding NIRSpec configurations, wavelength coverage, spectral resolution, lensing magnification, and spectroscopic-redshift range.

\begin{deluxetable*}{lclcccc}
  \tablecaption{JWST/NIRSpec surveys contributing to the final galaxy sample.\label{tab:survey_summary}}
  \tablewidth{0pt}
  \tablehead{\colhead{Survey} & \colhead{$N_{\rm unique}$} & \colhead{Disperser(s)} & \colhead{$\lambda_{\rm obs}$ ($\mu$m)} & \colhead{$R$} & \colhead{$\mu$ range} & \colhead{$z_{\rm spec}$ range}}
  \startdata
  DREAMS & 19 & G140M/F070LP + G395M/F290LP & $0.70$--$1.27$, $2.87$--$5.27$ & $\sim1000$ & $1.00$--$38.83$ & $5.88$--$9.51$ \\
  SPURS & 29 & G140M/F100LP + G235M/F170LP + G395M/F290LP & $0.97$--$5.27$ & $\sim1000$ & $1.00$--$10.19$ & $4.98$--$9.56$ \\
  GLIMPSE-D & 51 & G395M/F290LP & $2.87$--$5.27$ & $\sim1000$ & $1.00$--$30.40$ & $5.01$--$8.58$ \\
  UNCOVER & 85 & PRISM/CLEAR & $0.60$--$5.30$ & $30$--$300$ & $1.40$--$17.63$ & $4.51$--$9.80$ \\
  CANUCS & 205 & PRISM/CLEAR & $0.60$--$5.30$ & $30$--$300$ & $1.03$--$93.29$ & $4.50$--$9.68$ \\
  GLASS & 11 & G140H/F100LP + G235H/F170LP + G395H/F290LP & $0.97$--$5.27$ & $\sim2700$ & $1.19$--$7.91$ & $4.80$--$7.88$ \\
  ERO & 5 & G235M/F170LP + G395M/F290LP & $1.66$--$5.27$ & $\sim1000$ & $1.00$--$3.23$ & $5.28$--$8.50$ \\
  All surveys & 405 & Mixed & $0.60$--$5.30$ & $30$--$2700$ & $1.00$--$93.29$ & $4.50$--$9.80$ \\
  \enddata
  \tablecomments{Counts refer to unique galaxies entering the PAll sample. The final sample spans $4.50\leq z_{\rm spec}\leq9.80$, has finite non-negative stellar masses and valid H$\beta$ measurements, and excludes BPT-selected, broad-line, and little-red-dot AGN. Wavelength intervals and resolving powers are nominal observed-frame values for the listed NIRSpec configurations.}
\end{deluxetable*}

\subsubsection{DREAMS}\label{sec:data:program:dreams}
We use data from the JWST program GO-4750, known as the DREAMS survey.
DREAMS focuses on the strong-lensing galaxy cluster MACS~J0416, allowing intrinsically faint galaxies to be observed through gravitational magnification.
Spectroscopic data were obtained with NIRSpec using the micro-shutter assembly in medium-resolution configurations, specifically G140M/F070LP and G395M/F290LP, corresponding to a resolving power of $R \sim 1000$.

We use the NIRSpec data reduced by the DREAMS survey team. An overview of the DREAMS survey and its data reduction is presented in \cite{nakajima_ultra-faint_2025}, with a complete description to be provided in a forthcoming publication (Nakajima et al., in preparation).
Within the DREAMS dataset, we select galaxies with spectroscopic redshift above 4.5.

We use the NIRCam imaging data of MACS J0416 obtained in the CANUCS program \citep{willott_near-infrared_2022}. The NIRCam imaging data are reduced in the same manner as described in \cite{harikane_comprehensive_2023} and \cite{nakajima_ultra-faint_2025}.
Gravitational lensing magnification factors are computed using lens models of MACS~J0416 constructed with the \textsc{glafic} model \citep{kawamata_sizeluminosity_2018}\footnote{\url{https://archive.stsci.edu/prepds/frontier/lensmodels/}}, following the methodology described in \cite{oguri_mass_2010} and \cite{oguri_hundreds_2021}.

\subsection{ERO and GLASS}\label{sec:data:program:ero_glass}
We also use data from the ERO and GLASS programs, which provide early JWST observations of the strong-lensing galaxy cluster SMACS J0723.3-7327 and Abell 2744. For photometry and spectroscopic data, we refer to the data used in \cite{harikane_comprehensive_2023} and \cite{nakajima_jwst_2023}. We also refer to the spectroscopic redshift and magnification factor of the objects in ERO and GLASS from \cite{nakajima_jwst_2023}.

\subsubsection{GLIMPSE}\label{sec:data:program:glimpse}
We also use data from the GLIMPSE survey, which provides deep JWST observations of the strongly lensed galaxy cluster AS1063.
The survey combines NIRCam imaging and NIRSpec spectroscopy, enabling the study of intrinsically faint galaxies through gravitational lensing.

The GLIMPSE program \citep[GO-3293; PI: H. Atek][]{atek_jwsts_2025} has obtained deep NIRCam imaging in the AS1063 field with a total integration time of 120 hours, covering seven broad-band filters (F090W, F115W, F150W, F200W, F277W, F356W, and F444W) and two medium-band filters (F410M and F480M). We use the reduced imaging data from DAWN JWST Archive \citep[DJA; ][]{valentino_atlas_2023}.\footnote{\url{https://dawn-cph.github.io/dja/}} The imaging data in DJA are reduced using the \texttt{grizli} pipeline \citep{brammer_grizli_2023}. Gravitational lensing magnification factors for the GLIMPSE sources are computed in the same manner as for DREAMS using \textsc{glafic} models.

For the NIRSpec spectroscopy, we use the reduced NIRSpec MOS spectra from the GLIMPSE-D survey \citep[DD-9223; PI: S. Fujimoto;][]{fujimoto_glimpse-d_2025}. The NIRSpec observations in the GLIMPSE-D survey are obtained using the micro-shutter array (MSA) in medium-resolution grating/filter pairs of G140M/F100LP, G235M/F170LP, and G395M/F290LP. The total on-source integration time for most of the targets is approximately 9-11 hours. We reduce the NIRSpec/MOS data using the \texttt{jwst} pipeline version 1.16.1 together with the Calibration Reference Data System (CRDS) context file \texttt{1303.pmap}. In addition to the default reduction procedures in the JWST pipeline, we apply several custom modifications to improve the quality of the reduced spectra. Further details of the data reduction process are described in \cite{nakajima_jwst_2023}.

\subsubsection{SPURS}\label{sec:data:program:go9214}
We also use spectroscopy from the SPURS program, which provides deep JWST/NIRSpec observations targeting theAbell 2744 cluster in November 2025. The targets in the SPURS program are observed in multi-object spectroscopy mode with medium-resolution grating/filter pairs of G140M/F100LP, G235M/F170LP, and G395M/F290LP. The exposure times are 29.47, 8.02 and 2.95 hours for the G140M, G235M and G395M gratings, respectively. We reduce the NIRSpec/MOS data in the same manner as for the GLIMPSE survey data. 

For imaging, we use NIRCam data of Abell 2744 from the GLASS survey, adopting the reduced products used in \citet{harikane_comprehensive_2023}. Gravitational lensing magnification factors for the SPURS sources are computed in the same manner as for DREAMS using \textsc{glafic} models.

\subsubsection{UNCOVER}\label{sec:data:program:uncover}
The UNCOVER program provides deep JWST observations of massive lensing galaxy clusters, designed to probe intrinsically faint galaxies at high redshift through gravitational magnification.

The survey combines NIRCam imaging with NIRSpec multi-object spectroscopy.
NIRSpec observations are obtained using the micro-shutter array (MSA) in the low-resolution prism mode, providing continuous wavelength coverage from $\sim0.6$ to $5.5~\mu$m at a spectral resolving power of $R \sim 100$.
Typical observations consist of multiple MSA configurations with individual exposure times of 2.4--4.4 hr, resulting in total on-source integration times of several kiloseconds.

Photometric and spectroscopic measurements used in this work are taken from the reduced UNCOVER data products.\footnote{\url{https://jwst-uncover.github.io/\#releases}}
Gravitational lensing magnification factors are derived from cluster lens models adopted by the UNCOVER and MEGA-Science \citep[GO-4111; PI: K. Suess;][]{suess_medium_2024} collaboration. In addition to the NIRCam broad-band imaging observations by the UNCOVER program, the MEGA-Science program has conducted deep (2.3--4.6 hr) JWST/NIRCam imaging observations in two short broad-band filters (i.e., F070W and F090W) and eleven medium-band filters (i.e., F140M, F162M, F182M, F210M, F250M, F300M, F335M, F360M, F430M, F460M, and F480M). For the details, we refer to the UNCOVER and MEGA-Science data release documentation \citep{bezanson_jwst_2024,suess_medium_2024}.

\subsubsection{CANUCS}\label{sec:data:program:canucs}
The CANUCS program provides deep JWST observations of multiple lensing galaxy clusters (i.e., A370, MACS0416, MACS0417, MACS1149, and MACS1423), designed to probe intrinsically faint galaxies at high redshift through gravitational magnification in a vast cosmological volume.

The survey combines NIRCam imaging with spectroscopy by NIRISS and NIRSpec. NIRCam imaging is obtained in seven broad-band filters (F090W, F115W, F150W, F200W, F277W, F356W, and F444W) and the medium-band filter F410M with a total integration time of 6.4 ks per filter.
NIRSpec observations in the CANUCS program are also conducted in multi-object spectroscopy mode with low-resolution prism. 

Photometric and spectroscopic measurements as well as the magnification factors of the objects used in this work are taken from the reduced CANUCS DR1 data release products.\footnote{\url{https://niriss.github.io/data_release1.html}} Details of the data products are provided in the CANUCS DR1 documentation \citep{sarrouh_canucstechnicolor_2026}.

\subsubsection{Photometry}\label{sec:data:photo}
For NIRCam imaging of objects in DREAMS, GLIMPSE-D, UNCOVER, SPURS, GLASS, and ERO, we obtain photometric measurement following the procedures described in \citet{harikane_comprehensive_2023}. In brief, we build multi-band source catalogs using a \texttt{SWarp} \citep{bertin_swarp_2010}  detection image constructed as a weighted mean stack image of bands redder than the Lyman break for each dropout selection. We PSF-match all bands and run \texttt{SExtractor} \citep{bertin_sextractor_1996} in dual-image mode for source detection and photometry. To calculate the total magnitude for each band-image, we quantify the aperture-correction factor by comparing the MAG\_AUTO photometry and $3^{\prime\prime}.0$ diameter aperture photometry values for the multi-band stacked image and apply the correction factor to the $3^{\prime\prime}.0$ diameter aperture photometry in each filter. We correct for Galactic extinction following \cite{schlegel_maps_1998} and \cite{schlafly_measuring_2011}. For further details, we refer to \citet{harikane_comprehensive_2023}.
For the CANUCS survey, we use the official NIRCam photometry measurements from the CANUCS DR1 \citep{sarrouh_canucstechnicolor_2026}.

\subsection{Visual Inspection and Emission Line Flux Measurement}\label{sec:data:lineflux}
After reducing NIRSpec spectra, we have visually checked all spectra for the data quality and Ly$\alpha$ break and emission lines such as [O\,{\sc ii}]\,\W\W3727,3729, H$\gamma$, H$\beta$, [O\,{\sc iii}]\,\W\W4959,5007, and H$\alpha$. For objects with identified emission lines, we measure emission line fluxes. Emission lines, including Ly$\alpha$, N\,{\sc iv}]\,\W\W1483,1486, C\,{\sc IV}\,\W\W1548,1551, He\,{\sc ii}\,\W1640, O\,{\sc iii}]\,\W\W1661,1666, C\,{\sc iii}]\,\W\W1907,1909, H$\alpha$, H$\beta$, H$\gamma$, and [O\,{\sc iii}]\,\W\W4959,5007, are fitted with Gaussian profiles, with the noise spectrum used to weight the fits. All wavelength measurements are performed on the vacuum wavelength scale. The uncertainty in each flux measurement is estimated by summing the noise levels of spectral bins in quadrature within a $\pm2\sigma$ window centered on the Gaussian peak, where $\sigma$ is derived from the fitted FWHM.

We perform simultaneous fitting of key emission lines to properly account for doublet degeneracies and to determine the systemic redshift.
We mask the wavelength regions affected by strong sky lines or other artifacts to avoid biases in the fitting. The best-fit parameters and their uncertainties are derived from the posterior probability distributions obtained through Markov Chain Monte Carlo (MCMC) sampling using the \texttt{emcee} module \citep{foreman-mackey_emcee_2013}. 

After visually inspecting the fitting results and excluding cases with unreliable fits, we adopt the systemic redshift inferred from the fit for each source. We require the systemic redshift to be determined by at least two emission lines. Using this systemic redshift as a prior, we then fit each emission-line complex separately to measure the line fluxes and related quantities. Because of the limited wavelength coverage within each fitting window, we fit the following line complexes simultaneously: N\,{\sc iv}]\,\W\W1483,1486, C\,{\sc iv}\,\W\W1548,1551, the N\,{\sc iii}]\,\W1750 quintet, Si\,{\sc iii}]\,\W\W1883,1892 + C\,{\sc iii}]\,\W\W1907,1909, [O\,{\sc ii}]\,\W\W3726,3729, H$\gamma$ + [O\,{\sc iii}]\,\W4363, H$\beta$ + [O\,{\sc iii}]\,$\lambda\lambda4959,5007$, H$\alpha$+[N\,{\sc ii}]\,\W\W6548,6583, He\,{\sc ii}\,\W4686 + [Fe\,{\sc iii}]\,\W4658 + [Ar\,{\sc iv}]\,\W4711, and [S\,{\sc ii}]\,\W\W6716,6731. For each emission line, we calculate the signal-to-noise ratio (S/N) from the ratio of the best-fit flux to its measurement uncertainty, and we define a line as detected when S/N$>5$.

We also fit H$\alpha$ and H$\beta$ with narrow-only and narrow-plus-broad Gaussian models using the same fitting framework. For H$\alpha$, we simultaneously fit the [N\,{\sc ii}]\,\W\W6548,6583 lines. We compare the models using the widely applicable information criterion (WAIC), defining $\Delta{\rm WAIC}\equiv{\rm WAIC}_{\rm single}-{\rm WAIC}_{\rm two-component}$. We consider a broad component to be detected when $\Delta{\rm WAIC}>10$, its FWHM exceeds $500~{\rm km\,s^{-1}}$, and its flux is detected at $S/N>5$. Objects satisfying these criteria in either H$\alpha$ or H$\beta$ are classified as broad-line AGN candidates and excluded as described in Section~\ref{sec:data:sample:sel}. For subsequent abundance estimates and continuum-based diagnostics, we resample from the posterior probability distributions of these fits to propagate degeneracies and measurement uncertainties consistently.

\subsection{SED Fitting}
We perform spectral energy distribution (SED) fitting to derive stellar masses, star-formation rates, dust attenuation, ionization parameters, and stellar metallicities using the Bayesian SED-fitting code \texttt{Bagpipes} \citep{carnall_inferring_2018}. We fit the multi-band photometry and spectroscopic emission-line fluxes simultaneously, fixing the redshift of each source to its spectroscopic redshift. This joint fitting allows the broad-band photometry and nebular line measurements to constrain the stellar and nebular components in a self-consistent manner.

For the stellar population model, we adopt the BPASS v2.2.1 binary stellar population synthesis models \citep{eldridge_binary_2017}, assuming the \texttt{imf135\_300} initial mass function. Nebular line and continuum emission are included using pre-computed \textsc{Cloudy} grids \citep{gunasekera_2025_2025,ferland_2017_2017}, with the ionization parameter allowed to vary over $-4 \leq \log U \leq 0$. We assume a Calzetti attenuation law \citep{calzetti_dust_2000}, with the $V$-band attenuation allowed to vary over $0 \leq A_V \leq 4$.

We adopted a five-bin non-parametric star-formation history (SFH) parameterized by the logarithmic SFR ratios between adjacent time bins, following the continuity-SFH parameterization discussed by \citet{leja_how_2019}. 
The bin edges were determined separately for each galaxy at its fixed spectroscopic redshift. 
The youngest bin spans $0$--$10\,\mathrm{Myr}$, while the remaining four bins are spaced uniformly in logarithmic lookback time between $10\,\mathrm{Myr}$ and the time elapsed from $z=30$ to the observed redshift. 
We adopted a uniform prior on the logarithm of the total formed stellar mass over $2 \leq \log_{10}(M_{\mathrm{formed}}/M_{\odot}) \leq 12$ and a log-uniform prior on the stellar metallicity over $0.005 \leq Z_{\ast}/Z_{\odot} \leq 5$. 
The logarithmic SFR ratios between adjacent bins were assigned broad uniform priors over $-5 \leq \Delta\log_{10}(\mathrm{SFR}) \leq 5$.

The input photometry is corrected for gravitational magnification, and the spectroscopic emission-line fluxes are corrected for both magnification and slit-loss effects before fitting. We include only galaxies with valid emission-line measurements in this analysis. The posterior distributions are sampled with the \texttt{nautilus} sampler \citep{lange_nautilus_2023}.

From the resulting posterior samples, we derive the stellar mass, formed stellar mass, ${\rm SFR}_{10}$, ${\rm SFR}_{100}$, specific star-formation rate, burstiness parameter defined as $\log[{\rm SFR}_{10}/{\rm SFR}_{100}]$, dust attenuation, ionization parameter, and stellar metallicity.

\subsection{Sample}\label{sec:data:sample}

\begin{figure}[htbp]
  \centering
  \includegraphics[width=\linewidth]{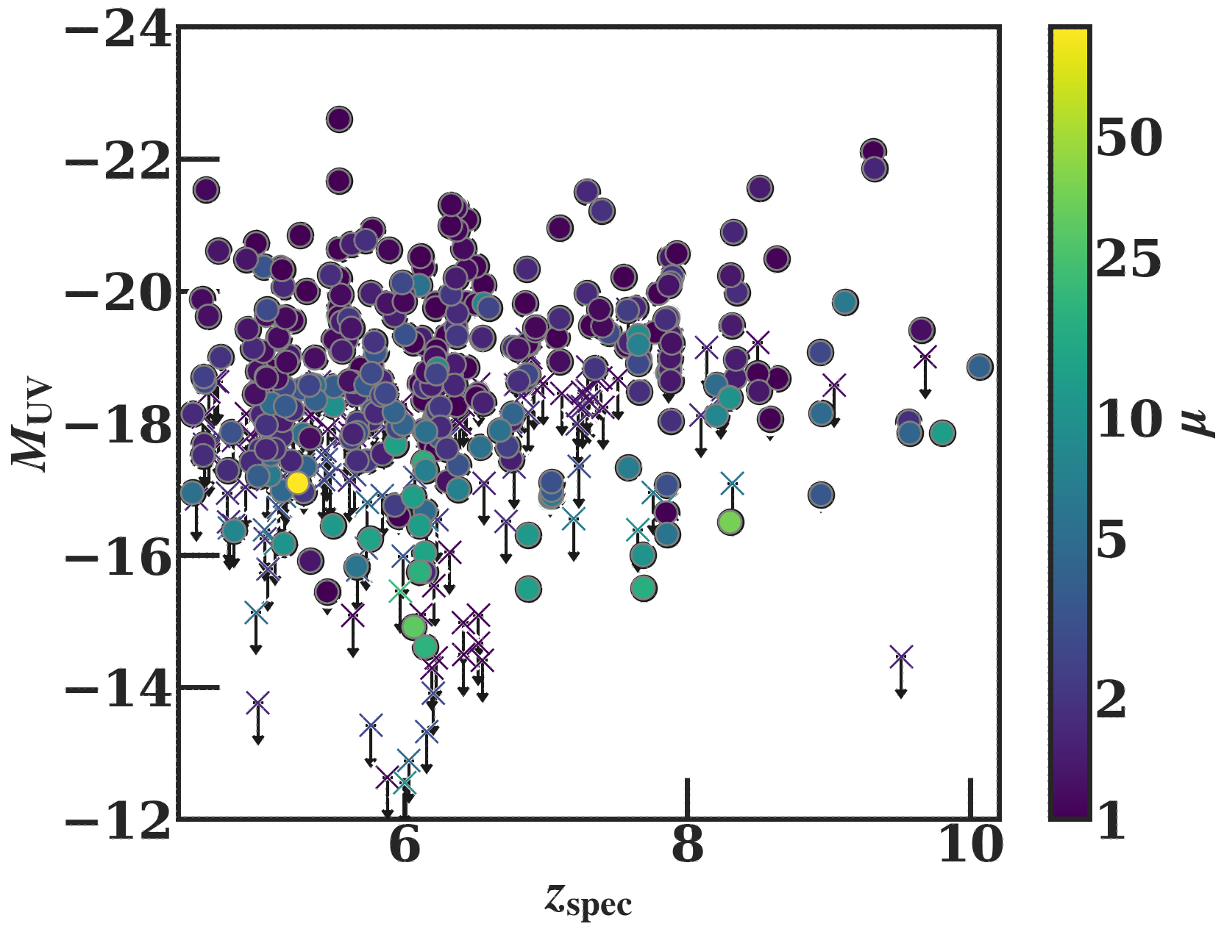}
  \caption{The distribution of UV magnitude by redshift for the sample after excluding AGN and LRD candidates. Circles and crosses correspond to the objects at $z<8$ ($z>8$) with F150W (F200W) band detection or non-detection, respectively. For the non-detection sources, we present the 3$\sigma$ upper limit. The colors indicate the magnification factor, with the color bar at left showing the corresponding scale.
  }
  \label{fig:muv_z} 
  \end{figure}
  
  \begin{figure}[htbp]
  \centering
  \includegraphics[width=\linewidth]{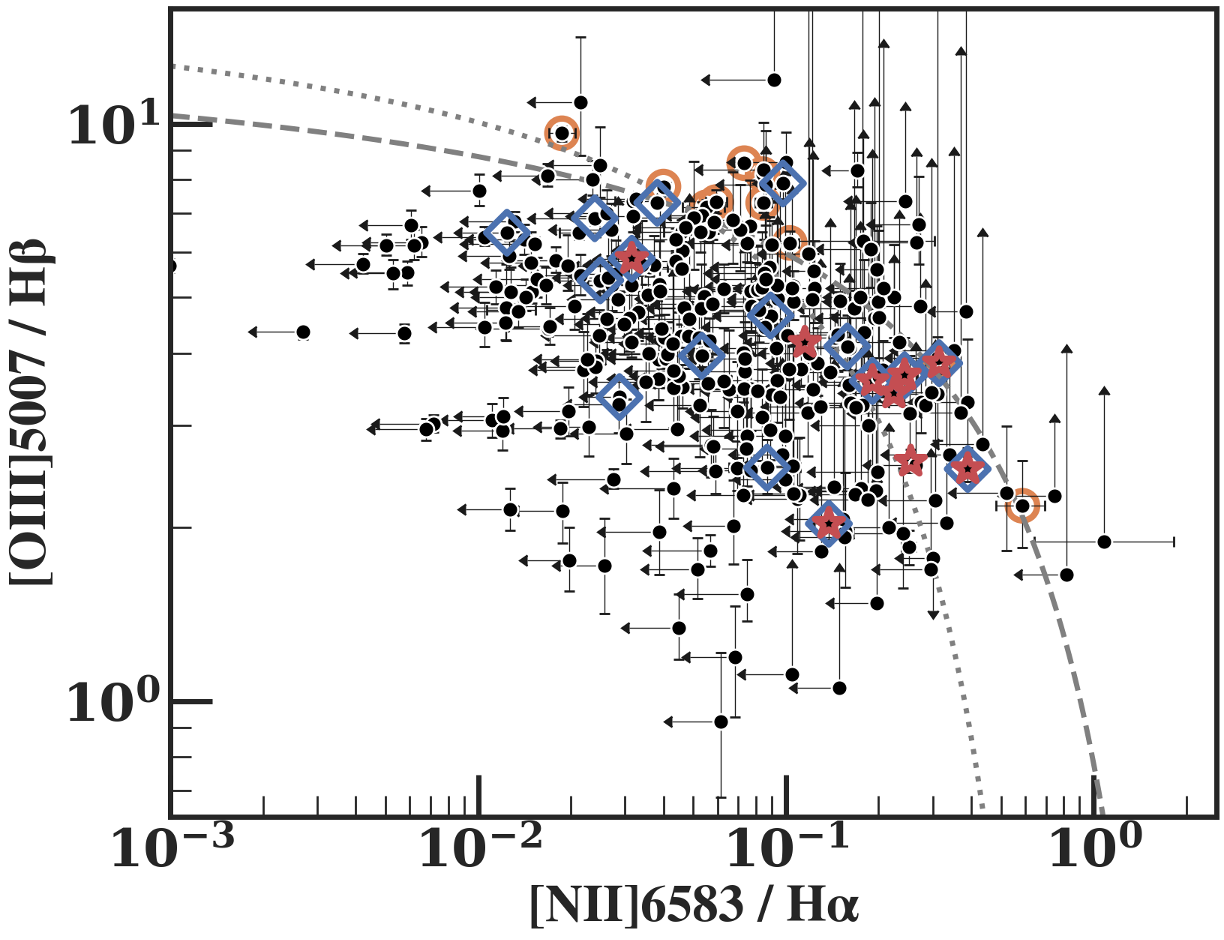}
  \caption{The BPT diagram. The objects encircled by blue diamonds, red circles, and yellow pentagons represent the AGN candidate selected from the broad line detection (Type-I AGN candidate), BPT selection, and red optical slope (i.e., LRD candidates).}
  \label{fig:bpt} 
  \end{figure}

\subsubsection{Sample Selection}\label{sec:data:sample:sel}
To mitigate AGN contamination, we exclude galaxies showing AGN-like line ratios using the classical BPT diagram based on [O\,{\sc iii}]/H$\beta$ and [N\,{\sc ii}]/H$\alpha$ where H$\alpha$ is within the NIRSpec wavelength coverage. We remove objects that satisfy the AGN selection criteria of \cite{kewley_optical_2001} and \cite{kauffmann_host_2003}. We show the BPT diagram and emission line ratios of the objects in Figure \ref{fig:bpt}. 
We exclude the broad-line AGN candidates identified from the H$\alpha$ and H$\beta$ fitting described in Section~\ref{sec:data:lineflux}. Moreover, we also exclude objects which satisfy the Little Red Dots (LRD) color selection adopted in \cite{kocevski_rise_2025}.
In the BPT diagram shown in Figure \ref{fig:bpt}, we indicate the objects identified as AGN candidates based on broad-line detection, BPT selection, and LRD color selection.
We thus exclude AGN candidates identified using the available BPT, broad-line, and LRD diagnostics. These criteria do not provide a complete census of AGN activity at high redshift: the classical low-redshift BPT demarcations may not cleanly separate star formation and AGN activity in high-redshift, low-metallicity galaxies, and broad-line identification is limited by the available spectral S/N. Residual weak or narrow-line AGN therefore cannot be ruled out. However, because we perform stacking analysis using the wavelength-by-wavelength median, a small number of residual objects with exceptionally strong AGN emission lines have substantially less leverage on the stacked spectrum than they would in a mean stack.
Our initial compilation contains 472 spectra at $4.5<z_{\rm spec}<10.1$. We exclude AGN and LRD candidates, spectra affected by severe reduction artifacts or unreliable spectral features, spectra that do not satisfy our data-quality requirements, and spectra without H$\beta$ coverage required for normalization. After these cuts, 405 spectra spanning $4.50\leq z_{\rm spec}\leq9.80$ are retained for the stacking analysis.

\subsubsection{Sample Properties}\label{sec:data:sample:prop}
We briefly explain the basic properties of our galaxy sample. In Figure \ref{fig:muv_z}, we show the redshift, UV magnitude, and magnification factor for the sample after excluding AGN and LRD candidates. Using the magnification factor, we calculate the absolute UV magnitude from the photometry of F150W and F200W bands for the objects at $z<8$ and $z>8$, respectively. We find that some of the objects are as faint as $M_{\rm UV}\gtrsim-12$. Around half of faint star-forming galaxies with $M_{\rm UV}>-17.5$ mag are magnified by more than a factor of 3, indicating the importance of gravitational lensing in probing the faint galaxy population at high redshift. 

\section{Stacking Analysis}\label{sec:data:stack}
To detect faint emission lines that are not individually detected in our sample, we perform a stacking analysis. The stacking sample consists of 405 spectra that satisfy the selection criteria described in Section~\ref{sec:data:sample:sel}. We conduct the stacking analysis for each stellar mass bin to investigate the mass dependence of emission line properties. We define the following four stellar mass bins: $5.5<\log(M_\star/M_\odot)<7.0$, $7.0 \leq \log(M_\star/M_\odot) < 8.0$, $8.0<\log(M_\star/M_\odot)<9.0$, and $9.0<\log(M_\star/M_\odot)<10.5$. We conduct the stacking procedure for grating spectra and prism spectra. For the prism-resolution stacks, we include both prism spectra and grating spectra convolved to the prism resolution using the line spread function provided by STScI.\footnote{\url{https://jwst-docs.stsci.edu/jwst-near-infrared-spectrograph/nirspec-instrumentation/nirspec-dispersers-and-filters}} Of the 405 spectra, 115 have usable grating data and constitute the grating-stack subsample. We describe our stacking methodology. We first shift each spectrum to the rest frame using the systemic redshift determined from the emission line fitting in a flux-conserving manner. 
Similar to \cite{isobe_jades_2026-1}, we interpolate both the grating and prism spectra onto a common rest-frame wavelength grid sampled at the wavelength-dependent FWHM of the line spread function provided by STScI, evaluated at the median redshift of each stack and assuming an effective resolving power 1.5 times higher than the nominal JWST/NIRSpec value for compact sources \citep[e.g.,][]{de_graaff_ionised_2024}.
We then normalize the spectra by the $\rm  H\beta$ flux divided by a factor of $(1+z)$ for the redshift correction, following the works of \cite{isobe_jades_2025} and \cite{nishigaki_dreamsii_2025}. We generate bootstrap realizations by randomly resampling the input spectra with replacement and recomputing the wavelength-by-wavelength median stack. In each realization, we also perturb the flux of every input spectrum according to its error spectrum. At each wavelength, we adopt the median of the resulting bootstrap-stacked spectra as the fiducial stacked spectrum and their 16th--84th percentile range as the bootstrap uncertainty. Thus, the reported line fluxes, including N\,{\sc iv}], are measured from the bootstrap-median spectrum, while the bootstrap uncertainty accounts for both measurement fluctuations and sample variance. Separately from the bootstrap, we perform Monte Carlo (MC) simulations in which we perturb each spectrum's flux according to its measurement uncertainties and recompute the median stack many times. The resulting distributions from MC simulations yield the measurement-error spectra of the median stacks. To evaluate the signal-to-noise ratio of emission lines in the stacked spectra, we use the propagated measurement errors, while we adopt the bootstrap-derived errors when fitting the lines. The number of objects in each mass bin is summarized in Table~\ref{tab:stack_prop}. We hereafter refer to the grating and prism-resolution stacks as the ``grating stack'' and ``prism-resolution stack,'' respectively, and name the stellar mass bins GM1, GM2, GM3, and GM4 (PM1, PM2, PM3, and PM4) in order of increasing stellar mass. We also construct full stacks without stellar-mass binning. The GAll stack binning contains the 115 spectra with usable grating data, whereas the PAll stack binning contains all 405 spectra at prism resolution, including grating spectra convolved to that resolution. The properties of the stacked spectra are summarized in Table~\ref{tab:stack_prop}. \par

For grating data, we also conduct the stacking analysis with a continuum-subtraction process to suppress the scatter in the continuum level and to enhance the S/N of emission lines. We mask out the wavelengths around the strong and notable emission lines and then calculate the moving median of the remaining continuum to estimate the continuum level, which is then subtracted from the original spectra before stacking. The window size for the moving median is set to 60 {\AA} in the rest frame, which is sufficiently larger than the typical width of emission lines to avoid biasing the continuum estimation. \par 

\begin{deluxetable*}{lccccccc}
  \tablecaption{Galaxy subsample properties for the stacked spectra.\label{tab:stack_prop}}
  \tablewidth{0pt}
  \tablehead{\colhead{Subsample} & \colhead{$N_{\rm gal}$} & \colhead{$z$ range} & \colhead{$z_{\rm med}$} & \colhead{$\log M_\star/M_\odot$ range} & \colhead{$\log M_\star/M_\odot$} & \colhead{$\log {\rm SFR_{10}}/M_\odot~{\rm yr}^{-1}$} & \colhead{$\log {\rm SFR_{10}/SFR_{100}}$}}
  \startdata
  GM1 & 22 & $4.80$--$9.51$ & $6.59$ & $5.69$--$6.97$ & $6.63_{-0.56}^{+0.28}$ & $-0.65_{-0.63}^{+0.55}$ & $1.00_{-0.51}^{+0.00}$ \\
  GM2 & 47 & $4.98$--$9.56$ & $6.49$ & $7.00$--$7.98$ & $7.50_{-0.31}^{+0.24}$ & $0.18_{-0.98}^{+0.36}$ & $0.85_{-1.19}^{+0.15}$ \\
  GM3 & 36 & $5.01$--$8.50$ & $6.51$ & $8.02$--$8.95$ & $8.28_{-0.16}^{+0.43}$ & $0.56_{-0.77}^{+0.51}$ & $0.20_{-0.63}^{+0.72}$ \\
  GM4 & 8 & $5.13$--$9.31$ & $6.95$ & $9.11$--$9.85$ & $9.42_{-0.15}^{+0.13}$ & $1.15_{-0.86}^{+0.36}$ & $-0.19_{-0.42}^{+0.49}$ \\
  GAll & 115 & $4.80$--$9.56$ & $6.55$ & $<4.40$--$9.85$ & $7.69_{-0.81}^{+0.85}$ & $0.17_{-1.02}^{+0.68}$ & $0.77_{-1.17}^{+0.23}$ \\
  PM1 & 92 & $4.50$--$9.80$ & $6.12$ & $5.69$--$7.00$ & $6.73_{-0.41}^{+0.18}$ & $-0.31_{-0.62}^{+0.22}$ & $1.00_{-0.00}^{+0.00}$ \\
  PM2 & 182 & $4.51$--$9.57$ & $6.21$ & $7.00$--$8.00$ & $7.52_{-0.34}^{+0.29}$ & $0.18_{-0.73}^{+0.41}$ & $0.95_{-1.12}^{+0.05}$ \\
  PM3 & 98 & $4.57$--$9.65$ & $6.49$ & $8.00$--$8.96$ & $8.31_{-0.23}^{+0.36}$ & $0.42_{-0.60}^{+0.67}$ & $0.09_{-0.73}^{+0.90}$ \\
  PM4 & 30 & $4.60$--$9.68$ & $6.49$ & $9.00$--$9.98$ & $9.35_{-0.27}^{+0.22}$ & $1.07_{-0.60}^{+0.35}$ & $-0.22_{-0.27}^{+0.52}$ \\
  PAll & 405 & $4.50$--$9.80$ & $6.27$ & $<4.40$--$10.71$ & $7.59_{-0.77}^{+0.89}$ & $0.10_{-0.72}^{+0.69}$ & $0.90_{-1.25}^{+0.10}$ \\
  \enddata
  \tablecomments{PAll contains all 405 spectra used in the stacking analysis, including prism spectra and grating spectra convolved to the prism resolution. GAll is the subset of 115 spectra with usable grating observations. Three PAll sources and two GAll sources lie outside the adopted stellar-mass binning range and therefore enter the full stacks but not PM1--PM4 or GM1--GM4. $\log {\rm SFR}_{10}$ represents the star-formation rate averaged over the past 10 Myr, while $\log({\rm SFR}_{10}/{\rm SFR}_{100})$ represents the ratio of the star-formation rates averaged over the past 10 and 100 Myr. By definition, $\log({\rm SFR}_{10}/{\rm SFR}_{100})$ cannot be larger than 1, at which all of the star-formation in the past 100 Myr is attributed to the star-formation in the recent 10 Myr. The $\log {\rm SFR_{10}}$ and $\log {\rm SFR_{10}/SFR_{100}}$ columns are re-aggregated from the individual-object catalog at $z_{\rm spec}>4.5$ using the same mass-bin and AGN/control masks used to define each subsample. The $\log M_\star$ column gives the stack-summary median with 16th--84th percentile ranges.}
\end{deluxetable*}

We conduct emission line flux measurements for the stacked spectra in the same manner as the individual galaxy spectra. During the fitting, we use the uncertainty obtained by the bootstrap resampling as the error spectrum to properly account for the sample variance in the stacked spectra. For the S/N calculation, we use the measurement error spectrum obtained from the Monte Carlo simulations to evaluate the significance of emission line detections in the stacked spectra. Because median stacking reduces the impact of false detections caused by shot noise, we adopt a slightly relaxed detection threshold of S/N$>3$ for the stacked spectra. We do, however, visually inspect the spectra for the artificial features. Via the visual inspection, we find that the fluxes around H$\gamma$ and [O\,{\sc iii}]\,\W4363 for the GM4 spectra are severely impacted by the wiggles caused by the small sample size, so we do not use the H$\gamma$ and [O\,{\sc iii}]\,\W4363 measurements for GM4 in the following analysis.\par

\begin{figure*}
\centering
\includegraphics[width=\linewidth]{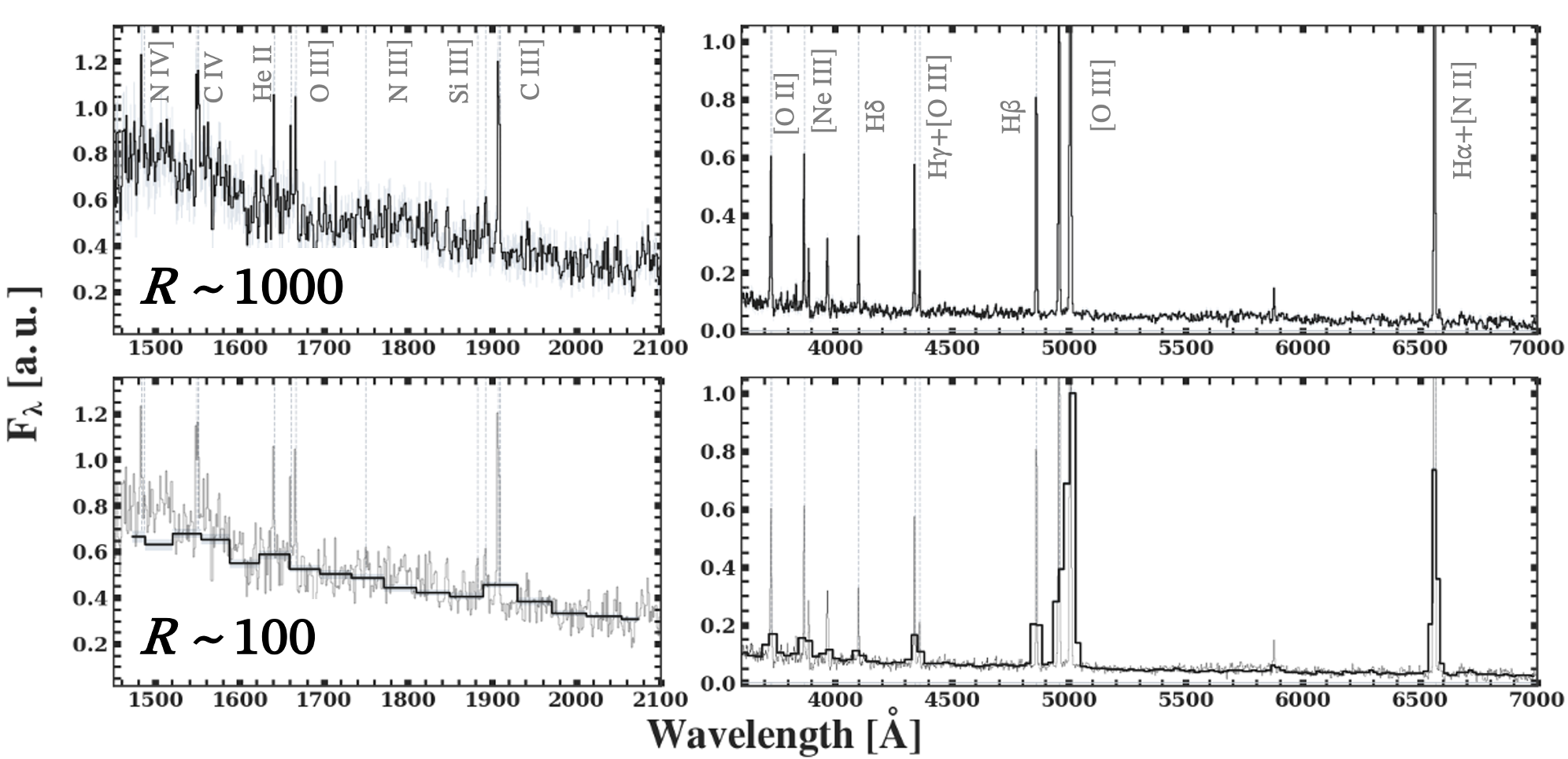}
\caption{The stacked spectra of the full star-forming galaxy sample at grating (GAll) and prism (PAll) resolutions are shown in the top and bottom panels, respectively. For both the GAll and PAll spectra, the UV (left) and optical (right) components are arbitrarily normalized for visualization. The wavelengths of key emission lines are marked by vertical dotted lines. The shaded regions indicate the $1\sigma$ uncertainties estimated through bootstrap resampling. For the $R\sim100$ PAll spectra, we overlay the corresponding $R\sim1000$ GAll spectra to confirm their consistency.}
\label{fig:stack_full}
\end{figure*}

\begin{figure*}
  \centering
  \includegraphics[width=\linewidth]{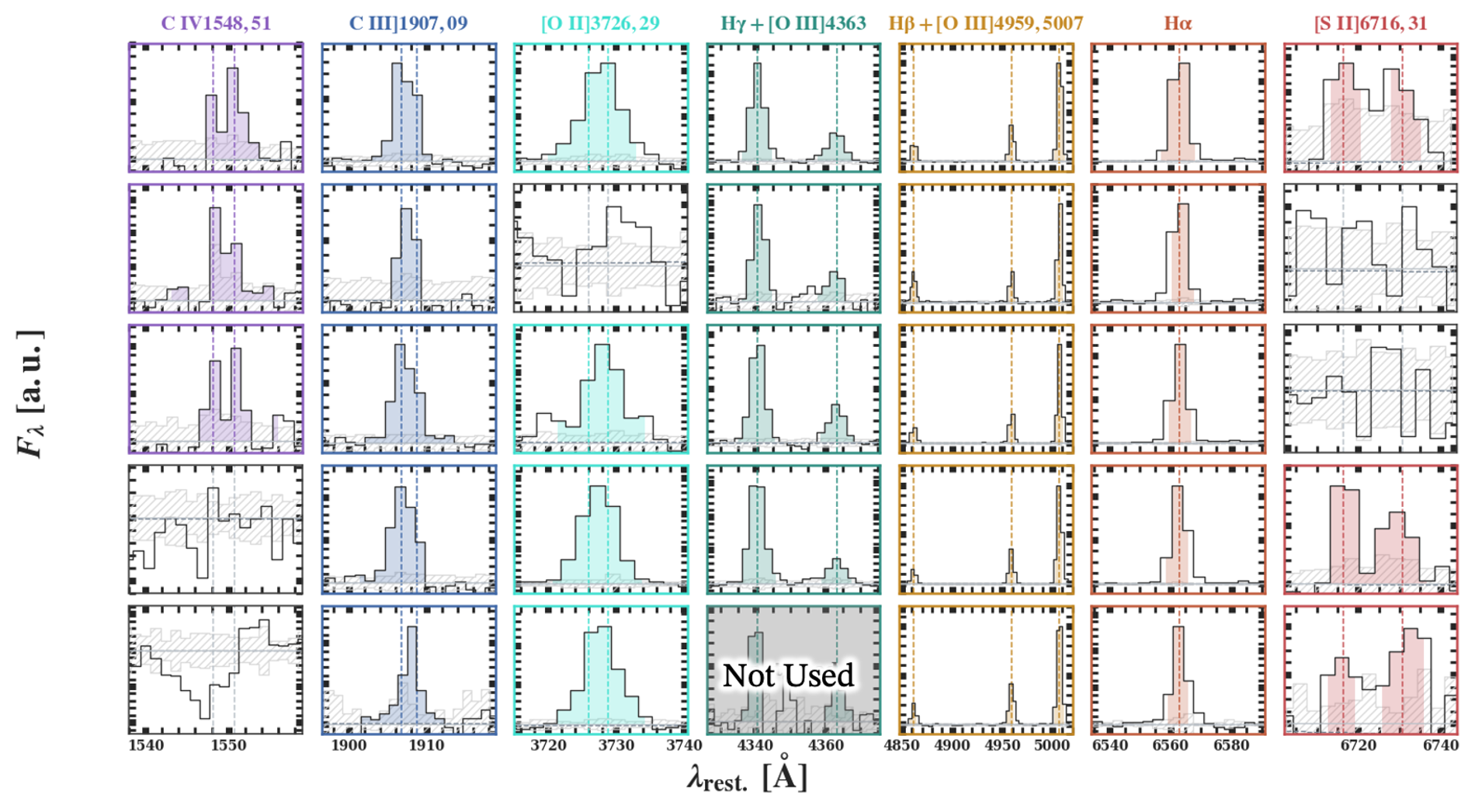}
  \caption{Zoomed-in stacked spectra around key UV and optical emission lines, including C\,{\sc iv}\,\W\W1548,1551, C\,{\sc iii}]\,\W\W1907,1909, [O\,{\sc ii}]\,$\lambda\lambda3727,3729$, H$\gamma$+[O\,{\sc iii}]\,$\lambda4363$, H$\beta$+[O\,{\sc iii}]\,$\lambda\lambda4959,5007$, H$\alpha$, and [S\,{\sc ii}]\,\W\W6716,6731. The spectra are shown in the rest frame and are arbitrarily scaled for visualization. The grey hatched regions indicate the $1\sigma$ uncertainties estimated through bootstrap resampling. The vertical dotted lines mark the central wavelengths of the emission lines. Colored and uncolored panels indicate detections and non-detections, respectively. From top to bottom, the spectra correspond to GAll, GM1, GM2, GM3, and GM4. The H$\gamma$+[O\,{\sc iii}]\,\W4363 spectrum for GM4 is excluded because of poor data quality.}
  \label{fig:stack_basic}
\end{figure*}

We present the full UV and optical stacked spectra for the unbinned full-stack subsamples in both $R\sim1000$ and $R\sim100$ resolutions (i.e., GAll and PAll) in Figure \ref{fig:stack_full}. Here, the stacked spectra based only on grating data have $R\sim1000$ because of the resampling in the stacking procedure. We do not find significant discrepancies between $R\sim1000$ and $R\sim100$ spectra in terms of the general emission line strength and the continuum shapes. In Figure \ref{fig:stack_basic}, we show the zoom-in view of the spectra around the key optical emission lines such as [O\,{\sc ii}]\,$\lambda\lambda3726,3729$, H$\beta$, H$\gamma$ + [O\,\textsc{iii}]\,$\lambda4363$, [{\rm O}\,{\sc iii}]\,$\lambda\lambda4959,5007$, and H$\alpha$ for each mass bin. In addition, we also show the C\,{\sc iv}\,\W\W1548,1551, C\,{\sc iii}]\,\W\W1907,1909, and [S\,{\sc ii}]\,\W\W6716,6731 lines that are sensitive to the electron density and ionization parameter of the H\,{\sc ii} regions. We find that the stacked spectra detect faint emission lines such as [O\,{\sc iii}]\,$\lambda4363$ even in the lowest stellar-mass subsample (i.e., GM1; representative $M_\star\simeq10^{6.6}\,M_\odot$), which allows us to constrain the characteristic electron temperature and metallicity of the ionized gas at lower masses than previous high-redshift studies \citep[e.g.,][]{nakajima_ultra-faint_2025,heintz_dilution_2023}.\par 

\begin{figure*}
  \centering
  \includegraphics[width=0.58\linewidth]{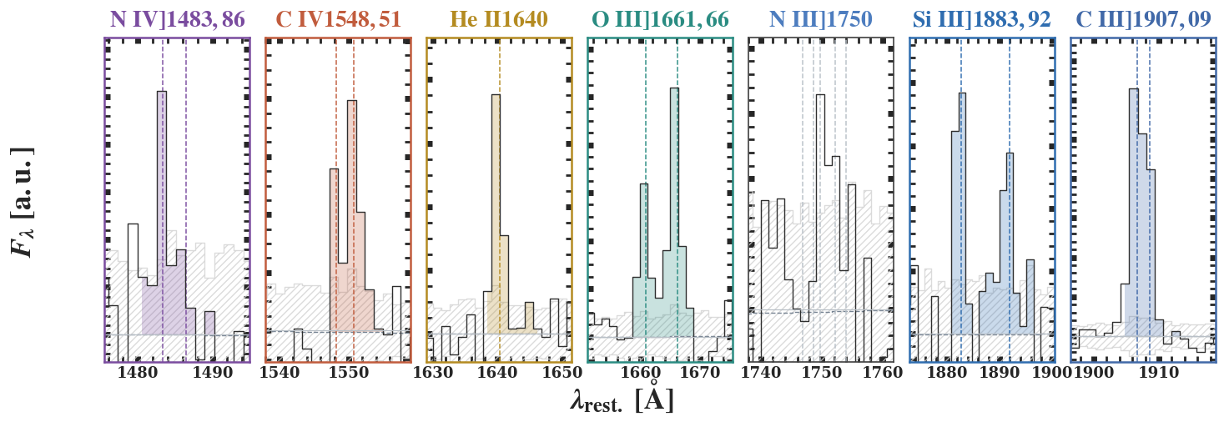}
  \hfill
  \includegraphics[width=0.40\linewidth]{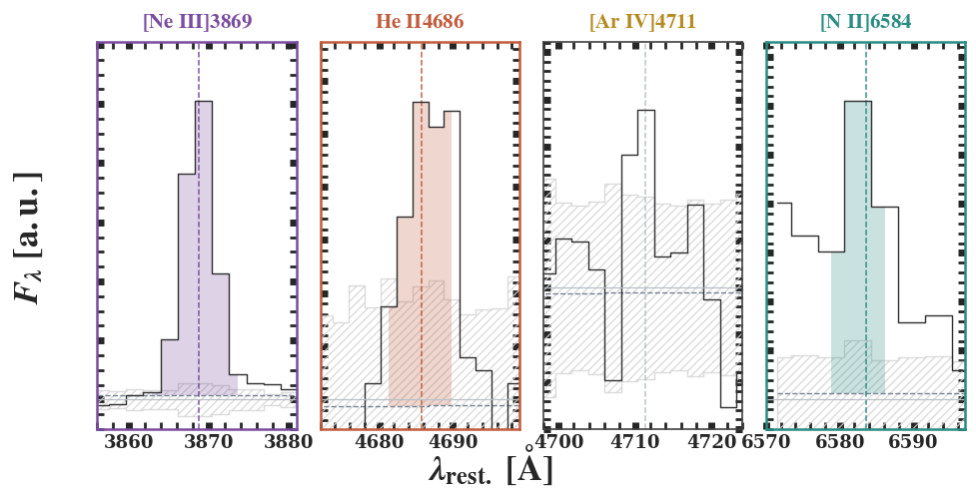}
  \caption{Zoomed-in GAll stacked spectra around key optical and UV emission lines. The left panel shows the UV spectra around N\,{\sc iv}]\,$\lambda\lambda1483,1486$, C\,{\sc iv}\,$\lambda\lambda1548,1551$, He\,{\sc ii}\,$\lambda1640$ + O\,{\sc iii}]\,$\lambda\lambda1661,1666$, the N\,{\sc iii}]\,$\lambda1750$ quintet, Si\,{\sc iii}]\,$\lambda\lambda1883,1892$, and C\,{\sc iii}]\,$\lambda\lambda1907,1909$. The right panel shows the optical spectra around [O\,{\sc ii}]\,$\lambda\lambda3726,3729$, H$\gamma$+[O\,{\sc iii}]\,$\lambda4363$, H$\beta$ + [O\,{\sc iii}]\,$\lambda\lambda4959,5007$, and H$\alpha$ + [N\,{\sc ii}]\,$\lambda\lambda6548,6583$. The symbols represent the same as in Figure \ref{fig:stack_basic}.}
  \label{fig:stack_spec}
\end{figure*}

In addition, we also measure the flux of other faint UV and optical lines to constrain the chemical abundances of the faint galaxies. We only consider the GAll stack for the nitrogen, carbon, argon, and neon abundance measurement because of the sensitivity limitation. Within the GAll stack, we detect the UV emission lines such as N\,{\sc iv}]\,$\lambda\lambda1483,1486$, C\,{\sc iv}\,$\lambda\lambda1548,1551$, He\,{\sc ii}\,$\lambda1640$ + O\,{\sc iii}]\,$\lambda\lambda1661,1666$, Si\,{\sc iii}]\,\W\W1883,1892, and C\,{\sc iii}]\,$\lambda\lambda1907,1909$ in the stacked spectra, which provide valuable diagnostics of the ionizing sources and chemical abundances in these galaxies. The zoomed-in views of the UV emission lines are shown in Figure \ref{fig:stack_spec}. The optical lines [Ne\,{\sc iii}]\,\W3869, He\,{\sc ii}\,\W4686, and [N\,{\sc ii}]\,\W6583 are also detected in the stacked spectra. In contrast, [Ar\,{\sc iv}]\,\W4711 is not detected, and we report a 3$\sigma$ upper limit on its flux, which is propagated into an upper limit on Ar/O. The measured fluxes of the emission lines in the stacked spectra are summarized in Table~\ref{tab:stackall_flux}. 

  \begin{deluxetable*}{lccccc}
  \tablecaption{H$\beta$-normalized emission-line fluxes for the grating stacks.\label{tab:gratstack_hbeta_lines}}
  \tabletypesize{\scriptsize}
  \tablewidth{0pt}
  \tablehead{
  \colhead{Line} & \colhead{GM1} & \colhead{GM2} & \colhead{GM3} & \colhead{GM4} & \colhead{GAll}\\
  }
  \startdata
  C\,{\sc iv}$\lambda\lambda1548,1551$ & $0.84_{-0.45}^{+0.42}$ & $0.32_{-0.12}^{+0.12}$ & $<0.23$ & $<0.27$ & $0.19_{-0.06}^{+0.06}$\\
  C\,{\sc iii}]$\lambda\lambda1907,1909$ & $0.53_{-0.19}^{+0.20}$ & $0.52_{-0.06}^{+0.06}$ & $0.43_{-0.05}^{+0.05}$ & $0.26_{-0.08}^{+0.09}$ & $0.45_{-0.03}^{+0.03}$\\
  {[O\,{\sc ii}]$\lambda\lambda3726,3729$} & $<0.21$ & $0.39_{-0.05}^{+0.05}$ & $0.96_{-0.12}^{+0.12}$ & $1.38_{-0.32}^{+0.32}$ & $0.60_{-0.07}^{+0.08}$\\
  H$\gamma$ & $0.42_{-0.05}^{+0.05}$ & $0.46_{-0.03}^{+0.03}$ & $0.46_{-0.03}^{+0.02}$ & $0.35_{-0.08}^{+0.08}$ & $0.44_{-0.02}^{+0.02}$\\
  {[O\,{\sc iii}]$\lambda4363$} & $0.12_{-0.04}^{+0.04}$ & $0.17_{-0.02}^{+0.02}$ & $0.11_{-0.02}^{+0.02}$ & $0.13_{-0.04}^{+0.05}$ & $0.15_{-0.01}^{+0.01}$\\
  H$\beta$ & $1.00_{-0.06}^{+0.06}$ & $1.00_{-0.03}^{+0.03}$ & $1.00_{-0.03}^{+0.03}$ & $1.00_{-0.10}^{+0.10}$ & $1.00_{-0.02}^{+0.02}$\\
  {[O\,{\sc iii}]$\lambda4959$} & $1.12_{-0.09}^{+0.09}$ & $1.67_{-0.08}^{+0.08}$ & $2.28_{-0.07}^{+0.07}$ & $2.13_{-0.17}^{+0.16}$ & $1.84_{-0.04}^{+0.04}$\\
  {[O\,{\sc iii}]$\lambda5007$} & $3.34_{-0.26}^{+0.27}$ & $4.99_{-0.24}^{+0.23}$ & $6.80_{-0.20}^{+0.21}$ & $6.35_{-0.51}^{+0.48}$ & $5.49_{-0.13}^{+0.13}$\\
  H$\alpha$ & $2.86_{-0.21}^{+0.21}$ & $2.72_{-0.11}^{+0.10}$ & $3.14_{-0.12}^{+0.13}$ & $2.79_{-0.81}^{+0.95}$ & $2.98_{-0.08}^{+0.08}$\\
  {[S\,{\sc ii}]$\lambda\lambda6716,6731$} & $<0.27$ & $<0.10$ & $0.23_{-0.06}^{+0.06}$ & $0.59_{-0.21}^{+0.25}$ & $0.08_{-0.03}^{+0.03}$\\
  \enddata
  \tablecomments{Entries are observed flux ratios relative to H$\beta$ with uncertainties given by 16th/84th percentiles. A detection requires ${\rm S/N}_{\rm meas,det} \ge 3$ for the grating stacks. Non-detections are reported as 3$\sigma$ upper limits.}
\end{deluxetable*}

  \begin{deluxetable*}{lc}
  \tablecaption{H$\beta$-normalized emission-line fluxes for GAll.\label{tab:stackall_flux}}
  \tabletypesize{\scriptsize}
  \tablewidth{0pt}
  \tablehead{
  \colhead{Line} & \colhead{GAll}
  }
  \startdata
  N\,{\sc iv}]$\lambda\lambda1483,1486$ & $0.12_{-0.04}^{+0.05}$\\
  C\,{\sc iv}$\lambda\lambda1548,1551$ & $0.19_{-0.06}^{+0.06}$\\
  He\,{\sc ii}$\lambda1640$ & $0.12_{-0.03}^{+0.04}$\\
  O\,{\sc iii}]$\lambda\lambda1661,1666$ & $0.28_{-0.04}^{+0.04}$\\
  N\,{\sc iii}]$\lambda1750$ & \nodata\\
  Si\,{\sc iii}]$\lambda\lambda1883,1892$ & $0.14_{-0.03}^{+0.03}$\\
  C\,{\sc iii}]$\lambda\lambda1907,1909$ & $0.45_{-0.03}^{+0.03}$\\
  {[O\,{\sc ii}]$\lambda\lambda3726,3729$} & $0.60_{-0.07}^{+0.08}$\\
  {[Ne\,{\sc iii}]$\lambda3869$} & $0.41_{-0.03}^{+0.03}$\\
  H$\gamma$ & $0.44_{-0.03}^{+0.03}$\\
  {[O\,{\sc iii}]$\lambda4363$} & $0.15_{-0.01}^{+0.01}$\\
  He\,{\sc ii}$\lambda4686$ & $0.03_{-0.01}^{+0.01}$\\
  {[Ar\,{\sc iv}]$\lambda4711$} & $<0.03$\\
  H$\beta$ & $1.00_{-0.02}^{+0.02}$\\
  {[O\,{\sc iii}]$\lambda4959$} & $1.84_{-0.04}^{+0.04}$\\
  {[O\,{\sc iii}]$\lambda5007$} & $5.49_{-0.13}^{+0.13}$\\
  H$\alpha$ & $2.98_{-0.08}^{+0.08}$\\
  {[N\,{\sc ii}]$\lambda6583$} & $0.09_{-0.02}^{+0.02}$\\
  {[S\,{\sc ii}]$\lambda\lambda6716,6731$} & $0.08_{-0.02}^{+0.03}$\\
  \enddata
  \tablecomments{Entries are observed flux ratios relative to H$\beta$ with uncertainties given by 16th/84th percentiles. A detection requires S/N$_{\rm meas,det}\ge3$ for the grating stacks. Non-detections are reported as 3$\sigma$ upper limits. The [S\,{\sc ii}] row uses the summed flux of $\lambda6716+\lambda6731$; the [Ar\,{\sc iv}] row uses only $\lambda4711$.}
\end{deluxetable*}

\section{Results}\label{sec:res}
In the following analysis, we investigate the stellar population properties, ionization structure, electron density, and gas-phase metallicity of our galaxy sample. We also derive the C/O, N/O, and Ne/O abundance ratios, place an upper limit on Ar/O, and examine their dependence on stellar mass and gas-phase metallicity. Finally, we compare our results with previous measurements at similar and lower redshifts and discuss the implications for galaxy evolution in the early universe. We propagate the uncertainties in the emission line fluxes and SED fitting to measure physical parameters using a Monte Carlo approach, which allows us to account for the uncertainties in the measurements of observables and to derive robust estimates of the physical properties of our galaxy sample.

\subsection{Stellar Population Properties}\label{sec:res:sp}
\begin{figure}[htbp]
  \centering
  \includegraphics[width=\linewidth]{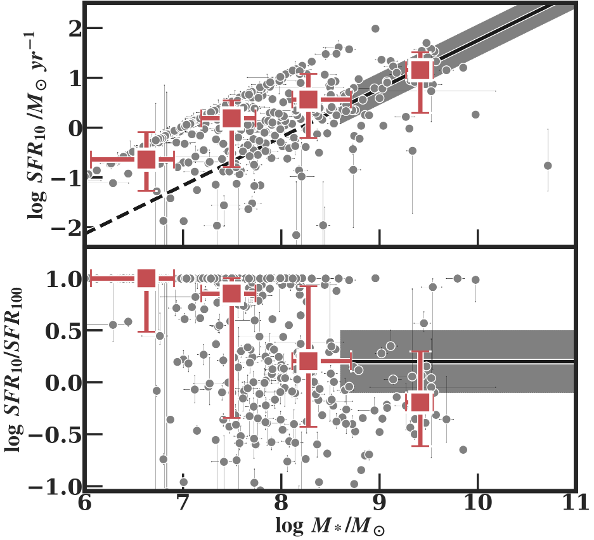}
  \caption{(Top): The stellar mass and star-formation rate relation for our galaxy sample. We adopt the star-formation within the last 10~Myr as the SFR, which is derived from the spectro-photometric SED fitting. (Bottom): The stellar mass and burstiness (i.e., $\rm \log SFR_{10}/SFR_{100}$) measurements for our sample.
  In both panels, black circles represent the measurements for the individual galaxies, whereas the red squares represent the representative value for our grating stacked spectra. The grey solid lines and grey regions represent the inferred relation and intrinsic scatter of main-sequence (top) or burstiness (bottom) for the $5<z<6$ galaxies from AURORA and JADES survey with the stellar mass at $M_\star>10^{8.5}~M_\odot$. \citep{clarke_star-forming_2025}}
\label{fig:sfms}
\end{figure}

We first examine the star-formation main sequence (SFMS) of our sample to assess whether the galaxies in our sample are representative of typical star-forming galaxies at their respective redshifts. We plot the stellar mass and star-formation rate (SFR) relation for our galaxy sample in Figure~\ref{fig:sfms}. We adopt the star formation over the last 10 Myr as the SFR, which is derived from the spectro-photometric SED fitting. The black circles represent the measurements for the individual galaxies, while the large black diamonds represent the median values of the individual measurements in each stellar mass bin. The grey solid line represents the SFMS regression derived from on photometric and spectroscopic data of $5<z<6$ galaxies with stellar mass above $10^{8.5}~M_\odot$ in AURORA and JADES surveys \citep{clarke_star-forming_2025}. We find that our sample broadly follows the SFMS relation at the stellar masses above $10^{8.5}~M_\odot$. We do see the increasing offset from the main sequence relation derived at $>10^{8.5}~M_\odot$ towards the lower stellar mass, as it has been found in \cite{clarke_star-forming_2025}. Many of the galaxies at $<10^{8.5}~M_\odot$ lie at the boundary that corresponds to the case where stars formed in the last 10 Myr account for the total stellar mass, indicating their very bursty nature. As shown in Figure \ref{fig:sfms}, we also check the burstiness of our sample by comparing the SFR within the last 10~Myr to the SFR averaged over the last 100~Myr. We find that the burstiness at $\log M_\star/M_\odot<8$ is generally above unity with increasing trend towards the lower stellar mass. The lowest-mass stacked spectrum with a representative stellar mass of $\log(M_\star/M_\odot)=6.63$ has the $\mathrm{\log SFR_{10}/SFR_{100}}\simeq1$, indicating that these faint systems are dominated by young stellar populations. Overall, our sample at the low stellar mass regime have generally higher contribution from the stellar population born within 10 Myr. Therefore, in the following analysis, we assume that the ionizing radiation from our galaxies is primarily powered by star formation, dominated by massive stars (i.e., O-type stars) with lifetimes below $\sim 10$ Myr.

\subsection{Ionization Structure}\label{sec:res:ion}

\begin{figure}[htbp]
  \centering
  \begin{subfigure}{\linewidth}
    \centering
    \includegraphics[width=0.95\linewidth]{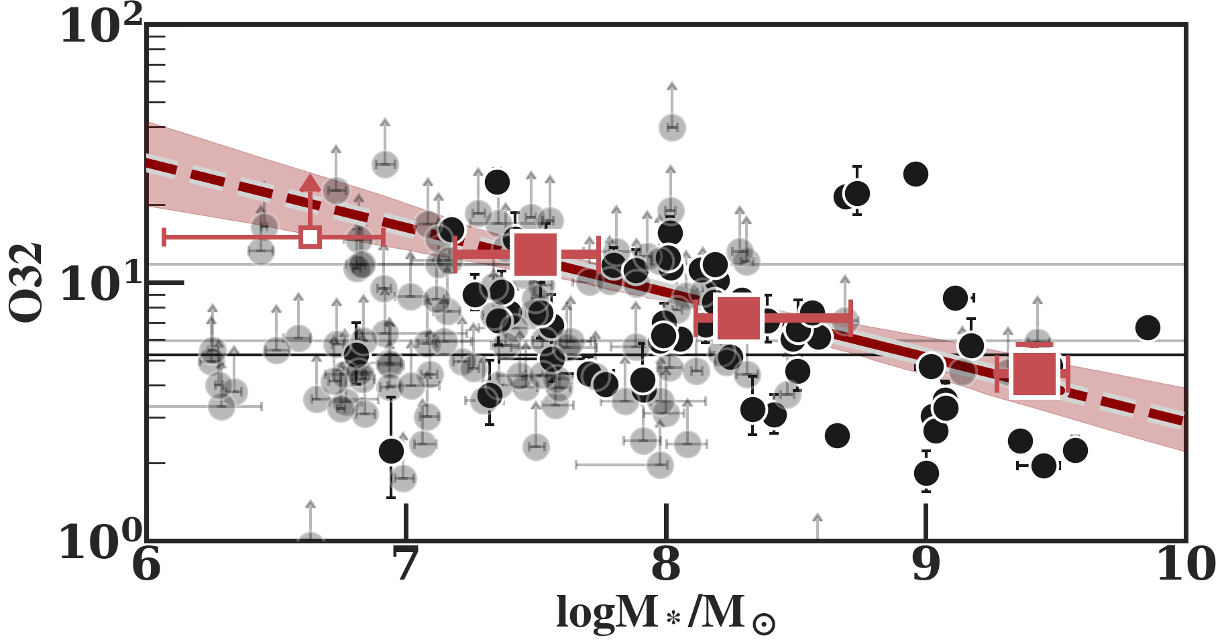}
    \subcaption{$O_{32}$.}
    \label{fig:o32}
  \end{subfigure}

  \vspace{0.5em}

  \begin{subfigure}{\linewidth}
    \centering
    \includegraphics[width=0.95\linewidth]{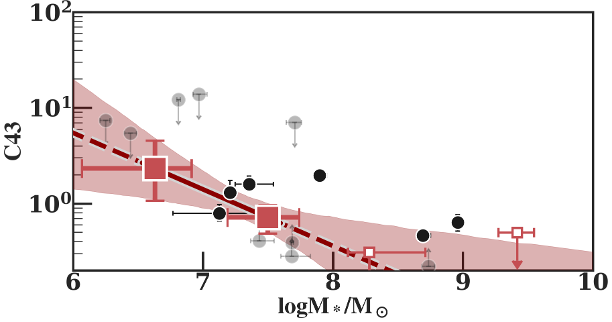}
    \subcaption{$C_{43}$.}
    \label{fig:c43}
  \end{subfigure}
  \caption{Stellar-mass dependence of (a) $O_{32}$ and (b) $C_{43}$. In panel (a), the solid black dots represent individual galaxies with detections of [O\,{\sc ii}]\,$\lambda\lambda3726,3729$ and [O\,{\sc iii}]\,$\lambda\lambda4959,5007$, while the transparent black dots represent galaxies with non-detections of [O\,{\sc ii}]\,$\lambda\lambda3726,3729$. In panel (b), the solid black dots represent individual galaxies with detections of C\,{\sc iv}\,$\lambda\lambda1548,1551$ and C\,{\sc iii}]\,$\lambda\lambda1907,1909$, while the transparent black dots represent galaxies with a non-detection of either C\,{\sc iv}\,$\lambda\lambda1548,1551$ or C\,{\sc iii}]\,$\lambda\lambda1907,1909$. In both panels, the red squares represent measurements from the stacked spectra of the grating sample, and the red lines show the best-fit linear relations between stellar mass and the logarithmic line ratio based on the stacked-spectrum measurements. The solid and dashed portions of the lines indicate the fitting and non-fitting ranges, respectively.}
  \label{fig:ionization_mass}
\end{figure}

Before deriving chemical abundances, we first assess the ionization structure of the interstellar medium by comparing emission-line ratios that probe different ionization states of the same element. We consider the $O_{32}$ and $C_{43}$ ratios, which are sensitive to the ionization parameters at relatively low and high ionized regions of the ISM, respectively.
The $O_{32}$ ratio is defined as
\begin{equation}
  O_{32} = \frac{\mathrm{[O\,\textsc{iii}]}\,\lambda5007}{\mathrm{[O\,\textsc{ii}]}\,\lambda\lambda3726,3729},
\end{equation}
and the $C_{43}$ ratio is defined as
\begin{equation}
  C_{43} = \frac{\mathrm{C\,\textsc{iv}}\,\lambda\lambda1548,1551}{\mathrm{C\,\textsc{iii}]}\,\lambda\lambda1907,1909}.
\end{equation}
We measure $O_{32}$ and $C_{43}$ for both individual and stacked spectra. We show the stellar-mass dependence of $O_{32}$ and $C_{43}$ in Figure~\ref{fig:ionization_mass}, panels~\subref{fig:o32} and \subref{fig:c43}, respectively. We summarize the fitting parameters in Table \ref{tab:fit}. We also perform linear regression between the stellar mass and $O_{32}$ and $C_{43}$ using the measurements of the $R\sim1000$ stacked spectra. We find that the $\log O_{32}$ can be expressed as $\log O_{32}=2.89\pm0.27-(0.243\pm0.004)\times\log M_\star/M_{\odot}$ and $\log C_{43}=4.33\pm8.26-(0.60\pm0.15)\times\log M_\star/M_{\odot}$. We generally find a higher ionization parameter at lower stellar mass, which is consistent with the findings of local extremely metal-poor galaxies \citep[e.g.,][]{kojima_extremely_2020}. We summarize the measurements of $O_{32}$ and $C_{43}$ for the stacked spectra in Table \ref{tab:stack_abundance}.

\begin{deluxetable*}{lccccccc}
  \tablecaption{Derived quantities for the grating stacks.\label{tab:stack_abundance}}
  \tabletypesize{\scriptsize}
  \tablewidth{0pt}
  \tablehead{
  \colhead{Stack} & \colhead{12+log(O/H)$_{\rm T_e}$} & \colhead{12+log(O/H)$_{Emp.}$} & \colhead{$O_{32}$} & \colhead{$C_{43}$} & \colhead{$\log~n_e($[S\,{\sc ii}])\,$/\mathrm{cm}^{-3}$} & \colhead{$\log~n_e($C\,{\sc iii}])\,$/\mathrm{cm}^{-3}$} & \colhead{$E(B-V)$}\\
  }
  \startdata
  GM1 & $7.27_{-0.14}^{+0.21}$ & $7.27_{-0.07}^{+0.06}$ & $>15.01$ & $1.45_{-0.73}^{+1.51}$ & $3.19_{-0.63}^{+0.73}$ & $4.75_{-0.65}^{+0.54}$ & $0.01_{-0.01}^{+0.07}$\\
  GM2 & $7.50_{-0.08}^{+0.08}$ & $7.51_{-0.03}^{+0.03}$ & $12.86_{-1.47}^{+2.17}$ & $0.61_{-0.21}^{+0.24}$ & \nodata & $3.77_{-0.46}^{+0.43}$ & $0.00_{-0.00}^{+0.00}$\\
  GM3 & $8.02_{-0.09}^{+0.10}$ & $7.90_{-0.07}^{+0.05}$ & $7.27_{-0.84}^{+1.06}$ & $<0.22$ & $2.87_{-0.63}^{+0.59}$ & $3.60_{-0.49}^{+0.36}$ & $0.10_{-0.04}^{+0.04}$\\
  GM4 & \nodata & $8.06_{-0.11}^{+0.08}$ & $4.44_{-0.79}^{+1.33}$ & $<-0.07$ & $3.25_{-0.59}^{+0.61}$ & $4.82_{-0.64}^{+0.47}$ & $0.00_{-0.00}^{+0.28}$\\
  GAll & $7.66_{-0.04}^{+0.06}$ & $7.67_{-0.04}^{+0.05}$ & $9.00_{-1.12}^{+1.31}$ & $0.41_{-0.15}^{+0.15}$ & $3.12_{-0.67}^{+0.64}$ & $3.66_{-0.61}^{+0.39}$ & $0.05_{-0.03}^{+0.03}$\\
  \enddata
  \tablecomments{For measurements based on detected emission lines, we report the median and the 16th/84th percentiles as the best-fit value and its uncertainty, respectively. Otherwise, the upper/lower limits correspond to the 16/84-th percentiles values, respectively.}
\end{deluxetable*}

\subsection{Electron Density}\label{sec:res:ne}
\begin{figure}
  \centering
  \includegraphics[width=\linewidth]{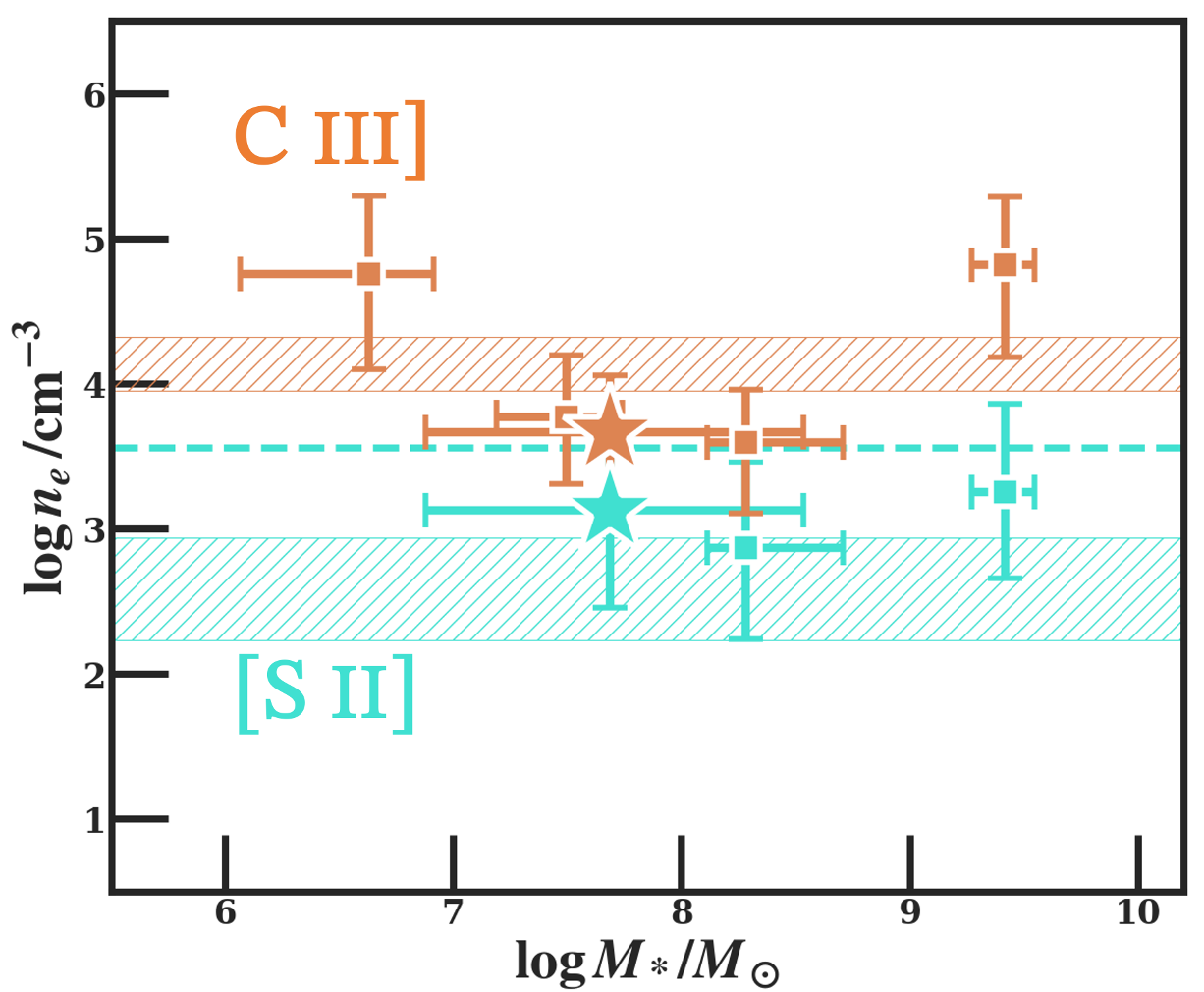}
  \caption{Electron density inferred from the [S\,{\sc ii}]\,\W\W6716,6731 and C\,{\sc iii}]\,\W\W1907,1909. The turquoise and orange symbols represent the electron density measurements based on the [S\,{\sc ii}] and  C\,{\sc iii}] ratios as a function of stellar mass, respectively. The squares and stars represent the measurements from the stacked spectra for each stellar mass bin (i.e., GM1 to GM4) and the all-galaxy stacked spectrum (i.e., GAll), respectively. The shaded regions indicate the literature measurements of electron density at similar redshifts from the AURORA survey \citep{topping_aurora_2025}. The dashed turquoise line indicates the critical density for [S\,{\sc ii}]\,\W6731, where the [S\,{\sc ii}] doublet line ratio starts to saturate. We note that the critical density for the C\,{\sc iii}]\,\W1909 line is around $10^8$~cm$^{-3}$ for the typical electron temperature in the H\,{\sc ii} region, which is much higher than the electron density of our sample.}
  \label{fig:ne}
\end{figure}

The electron density of the ISM is also a key parameter for the chemical abundance measurement.
The electron densities can be derived from the line ratios of the forbidden lines because the collisional de-excitation rate depends on the electron density, which affects the relative strength of the emission lines. Commonly used diagnostics include the [S\,{\sc ii}]\,$\lambda6716/\lambda6731$ and the C\,{\sc iii}]\,$\lambda1907/\lambda1909$ ratio. Because ${\rm S}^{+}$ and ${\rm C}^{+2}$ have ionization potentials of 10.36~eV and 24.38~eV, respectively, the electron density derived from these diagnostics probes different ionization zones of the ISM. We derive the electron density using the line ratios of [S\,{\sc ii}]\,$\lambda6716/\lambda6731$  and C\,{\sc iii}]\,$\lambda1907/\lambda1909$. We use \texttt{pyneb} to compute the electron density from these line ratios, assuming an electron temperature of $T_{\rm e}=10^4$~K.
We show the electron density measurements from the [S\,{\sc ii}] and C\,{\sc iii}] diagnostics as a function of stellar mass in Figure~\ref{fig:ne}. We also compare the electron number density measurements derived for the stacked spectra from the grating sample.
We find that the electron density derived from the [S\,{\sc ii}] and C\,{\sc iii}] ratios are generally around $n_{\rm e}\sim 10^3$~cm$^{-3}$ and $\sim 10^4$~cm$^{-3}$, which are below the critical densities of the respective diagnostics, and that electron density of the ${\rm C}^{+2}$-abundant region is generally higher than that of the ${\rm S}^{+}$-abundant region. These values are consistent with the measurements of $2<z<10$ galaxies from the AURORA survey \cite{topping_aurora_2025} \citep[see also][]{isobe_redshift_2023}.
Because we do not find a significant correlation between the electron density and stellar mass, we adopt a formulation of electron density as a function of redshift presented in \cite{topping_aurora_2025} for the chemical abundance measurement.

\subsection{Gas-Phase Metallicity} \label{sec:res:metal}
\subsubsection{Direct-Method}\label{sec:res:metal:direct}
We derive gas-phase oxygen abundances using the direct ($T_{\rm e}$) method, following the same approach as in previous JWST studies. The electron temperature of the high-ionization zone, $T_{\rm e}(\mathrm{[O\,\textsc{iii}]})$, is estimated from the auroral-to-nebular line ratio, $\mathrm{[O\,\textsc{iii}]}\,\lambda4363/\lambda5007$, when available \citep{osterbrock_astrophysics_2006}. We adopt the electron density of the ${\rm O}^{+2}$ abundant region from the empirical electron density derived from $\mathrm{C\,\textsc{iii}]}\,\lambda\lambda1907,1909$ ratios in \cite{topping_aurora_2025}. We correct the observed line fluxes for dust attenuation using the Balmer decrement (e.g., H$\alpha$/H$\beta$) when at least two Balmer lines are detected, and adopting the \cite{calzetti_dust_2000} attenuation curve.

The electron temperature in the low-ionization zone, $T_{\rm e}(\mathrm{[O\,\textsc{ii}]})$, is required to estimate the $\mathrm{O^+/H^+}$ abundance. We adopt the commonly used empirical relation \citep[e.g.,][]{campbell_stellar_1986,garnett_electron_1992} between $T_{\rm e}(\mathrm{[O\,\textsc{ii}]})$ and $T_{\rm e}(\mathrm{[O\,\textsc{iii}]})$:
\begin{equation}
T_{\rm e}(\mathrm{[O\,\textsc{ii}]}) = 0.7 \times T_{\rm e}(\mathrm{[O\,\textsc{iii}]}) + 3000~\mathrm{K}.
\end{equation}

Ionic abundances are computed using \texttt{pyneb}, assuming a two-zone ionization structure. The $\mathrm{O^{+2}/H^+}$ abundance is derived from the dust-corrected $\mathrm{[O\,\textsc{iii}]}\,\lambda5007/\mathrm{H}\beta$ ratio using $T_{\rm e}(\mathrm{[O\,\textsc{iii}]})$, while $\mathrm{O^{+}/H^+}$ is estimated from $\mathrm{[O\,\textsc{ii}]}\,\lambda\lambda3726,3729/\mathrm{H}\beta$ with $T_{\rm e}(\mathrm{[O\,\textsc{ii}]})$. The total oxygen abundance is then obtained by summing the two major ionic components:
\begin{equation}
\frac{\mathrm{O}}{\mathrm{H}} = \frac{\mathrm{O^{+}}}{\mathrm{H^{+}}} + \frac{\mathrm{O^{+2}}}{\mathrm{H^{+}}}.
\end{equation}
We neglect contributions from higher ionization states (e.g., $\mathrm{O^{+3}}$), which may introduce a systematic uncertainty for systems with hard ionizing spectra.
The resulting metallicities span $12+\log(\mathrm{O/H}) \sim 7.1$--$8.2$, indicating that our sample probes predominantly metal-poor galaxies at high redshift.

\subsubsection{Strong-Line Calibrations}\label{sec:res:metal:strong}
\begin{figure*}[htbp]
  \centering
  \begin{minipage}{0.4\linewidth}
    \centering
    \includegraphics[width=\linewidth]{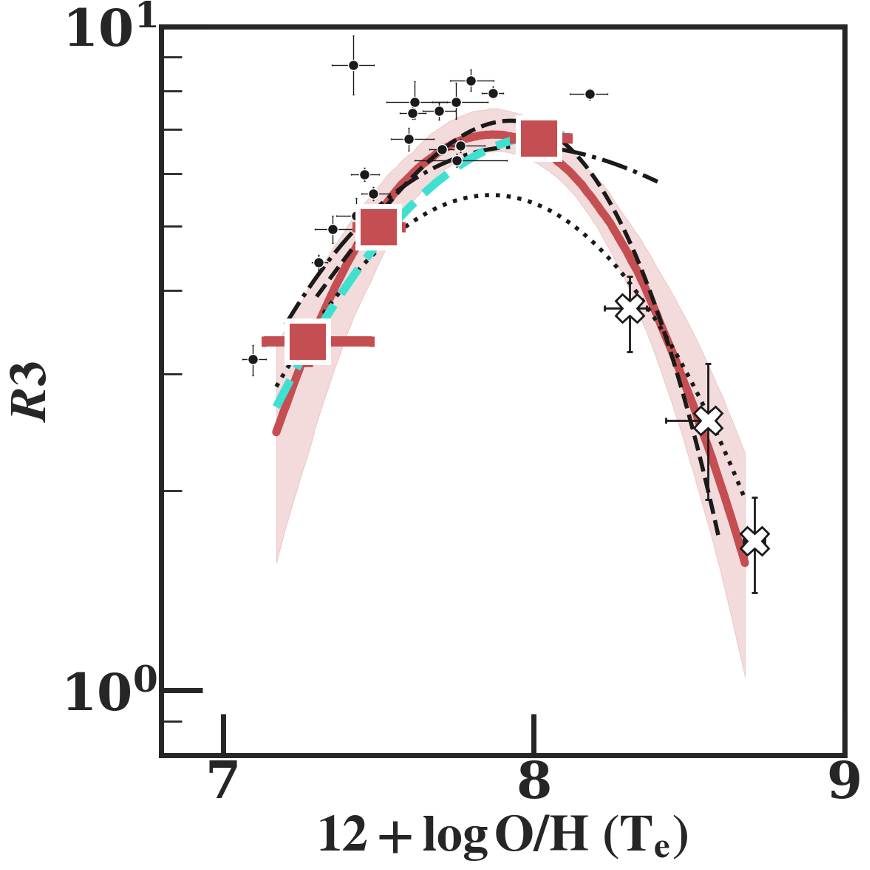}
    \par\centering (a)
  \end{minipage}\hfill
  \begin{minipage}{0.4\linewidth}
    \centering
    \includegraphics[width=\linewidth]{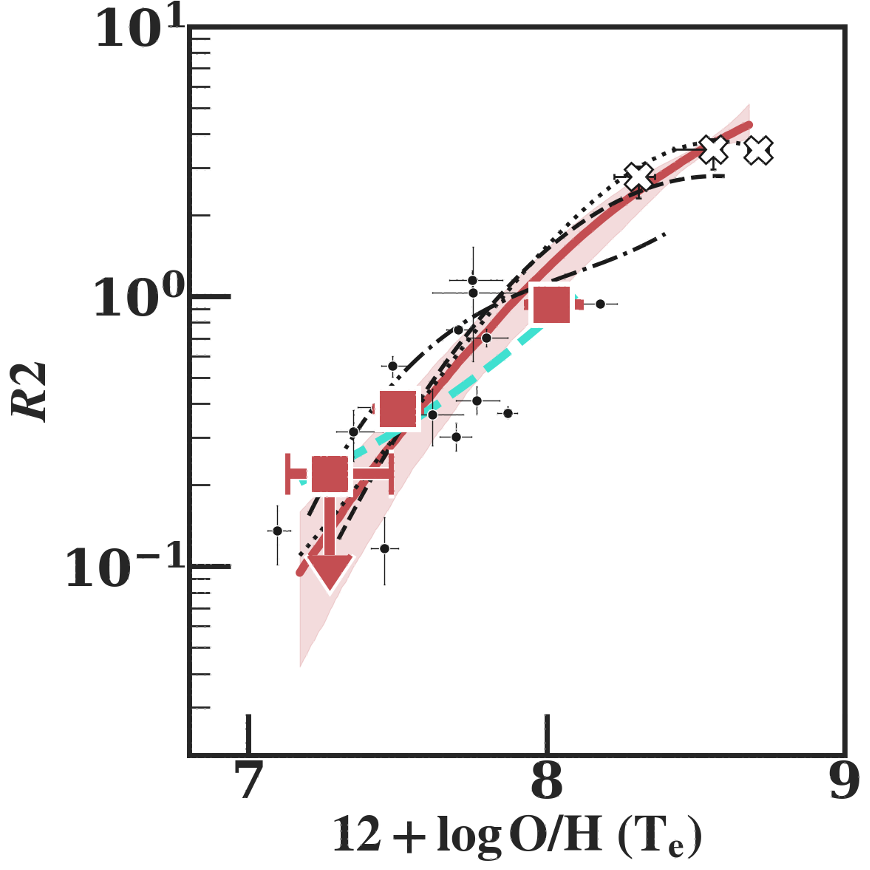}
    \par\centering (b)
  \end{minipage}
  \caption{Empirical calibrations between strong-line ratios and direct-method metallicity. (a) The relation between $R3$ and $12+\log({\rm O/H})$. (b) The relation between $R2$ and $12+\log({\rm O/H})$. The red squares and black circles represent measurements from our grating stacked spectra and individual galaxies with auroral-line detections, respectively. The white crosses represent measurements from the stacked spectra of local high-specific-star-formation galaxies from \cite{andrews_mass-metallicity_2013}. The solid red lines represent the best-fit polynomial relations, and the shaded regions indicate their 16th--84th-percentile uncertainties. The dotted, dashed, and dash-dotted black lines represent the empirical calibrations of \cite{isobe_jades_2026-1}, \cite{sanders_aurora_2026}, and \cite{chakraborty_unveiling_2025}, respectively. The turquoise dashed line represents the calibration of \cite{nakajima_empress_2022} for $z\sim0$ high-redshift analog galaxies with high-equivalent-width emission lines.
  }
  \label{fig:metal_calib}
\end{figure*}

\begin{figure}[htbp]
  \centering
  \includegraphics[width=0.8\linewidth]{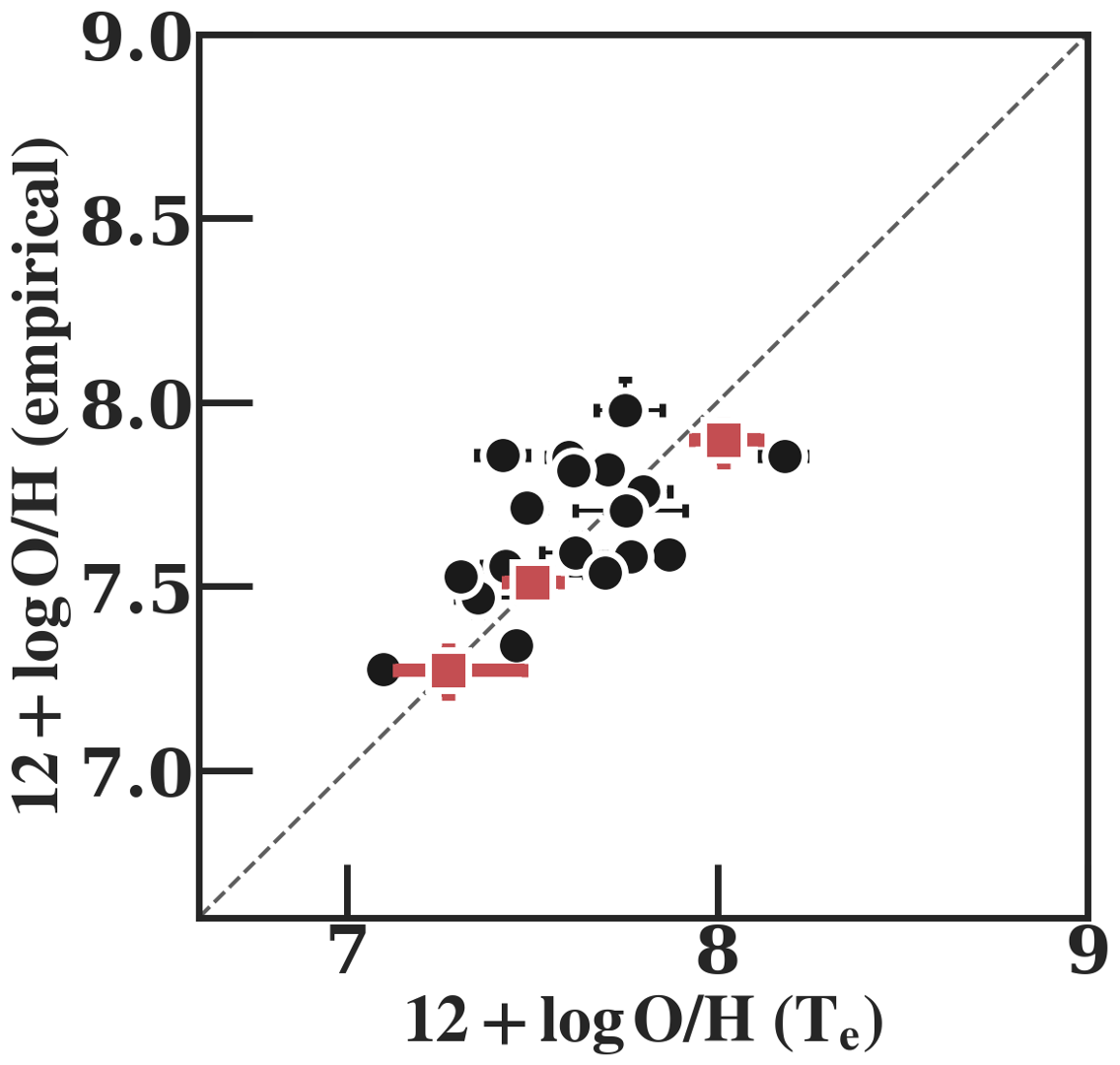}
  \caption{The comparison between the gas-phase metallicities measured from the direct method and the empirical strong line relations. The black dots represent the measurements for the grating galaxy spectra with [O\,{\sc ii}]\,$\lambda\lambda3726,3729$, [O\,{\sc iii}]\,$\lambda4363$, H$\beta$, and [O\,{\sc iii}]\,$\lambda\lambda4959,5007$ detections. The red squares represent the measurements for the stacked spectra from the grating sample. The gray dashed line corresponds to where the measurements from direct and empirical strong line methods are equal.}
  \label{fig:direct_ratio}
\end{figure}

For the subset of galaxies without auroral line detections, we estimate gas-phase metallicities using strong-line calibrations. We use $R3$ and $R2$ diagnostics, which are sensitive to metallicity and ionization parameter, respectively. The $R3$ ratio is defined as
\begin{equation}
  R3 = \mathrm{[O\,\textsc{iii}]}\,\lambda5007/{\mathrm{H}\beta}.
\end{equation}
Because the $R3$ diagnostic is double-valued with respect to metallicity, we also use the $R2$ ratio (i.e., [O\,{\sc ii}]\,\W\W3726,3729/H$\beta$) to break the degeneracy, motivated by the empirical relation between $R2$ and $12+\log(\mathrm{O/H})$ \citep[e.g.,][]{isobe_jades_2026-1,sanders_aurora_2026,nakajima_empress_2022}.
Instead of adopting a calibration from the literature, we derive empirical relations between the strong-line ratios and direct-method metallicity following a methodology similar to that of \cite{isobe_jades_2026-1}. The fit uses our grating stacked spectra and individual measurements with auroral-line detections. To cover the high-metallicity end, we additionally include stacked-spectrum measurements of local high-z analogs with high specific star formation from \cite{andrews_mass-metallicity_2013}. Because \cite{andrews_mass-metallicity_2013} construct galaxy stacked spectra in numerous bins, we take the median and the 16th and 84th percentiles of the measurements within 0.2-dex metallicity bins. We perform polynomial fitting in the form $f(x)=\sum_{n=0}^{N}c_nx^n$ for $x>0$ and $f(x)=c_0+c_1x$ for $x\le0$, where $x=12+\log({\rm O/H})-7$. We calculate the maximum-likelihood estimate using the likelihood function adopted in \cite{isobe_jades_2026-1}, and use the Akaike information criterion (AIC) to determine the optimal polynomial order. We adopt the best-fit result with the lowest AIC value that does not have an inflection point near the edge of the fitting range. We show the fitting result in Figure \ref{fig:metal_calib}. The best-fit polynomial order is $N=2$ for both the $\log R3$ and $\log R2$ calibrations. We summarize the best-fit coefficients in Table \ref{tab:fit}. The calibration of \cite{isobe_jades_2026-1} is derived from mass-binned stacks extending down to a representative stellar mass of $\log(M_\star/M_\odot)\simeq7.3$. Although these measurements are not included in our fit, our best-fit relations generally agree within the uncertainties at the low-metallicity end (i.e., $12+\log({\rm O/H})<7.5$) with literature calibrations based on high-redshift galaxies, including \cite{isobe_jades_2026-1}, \cite{sanders_aurora_2026}, and \cite{chakraborty_unveiling_2025}, as well as with those calibrated using high-redshift analogs with high H$\beta$ line equivalent width (i.e., $>200$\,{\AA}) \citep{nakajima_empress_2022}.
Based on the newly derived calibration, we derive strong-line based empirically calibrated metallicities (i.e., 12+$\rm \log (O/H)_{\rm Emp.}$) for individual galaxies and stacked spectra using the observed $R3$ and $R2$ values. For the spectra lacking [O\,{\sc ii}]\,\W\W3726,3729 line coverage, we instead infer $R2$ from the observed $R3$ and the stellar-mass--$O_{32}$ relation summarized in Table \ref{tab:fit}, propagating the uncertainty in the relation into the metallicity estimate. Based on the empirical relations using the $R3$ and $R2$ indices, we formulate the following likelihood function for the metallicity:
\begin{eqnarray}
-2\log \mathcal{L}(Z) & = & \frac{(\log R3_{{\rm obs}}-\log R3_{{\rm model}}(Z))^2}{\sigma_{\log R3}^2} \\ \nonumber
& & + \frac{(\log R2_{{\rm obs}}-\log R2_{{\rm model}}(Z))^2}{\sigma_{\log R2}^2}
\label{eq:likelihood_z}
\end{eqnarray}
We adopt the gas-phase metallicity which maximizes the likelihood. We present the comparison between the metallicity measurements derived using direct and empirical methods in Figure \ref{fig:direct_ratio}.\par

The empirical and direct methods yield broadly consistent metallicities for the grating stacked spectra, including the lowest-mass bin with a representative stellar mass of $\log(M_\star/M_\odot)=6.63$, demonstrating the internal consistency of the calibration over the metallicity range probed by our sample.

\subsection{Mass-Metallicity Relation}\label{sec:res:mzr}
\begin{figure*}[htbp]
  \centering
  \includegraphics[width=0.9\linewidth]{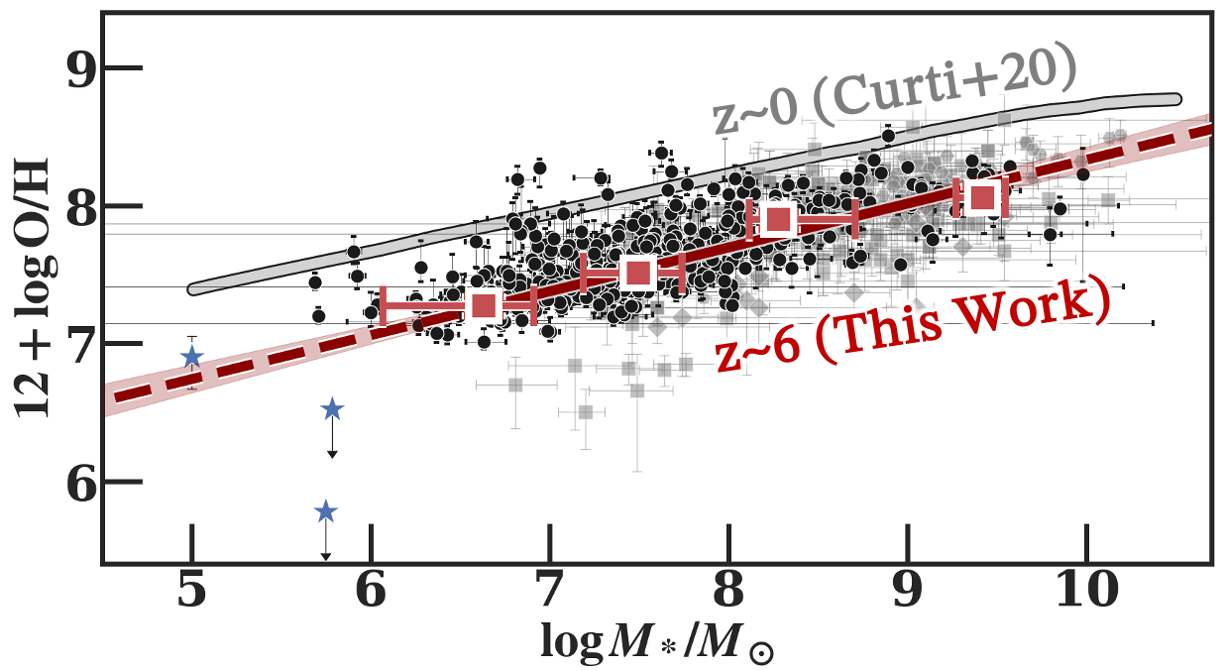}
  \caption{The stellar-mass and gas-phase metallicity relation (MZR) at $z\sim6$ (i.e., $z\simeq4.5$--$9$) based on the strong-line calibrations. The black circles represent measurements from individual galaxies with H$\beta$ or [O\,{\sc iii}]\,$\lambda\lambda4959,5007$ detections. The large red squares represent measurements from the grating stacks in individual stellar-mass bins. The gray dots represent measurements from the literature \citep{nakajima_jwst_2023,curti_jades_2024,hsiao_sapphires_2025,chemerynska_extreme_2024,asada_glimpse-ddt_2026}. The red dashed line and shaded region represent the fiducial two-parameter linear MZR of Equation~\ref{eq:mzr} and its fit uncertainty, respectively. The blue stars represent measurements of Pop~III galaxy candidates from the literature \citep{fujimoto_glimpse-d_2025,morishita_pristine_2025,cai_metal-free_2025}. The gray solid line represents the best-fit MZR for local galaxies from \cite{curti_massmetallicity_2020}.}
  \label{fig:mzr_strong}
  \end{figure*}

\begin{figure}[htbp]
    \centering
    \includegraphics[width=\linewidth]{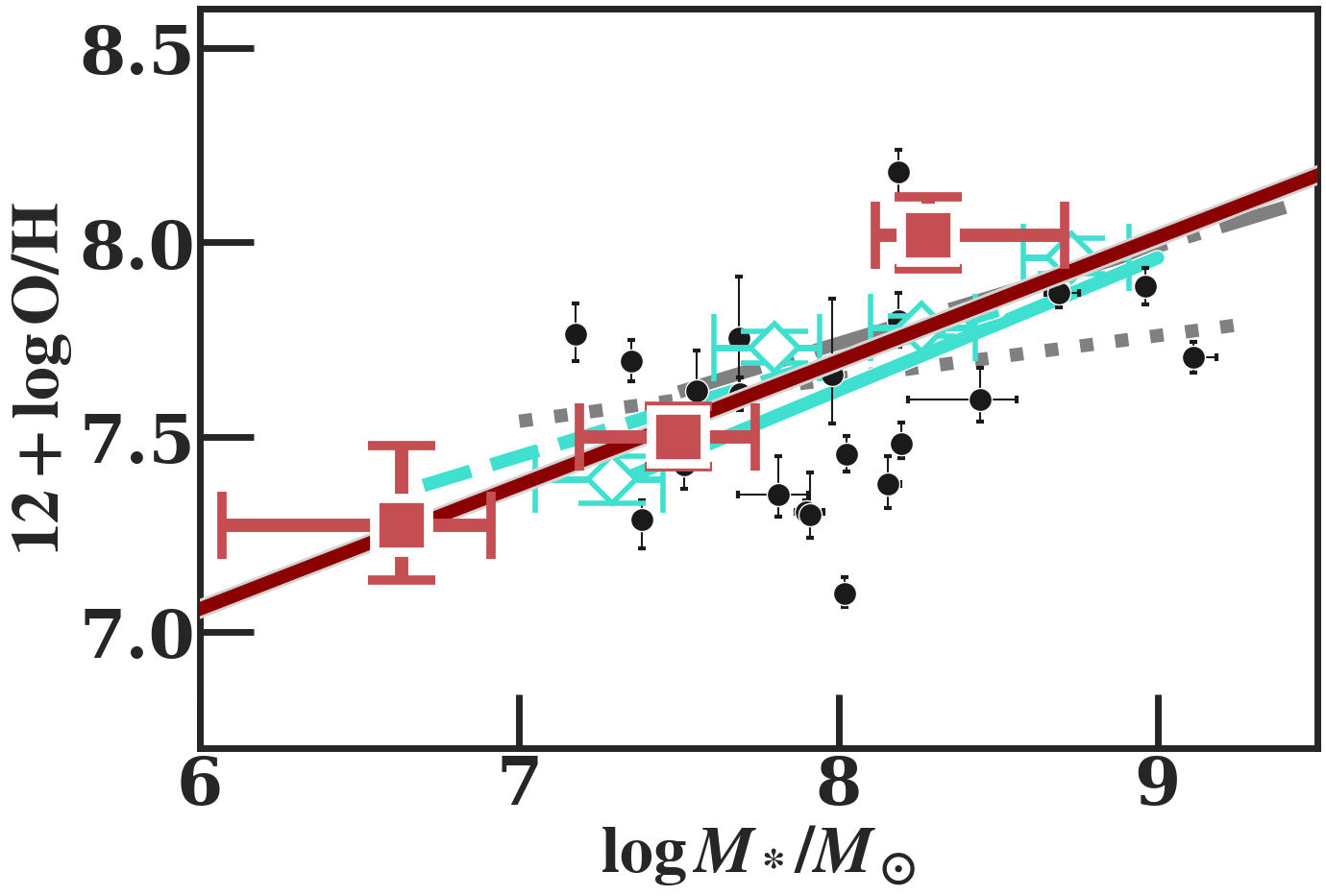}
    \caption{The stellar-mass and gas-phase metallicity relation (MZR) based on the direct method. The black circles represent measurements from individual galaxies with [O\,{\sc iii}]\,$\lambda4363$ detections. The red squares represent measurements from the grating stacks in individual stellar-mass bins. The turquoise diamonds represent measurements from the stacked spectra of \cite{isobe_jades_2026-1}. The red solid line and surrounding shaded region represent the fiducial linear MZR fitted to the empirically calibrated metallicities of the grating stacks (Equation~\ref{eq:mzr}) and its fit uncertainty, respectively. The turquoise solid and dashed lines represent the MZR fitting results from \cite{isobe_jades_2026-1} and \cite{hsiao_glimpse_2026}, respectively, over their corresponding fitting ranges. The gray dotted and dash-dotted lines represent the best-fit MZRs from \cite{curti_jades_2024} and \cite{nakajima_jwst_2023}, respectively.}
    \label{fig:mzr_direct}
  \end{figure}

\begin{figure}[htbp]
    \centering
    \includegraphics[width=\linewidth]{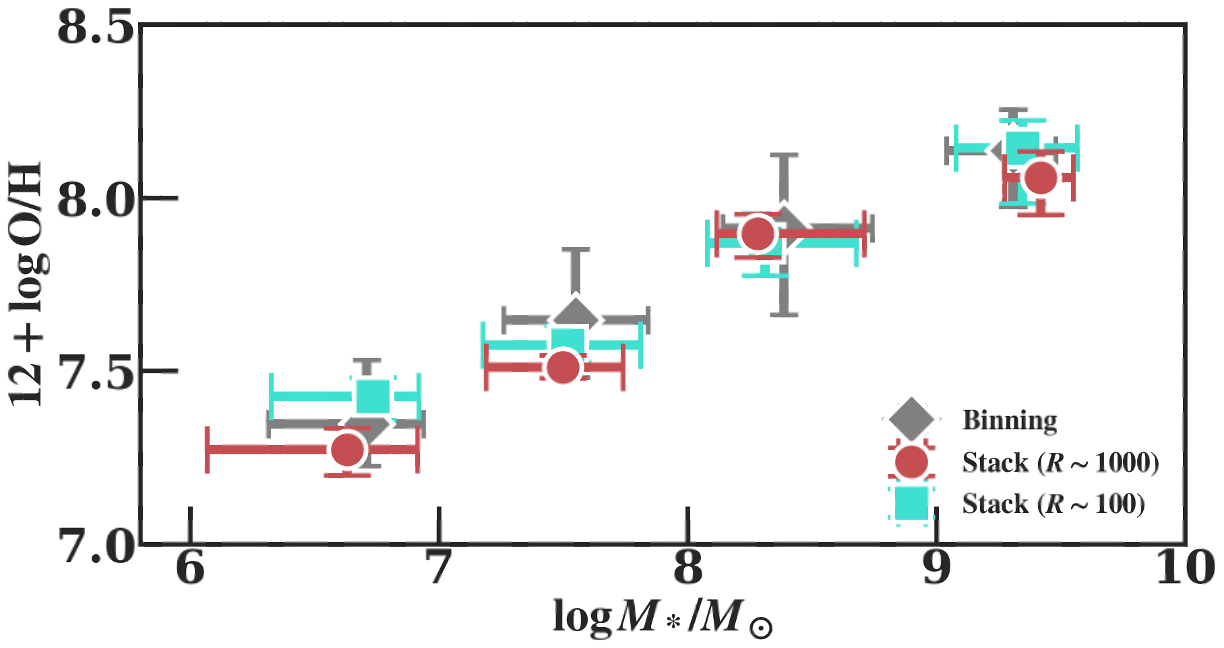}
    \caption{The comparison between the MZR measurement based on the $R\sim1000$ stacked spectra (red circles), $R\sim100$ stacked spectra (turquoise squares), and the binned measurements from individual galaxies (grey diamonds).
    }
    \label{fig:mzr_strong_compare}
    \end{figure}

We examine the mass--metallicity relation (MZR) using gas-phase metallicities derived from our emission-line analysis and stellar masses from the SED fitting. Figure~\ref{fig:mzr_strong} and \ref{fig:mzr_direct} show $12+\log(\mathrm{O/H})$ derived from the strong line calibration and direct method, respectively, as a function of $\log(M_\star/M_\odot)$ for our sample, with representative uncertainties from the posterior distributions described in Section~\ref{sec:data:lineflux}. We also present the measurements from the grating stacked spectra. We find that the stack measurements are generally consistent with the individual measurements for both direct and empirically calibrated metallicity measurements. We, however, find that measurements from \cite{hsiao_sapphires_2025} at $M_\star\simeq10^{7}-10^{8}~M_\odot$ are slightly lower in metallicity than our lensing sample, reaching below 12+$\rm \log O/H=7$. Because the \cite{hsiao_sapphires_2025} sample is selected from slitless spectroscopy of field galaxies, it may be biased toward galaxies with high specific star-formation rates and correspondingly low gas-phase metallicities \citep[e.g.,][]{lilly_gas_2013}, whose large emission-line equivalent widths make them preferentially detectable. In Figure \ref{fig:mzr_strong}, we also compare the local MZR from \cite{curti_massmetallicity_2020} and find that the $z\sim6$ is systematically lower than the local relation. In Figure \ref{fig:mzr_direct}, we compare our MZR measurement based solely on direct method with those from based on literature. We compare with the literature $z\sim6$ MZR from \cite{nakajima_jwst_2023} and \cite{curti_jades_2024} together with those derived solely from the direct method \cite{isobe_jades_2026-1,hsiao_glimpse_2026,gimenez-alcazar_shape_2026}. We find consistency with previous studies within the uncertainties at $M_\star\sim10^7$--$10^8~M_\odot$, while our lowest-mass bin at $M_\star\sim10^{6.6}~M_\odot$ lies along the extrapolation of the literature MZRs. The direct-method stack metallicities are slightly lower than literature MZRs measured above $\log(M_\star/M_\odot)\simeq7.5$ \citep[e.g.,][]{curti_chemical_2023,nakajima_jwst_2023}, while remaining broadly consistent with recent low-mass measurements \citep[e.g.,][]{isobe_jades_2026-1,hsiao_glimpse_2026}. To further confirm the consistency between the stack and individual measurements, we plot the binned mass-metallicity measurements based on individual galaxies together with the median stack measurements based on $R\sim1000$ and $R\sim100$ spectra (i.e., grating and prism-resolution stacked spectra, respectively) as shown in Figure \ref{fig:mzr_strong_compare}. For the empirically calibrated metallicities, the stack measurements in both low and high resolution are mostly consistent with the binned values based on the individual measurements within the measurement uncertainties. We do see a slight downward offset for the stacked spectra measurements compared to the binned values, which could be explained by the improved detection limit in the stacked spectra and highlights the importance of using stacking to evaluate the emission line strength at the faint end.

Given the limited number of stacked measurements, we quantify the MZR using a two-parameter linear relation pivoted at $\log(M_\star/M_\odot)=8$:
\begin{equation}
12+\log(\mathrm{O/H}) = Z_8 + \alpha\left[\log(M_\star/M_\odot)-8\right],
\label{eq:mzr}
\end{equation}
where $Z_8$ is the metallicity at $\log(M_\star/M_\odot)=8$ and $\alpha$ is the logarithmic stellar-mass slope over the mass range probed here. We fit the relation using the four grating-stack metallicity measurements based on the empirical calibration and obtain $Z_8=7.694\pm0.028$ and $\alpha=0.318\pm0.037$, with $\chi^2/{\rm dof}=4.879/2$, as summarized in Table~\ref{tab:fit}. We adopt this relation as the simplest fiducial description of the current measurements; the limited number of stacked measurements does not warrant a more flexible fit or allow us to constrain a high-mass turnover. We also find that the Pop III candidates with stellar masses of $M_\star\sim10^5~M_\odot$ \citep{morishita_pristine_2025,cai_metal-free_2025} lie below the extrapolation of our MZR fit, whereas the candidate with [O\,{\sc iii}]\,\W\W5007 emission detected in follow-up observations \citep{fujimoto_glimpse-d_2025} is consistent with the extrapolated relation \citep[see also][]{nakajima_ultra-faint_2025}. Because these candidates lie below the mass range of our stacked measurements, this comparison should be regarded as an extrapolation. In Section~\ref{sec:disc:mzr}, we compare the measured slope and normalization with previous studies and cosmological simulations.

We quantify the intrinsic scatter of MZR following \citet{curti_jades_2024} by subtracting in quadrature the mean metallicity uncertainty from the observed dispersion of individual galaxy metallicities within each stellar-mass bin. For the stellar mass bins of $10^{5.5}-10^{7.0}~M_\odot$, $10^{7.0}-10^{8.0}~M_\odot$, and $10^{8.0}-10^{9.0}~M_\odot$, we find intrinsic scatters of $\sigma_{\rm int}=0.22$, 0.18, and 0.20, respectively. We do not find clear evidence that the intrinsic scatter increases toward the lower-mass regime.

\subsection{Chemical Abundance}\label{sec:res:chem}
\subsubsection{Nitrogen Abundance from Different Nitrogen Lines}\label{sec:res:chem:nit}
Motivated by recent reports of nitrogen overabundance based on rest-frame UV spectra of high-redshift galaxies, we determine the nitrogen abundance using multiple nitrogen diagnostics, including N\,{\sc iv}]\,\W\W1483,1486 and the optical [N\,{\sc ii}]\,\W6583 lines. For the UV diagnostics, we calculate the $\mathrm{N^{+3}/O^{+2}}$ ionic abundances from the relative strengths of O\,{\sc iii}]\,\W\W1661,1666 with respect to  N\,{\sc iv}]\,\W\W1483,1486.
We do not use  the N\,{\sc iii}]\,\W1750 quintet because the quintet-line fluxes are not detected in our stacked spectra. We note that high-z N-enhanced galaxies compiled in \cite{martinez_under_2025} generally show weaker N\,{\sc iii}]\,\W1750 quintet lines fluxes than N\,{\sc iv}]\,\W\W1483,1486 line fluxes. We derive the ionic abundance ratio, $\mathrm{N^{+3}/O^{+2}}$, from the observed N\,{\sc iv}]\,\W\W1483,1486 and O\,{\sc iii}]\,\W\W1661,1666 line fluxes using \texttt{PyNeb}, obtaining $\log(\mathrm{N^{+3}/O^{+2}})=-0.77_{-0.19}^{+0.17}$ for the stacked spectrum. We independently infer the total N/O abundance directly from the N\,{\sc iv}]/O\,{\sc iii}] line ratio using the photoionization-model relation described in Appendix~\ref{app:icf}, with $C_{43}$ accounting for its dependence on the ionization state. Previous studies applying model-based ionization corrections to similar $\log \mathrm{N^{+3}/O^{+2}}$ (i.e., $\simeq-0.7$) ratios obtain comparable N/O abundances of $[\rm N/O]\simeq0.4$--0.6 \citep[e.g.,][]{martinez_under_2025,berg_fleeting_2026}, suggesting that our direct line-ratio calibration does not substantially overestimate N/O. We adopt the abundance inferred from N\,{\sc iv}]\,\W\W1483,1486 as our fiducial UV-based nitrogen abundance and denote it as $\mathrm{N/O}_{\rm UV}$.

The N\,{\sc iv}] flux is measured from the bootstrap-median GAll spectrum, so the fiducial measurement incorporates resampling of the contributing galaxies rather than relying on a single realization of the sample. Similar N\,{\sc iv}] features have also been recovered in large prism-based JWST composites \citep{hayes_average_2025}. Subsequent work using a separate $R\sim1000$ median-stacking analysis has likewise reported N\,{\sc iv}] emission at similar redshifts \citep{rai_unveiling_2026}.

\begin{deluxetable*}{lc}
  \tablecaption{Derived quantities for the grating stack GAll.\label{tab:stackall_abundance}}
  \tabletypesize{\scriptsize}
  \tablewidth{0pt}
  \tablehead{
  \colhead{Quantity} & \colhead{GAll}\\
  }
  \startdata
  12+log(O/H)$_{\rm T_e}$ & $7.66_{-0.04}^{+0.06}$\\
  12+log(O/H)$_{\rm Emp.}$ & $7.67_{-0.04}^{+0.05}$\\
  $O_{32}$ & $9.00_{-1.12}^{+1.31}$\\
  $C_{43}$ & $0.41_{-0.15}^{+0.15}$\\
  $\log n_e($[S\,{\sc ii}]\,$/\mathrm{cm}^{-3}$) & $3.12_{-0.67}^{+0.64}$\\
  $\log n_e($C\,{\sc iii}]\,$/\mathrm{cm}^{-3}$) & $3.66_{-0.61}^{+0.39}$\\
  $\rm [N/O]_{\rm UV}$ & $0.66_{-0.24}^{+0.24}$\\
  $\rm [N/O]_{\rm Opt}$ & $-0.73_{-0.10}^{+0.11}$\\
  $\rm [C/O]_{\rm UV}$ & $-0.47_{-0.06}^{+0.07}$\\
  $\rm [Ne/O]_{Opt}$ & $0.11_{-0.04}^{+0.04}$\\
  $\rm [Ar/O]_{\rm Opt}$ & $<0.38$\\
  $E(B-V)$ & $0.05_{-0.03}^{+0.03}$\\
  $\log({\rm C}^{+2}/{\rm O}^{+2})_{\rm UV}$ & $-0.66_{-0.06}^{+0.07}$\\
  $\log({\rm C}^{+3}/{\rm O}^{+2})_{\rm UV}$ & $-1.53_{-0.18}^{+0.14}$\\
  $\log({\rm N}^{+3}/{\rm O}^{+2})_{\rm UV}$ & $-0.77_{-0.19}^{+0.17}$\\
  $\log({\rm N}^{+}/{\rm O}^{+})_{\rm Opt}$ & $-0.93_{-0.11}^{+0.11}$\\
  $\log({\rm Ne}^{+2}/{\rm O}^{+2})_{\rm Opt}$ & $-0.75_{-0.04}^{+0.03}$\\
  $\log({\rm Ar}^{+3}/{\rm O}^{+2})_{\rm Opt}$ & $<-2.41$\\
  \enddata
  \tablecomments{For measurements based on detected emission lines, we report the median and the 16th/84th percentiles as the best-fit value and its uncertainty, respectively. Otherwise, the upper/lower limits correspond to the 16/84-th percentiles values, respectively. In particular, the Ar/O constraint is a 1$\sigma$ upper limit propagated from the 3$\sigma$ upper limit on the [Ar\,{\sc iv}]\,$\lambda4711$ flux. The 12+log(O/H)$_{\rm T_e}$ value is derived using the $T_e$ method, while the 12+log(O/H)$_{\rm Emp.}$ value is derived using the newly derived empirical calibration described in Section \ref{sec:res:metal:strong}.}
\end{deluxetable*}

We also derive the optical nitrogen abundance from the [N\,{\sc ii}]\,\W6583 lines using the calibration of \cite{belfiore_sdss_2017} \citep[see also][]{pagel_primordial_1992}, which is supported by five-atom calculations and other empirical calibrations \citep[e.g.,][]{pilyugin_new_2010}. The [N\,{\sc ii}]\,\W6583/[O\,{\sc ii}]\,\W\W3726,3729 ratio is generally insensitive to the ionization parameter. In Figure~\ref{fig:no_compare}, we compare $\mathrm{N/O}_{\rm UV}$ from N\,{\sc iv}]\,\W\W1483,1486 with the optical estimate, $\mathrm{N/O}_{\rm Opt}$, from [N\,{\sc ii}]\,\W6583. We find that $\mathrm{N/O}_{\rm UV}$ and $\mathrm{N/O}_{\rm Opt}$ do not agree in GAll, with the N\,{\sc iv}]-based measurement yielding 1.4 dex higher N/O value than the optical diagnostic. We also plot literature measurements of individual local and high-redshift galaxies for which both UV- and optical-based nitrogen abundances are available \citep[e.g.,][]{ji_ga-nifs_2024,pascale_nitrogen-enriched_2023,martinez_under_2025}. These individual galaxies generally show the same direction of offset, with higher $\rm [N/O]_{UV}$ than $\rm [N/O]_{Opt}$. We further investigate this discrepancy and discuss the potential origin of this discrepancy in Section~\ref{sec:disc:nuv_opt}.\par

\begin{figure*}[htbp]
    \centering
    \begin{minipage}{0.49\linewidth}
      \centering
      \includegraphics[width=\linewidth]{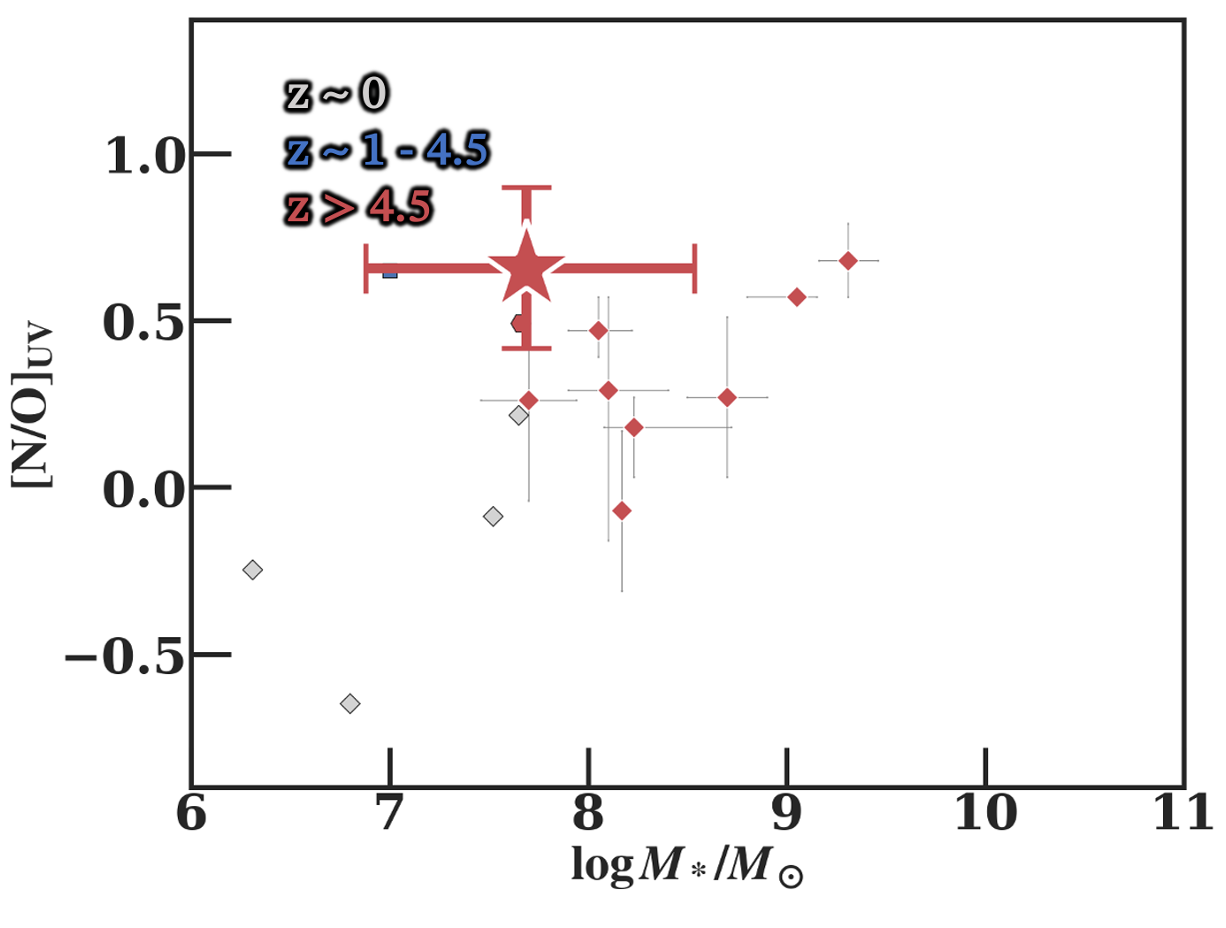}
      \par\centering (a)
    \end{minipage}\hfill
    \begin{minipage}{0.49\linewidth}
      \centering
      \includegraphics[width=\linewidth]{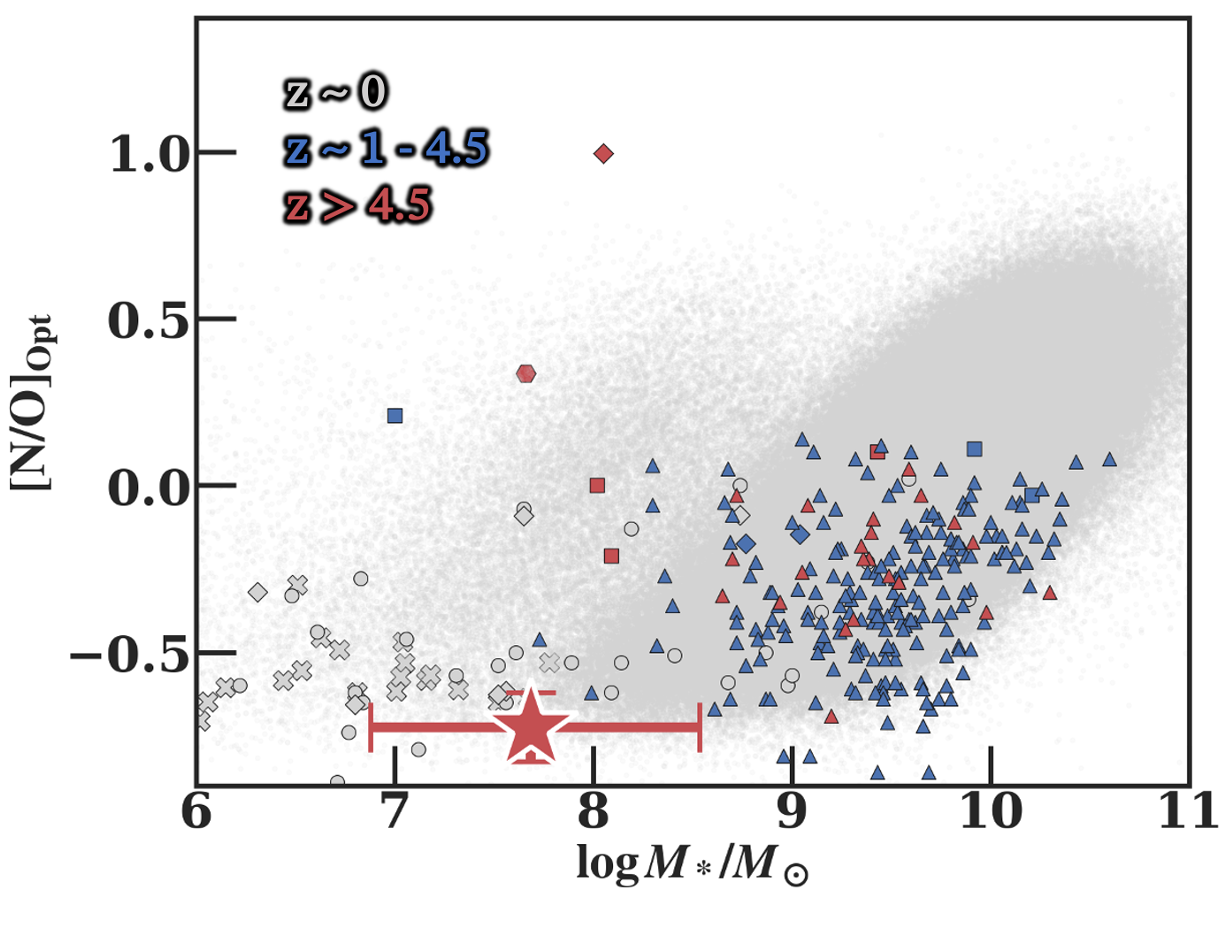}
      \par\centering (b)
    \end{minipage}

    \caption{(a) The [N/O] abundance ratio derived using N\,{\sc iv}]\,\W\W1483,1486 and O\,{\sc iii}]\,\W\W1661,1666 as a function of stellar mass. The red star represents our measurement for the GAll stack. The grey, blue, and red symbols correspond to the measurements for $z\sim0$ \citep{martinez_under_2025}, 1--4.5 \citep{welch_sunburst_2025}, and $>4.5$ \citep{isobe_jwst_2023,castellano_jwst_2024,berg_fleeting_2026,curti_jades_2025,marques-chaves_extreme_2024,topping_metal-poor_2024,schaerer_discovery_2024,naidu_cosmic_2026} galaxies. (b) The [N/O] abundance ratio derived using optical lines (e.g., [O\,{\sc ii}]\,\W\W3726,3729 and [N\,{\sc ii}]\,\W6583) as a function of stellar mass. The red star represents the measurement for the stacked spectra from all galaxy spectra.
    The grey, blue, and red symbols correspond to the measurements for $z\sim0$ \citep{martinez_under_2025,berg_chemical_2019}, 1--4.5 \citep{welch_sunburst_2025,cameron_jades_2026}, and $>4.5$ \citep{cameron_jades_2026,berg_fleeting_2026,zhang_potential_2026,arellano-cordova_jwst_2025,james_classy_2026} galaxies. We note that local N/O measurements based on DESI data \citep{zou_large_2024} are shown in the small grey dots.
    }
    \label{fig:no_mass_all}
  \end{figure*}

We plot the stellar mass dependence of $\log(\mathrm{N/O})$ based on UV and optical lines separately in Figure~\ref{fig:no_mass_all}. Besides our measurements, we also plot literature values for $z\sim0$ \citep{james_classy_2026,martinez_under_2025}, $z\sim1$--$4.5$ \citep{martinez_under_2025,zhang_potential_2026,pascale_nitrogen-enriched_2023,berg_chemical_2019}, and $>4.5$ galaxies \cite{isobe_jwst_2023,cameron_jades_2026,ji_ga-nifs_2024,schaerer_nitrogen_2026,arellano-cordova_jwst_2025,castellano_jwst_2024}. We also use the data of 2,856,369 local galaxies from the Dark Energy Spectroscopic Instrument (DESI) observations \citep{desi_collaboration_early_2024}. We refer to the value-added catalog described in \cite{zou_large_2024}.\footnote{\url{https://data.desi.lbl.gov/doc/releases/dr1/vac/stellar-mass-emline/}} We use [N\,{\sc ii}]\,\W6583 and [O\,{\sc ii}]\,\W\W3726,3729 line fluxes to derive $\rm [N/O]_{\rm Opt}$ under the empirical relation of \cite{belfiore_sdss_2017}. We also derive the gas-phase oxygen abundance from [O\,{\sc ii}]\,\W\W3726,3729, [O\,{\sc iii}]\,$\lambda\lambda4959,5007$, and H$\beta$ using the same empirical strong-line method described in Section~\ref{sec:res:metal:strong}, but adopting the local-galaxy calibrations from \cite{nakajima_empress_2022}. Here, we omit AGN candidates using the BPT diagram, correct for the dust attenuation using the H$\alpha$/H$\beta$ under \cite{calzetti_dust_2000}'s attenuation law, and assume the electron temperature of $T_{e}({\rm O\,\textsc{ii}})\sim15000~{\rm K}$.

For N/O measurements derived from optical lines, we find that N/O values are subsolar and show a weak positive correlation with stellar mass. However, the N/O measurements show a different trend, with the stacked spectra for all galaxies showing a super-solar N/O abundance. Neither the UV nor optical N/O measurements show significant stellar mass dependence, although the N/O measurements from the UV diagnostics are still limited to a narrow stellar-mass regime (i.e., $7.5\lesssim \log M_\star/M_\odot \lesssim 8.5$). We further investigate the origin of the nitrogen enhancement in Section \ref{sec:disc}.\par

\begin{figure*}[htbp]
  \centering
  \begin{minipage}{0.49\linewidth}
    \centering
    \includegraphics[width=\linewidth]{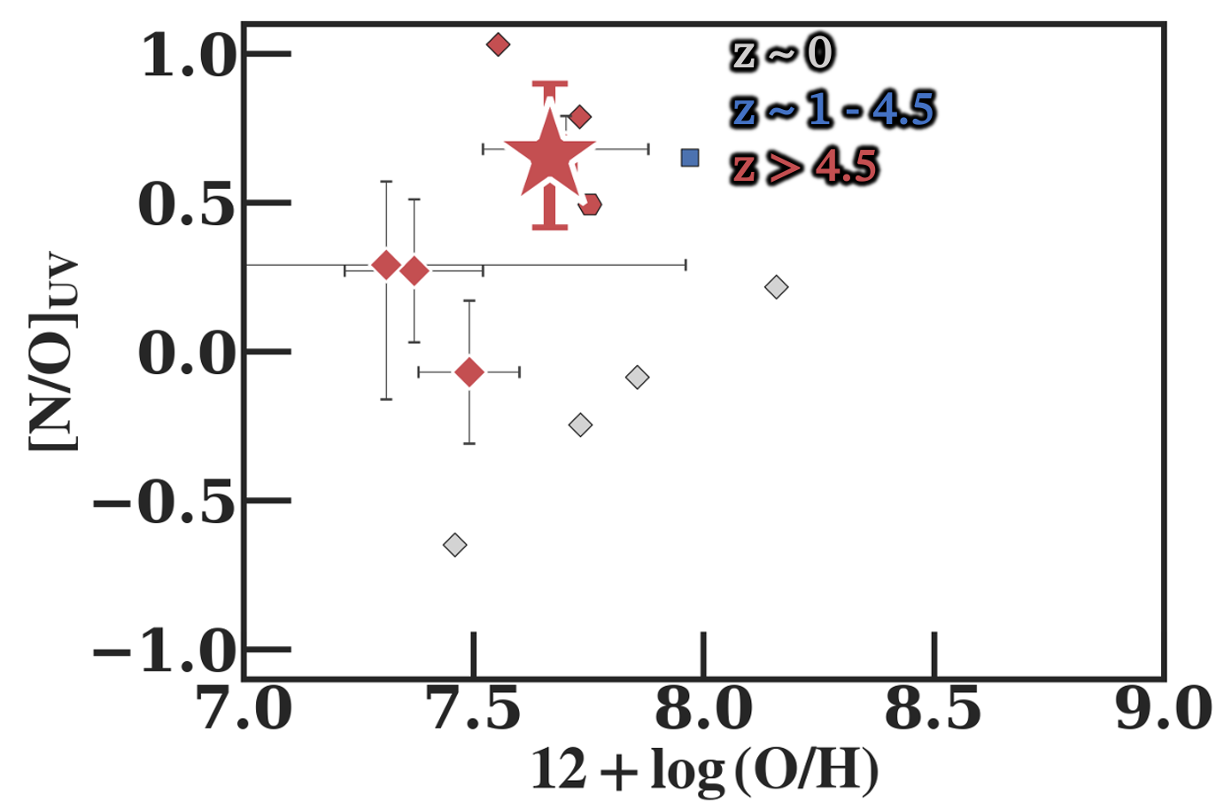}
    \par\centering (a)
  \end{minipage}\hfill
  \begin{minipage}{0.49\linewidth}
    \centering
    \includegraphics[width=\linewidth]{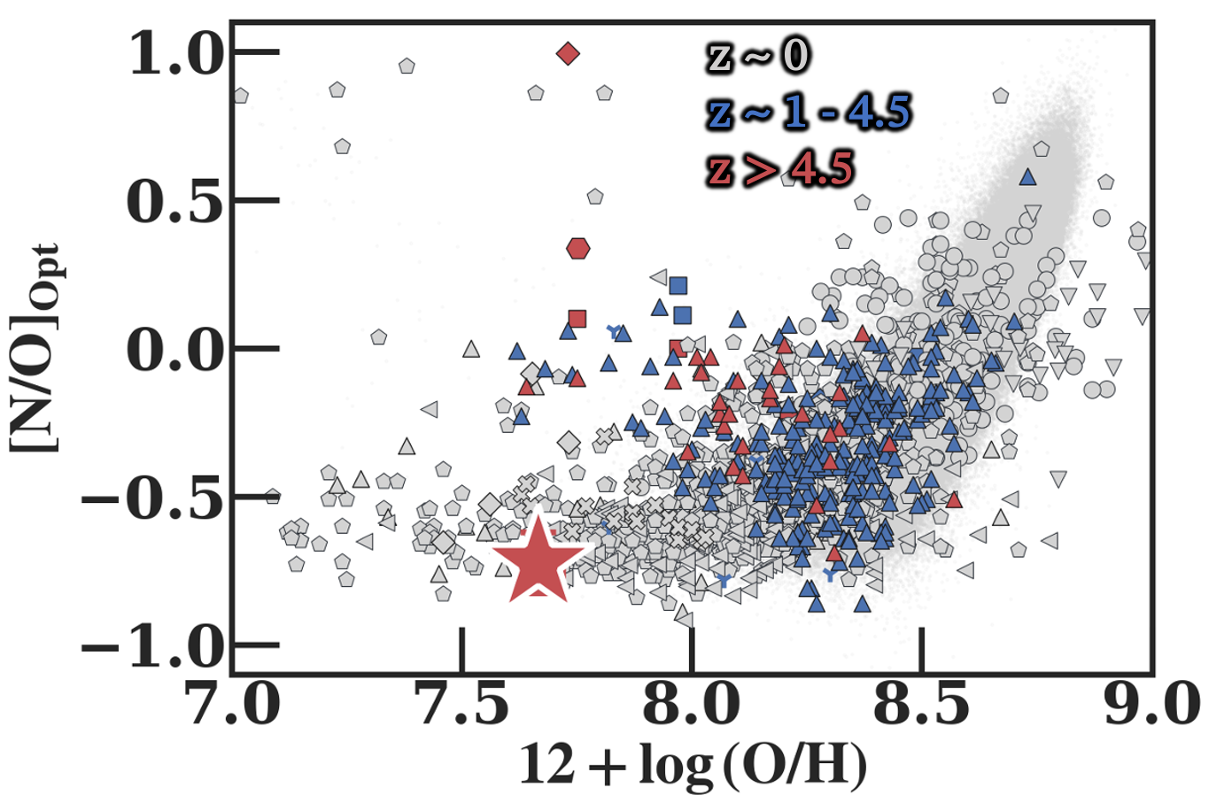}
    \par\centering (b)
  \end{minipage}

  \caption{(a) The [N/O] abundance ratio derived using N\,{\sc iv}]\,\W\W1483,1486 and O\,{\sc iii}]\,\W\W1661,1666 as a function of gas-phase metallicity. The red star represents the our measurement for GAll stack. The grey, blue, and red symbols correspond to the measurements for $z\sim0$ \citep{martinez_under_2025}, 1--4.5 \citep{welch_sunburst_2025}, and $>4.5$ \citep{isobe_jwst_2023,castellano_jwst_2024,berg_fleeting_2026,curti_jades_2025,marques-chaves_extreme_2024,topping_metal-poor_2024,schaerer_discovery_2024,naidu_cosmic_2026} galaxies. (b) The [N/O] abundance ratio derived using optical lines (e.g., [O\,{\sc ii}]\,\W\W3726,3729 and [N\,{\sc ii}]\,\W6583) as a function of gas-phase metallicity. The red star represents the measurement for the stacked spectra from all galaxy spectra.
  The grey, blue, and red symbols correspond to the measurements for $z\sim0$ \citep{martinez_under_2025,berg_chemical_2019,berg_chaos_2020,izotov_chemical_2006,vale_asari_bond_2016,tsamis_heavy_2003,esteban_reappraisal_2004,esteban_keck_2009,esteban_carbon_2014,garciarojas_chemical_2004,garcia-rojas_deep_2005,garcia-rojas_chemical_2007,peimbert_chemical_2005,toribio_san_cipriano_carbon_2016,toribio_san_cipriano_carbon_2017}, 1--4.5 \citep{welch_sunburst_2025,cameron_jades_2026,erb_physical_2010,christensen_gravitationally_2012,bayliss_physical_2014,james_testing_2014,steidel_reconciling_2016,matthee_x-shooter_2021}, and $>4.5$ \citep{cameron_jades_2026,berg_fleeting_2026,zhang_potential_2026,arellano-cordova_jwst_2025,james_classy_2026} galaxies. We note that local N/O measurements based on DESI data \citep{zou_large_2024} are shown in the small grey dots.
  }
  \label{fig:no_oh_all}
\end{figure*}

We also examine the relation between N/O and $12+\log(\mathrm{O/H})$ by UV and optical measurements. Figure~\ref{fig:no_oh_all} shows N/O as a function of $12+\log(\mathrm{O/H})$ for our sample, with measurements from both optical and UV diagnostics. We compare our measurements to local reference samples including local galaxies and H\,{\sc ii} regions \citep[e.g.,][]{izotov_chemical_2006,vale_asari_bond_2016}, $z\sim$1--4.5 galaxies \citep[e.g.,][]{welch_sunburst_2025,cameron_jades_2026}, and $z>4.5$ N emitters \citep[e.g.,][]{berg_fleeting_2026,isobe_jwst_2023,arellano-cordova_jwst_2025}. We find that the [N/O]$_{\rm Opt}$ measurements lie on the N/O--O/H trend of local galaxies and H\,{\sc ii} regions, while the [N/O]$_{\rm UV}$ measurements show a significant offset towards higher N/O at fixed O/H, consistent with some of the measurements for globular clusters and blue compact dwarf galaxies. Such nitrogen enrichment aligns with the recent discoveries of nitrogen-enhanced galaxies at high redshift \citep[e.g.,][]{cameron_nitrogen_2023,isobe_jwst_2023,watanabe_chemical_2026,tripodi_collective_2026}, also shown in Figure \ref{fig:no_oh_all}, suggesting that nitrogen enhancement inferred from the UV emission lines could be a common feature among the high redshift galaxies, not limited to several bright galaxies.

\subsubsection{Carbon Abundance}\label{sec:res:chem:car}

We measure the carbon-to-oxygen abundance ratio using UV-detected lines including C\,{\sc iii}]\,\W\W1907,1909 and O\,{\sc iii}]\,\W\W1661,1666. We measure both $\mathrm{C^{+2}/O^{+2}}$ and $\mathrm{C^{+3}/O^{+2}}$ abundances using \texttt{PyNeb}, and use the photoionization-model abundance calibration described in Appendix \ref{app:icf} to estimate the total C/O abundance ratios. Because we use $C_{43}$ to derive C/O, we implicitly consider contributions from both $\mathrm{C^{+2}/O^{+2}}$ and $\mathrm{C^{+3}/O^{+2}}$. We summarize the C/O values for the grating stacks in Table~\ref{tab:stack_abundance}.

\begin{figure*}[htbp]
  \centering
  \begin{minipage}{0.48\linewidth}
    \centering
    \includegraphics[width=\linewidth]{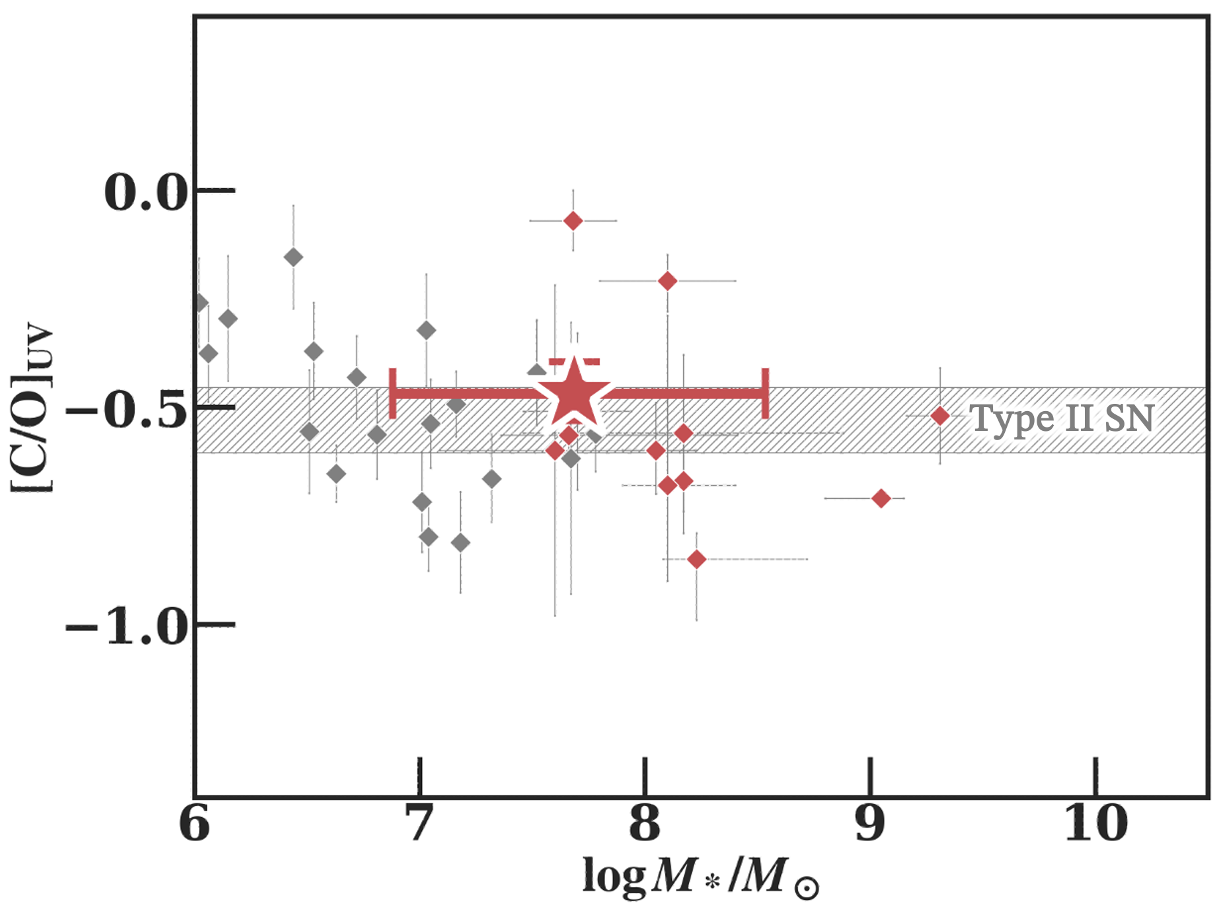}
    \par\centering (a)
  \end{minipage}\hfill
  \begin{minipage}{0.50\linewidth}
    \centering
    \includegraphics[width=\linewidth]{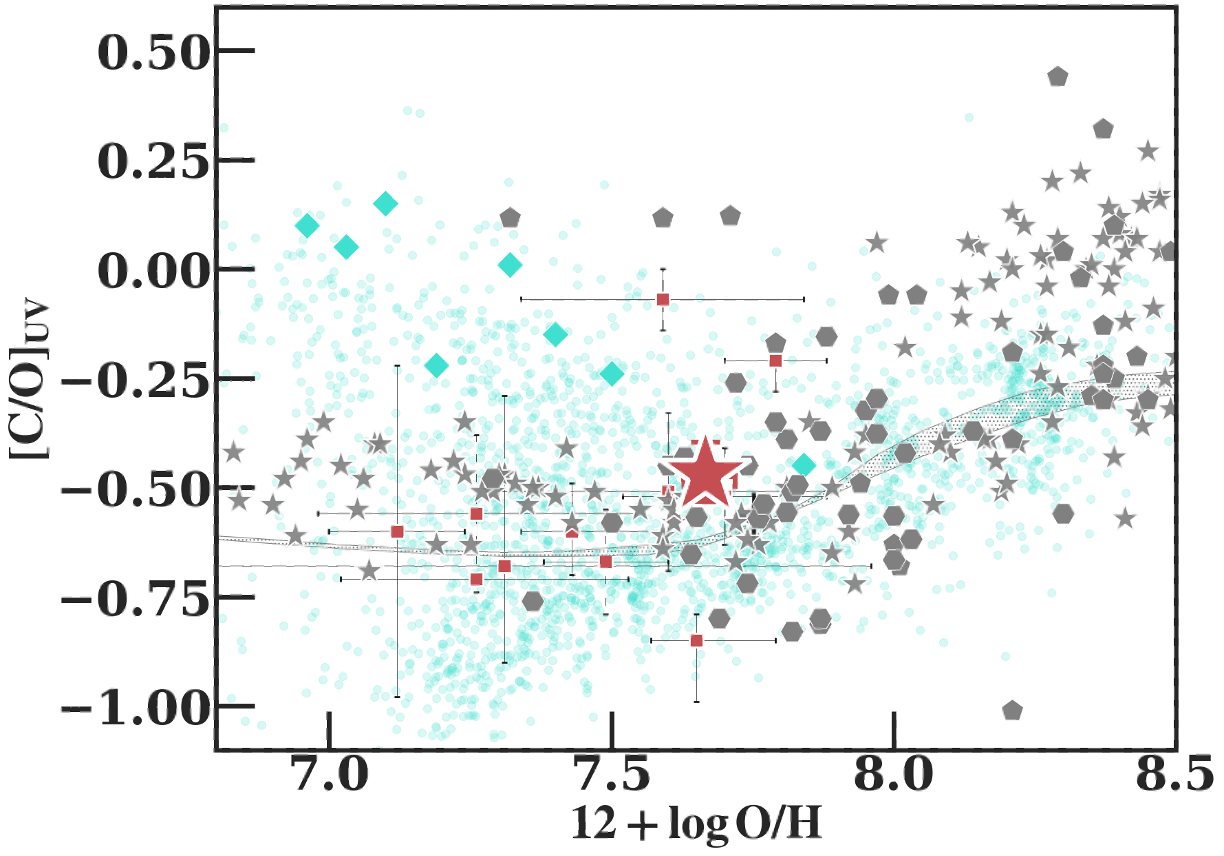}
    \par\centering (b)
  \end{minipage}

  \caption{(a) The [C/O] abundance ratio derived using UV spectra as a function of stellar mass. We show the measurements of the $z\sim2$ galaxies from \cite{berg_chemical_2019} in grey diamonds and the literature $z>6$ N-enhanced galaxies in red diamonds \citep[e.g.,][]{isobe_jwst_2023,arellano-cordova_jwst_2025,ji_ga-nifs_2024,schaerer_nitrogen_2026}. The grey horizontal band represents the C/O yield from the typical CCSNe calculated by \cite{tominaga_supernova_2007}. (b) The [C/O] abundance ratio derived using UV spectra as a function of gas-phase metallicity. We show the literature values for Milky Way stars \citep[grey stars][]{gustafsson_origin_1998,akerman_evolution_2004,fabbian_co_2009,nissen_carbon_2014}, local galaxies and H\,{\sc ii} regions \citep[grey pentagons][]{tsamis_heavy_2003,esteban_reappraisal_2004,esteban_keck_2009,
  esteban_carbon_2014,garciarojas_chemical_2004,garcia-rojas_deep_2005,garcia-rojas_chemical_2007,peimbert_chemical_2005,lopezsanchez_localized_2007,toribio_san_cipriano_carbon_2016,toribio_san_cipriano_carbon_2017,senchyna_ultraviolet_2017,berg_chaos_2020,vale_asari_bond_2016},
  $z\sim2$ galaxies \citep[grey hexagons][]{erb_physical_2010,christensen_gravitationally_2012,bayliss_physical_2014,james_testing_2014,stark_ultraviolet_2014,mainali_relics_2020,iani_scrutiny_2022}, and the stellar abundances of globular clusters \citep[cyan diamonds and circles][]{carretta_abundances_2005,schiavon_apogee_2024}. We also present the measurements for the $z>6$ N-enhanced galaxies from the literature \citep[red diamonds][]{cameron_nitrogen_2023,isobe_jwst_2023,topping_metal-poor_2024,arellano-cordova_jwst_2025}.
  The grey band represents the O/H--C/O relation predicted by the solar-neighborhood Galactic chemical-evolution model of \cite{kobayashi_origin_2020}.
  }
  \label{fig:co_all}
\end{figure*}

We plot the stellar mass dependence of $\log(\mathrm{C/O})$ in Figure~\ref{fig:co_all}. We find that $\log(\mathrm{C/O})$ shows only a tentative stellar mass dependence, and the measurements are generally consistent with a constant subsolar C/O ratio across the stellar mass range probed by our sample. We compare with the theoretical yields by Type II supernovae calculated by \cite{tominaga_supernova_2007}. We find that the C/O measurement for our stacked spectrum generally aligns with the Type II SN predictions. We also show $\log(\mathrm{C/O})$ as a function of $12+\log(\mathrm{O/H})$ in Figure~\ref{fig:co_all}.
Our C/O measurements align with the previous measurements of Milky Way stars, $z\sim2$ galaxies, and the globular clusters, which show a subsolar C/O ratio at low metallicity. We compare the observed C/O values with the solar-neighborhood Galactic chemical-evolution (GCE) model of \cite{kobayashi_origin_2020} \citep[see also][]{kobayashi_history_2000,kobayashi_chemo-dynamical_2023,kobayashi_nucleosynthesis_2025}, which includes enrichment from AGB stars, core-collapse supernovae (CCSNe), neutron-star mergers, and Type Ia SNe. While we do find slightly enhanced C/O for our stacked spectra, the difference between our stack measurements and GCE model predictions is within 0.2 dex. We further investigate the origin of the C/O enhancement at low metallicity in Section \ref{sec:disc}.

\subsection{Argon and Neon Abundances}\label{sec:res:chem:ar_ne_si}
To help constrain the origin of the nitrogen, carbon, and oxygen abundance patterns, we also consider the abundance of other elements such as argon and neon. The [Ar\,{\sc iv}]\,\W4711 line is not detected, so we adopt a 3$\sigma$ upper limit on its flux. Using the photoionization-model abundance calibration described in Appendix \ref{app:icf}, this flux constraint corresponds to a 1$\sigma$ upper limit of $\mathrm{[Ar/O]}<0.38$. For neon, we use the detected [Ne\,{\sc iii}]\,\W3869 and [O\,{\sc iii}]\,\W\W4959,5007 lines to measure the Ne/O abundance ratio. Because the ionization potentials of Ne$^{+2}$ and O$^{+2}$ are similar, the predicted [Ne\,{\sc iii}]/[O\,{\sc iii}] ratio changes only weakly with the ionization conditions. Nevertheless, we derive Ne/O from [Ne\,{\sc iii}]\,$\lambda3869$/[O\,{\sc iii}]\,$\lambda5007$ using the photoionization models described in the Appendix. We summarize the Ne/O measurement and Ar/O upper limit for the GAll stacked spectra in Table~\ref{tab:stack_abundance}. We compare Ne/O with theoretical yields from nucleosynthesis models and measurements from local galaxies, while using the Ar/O upper limit only as a consistency constraint.

  \begin{figure}[htbp]
    \centering
    \includegraphics[width=0.7\linewidth]{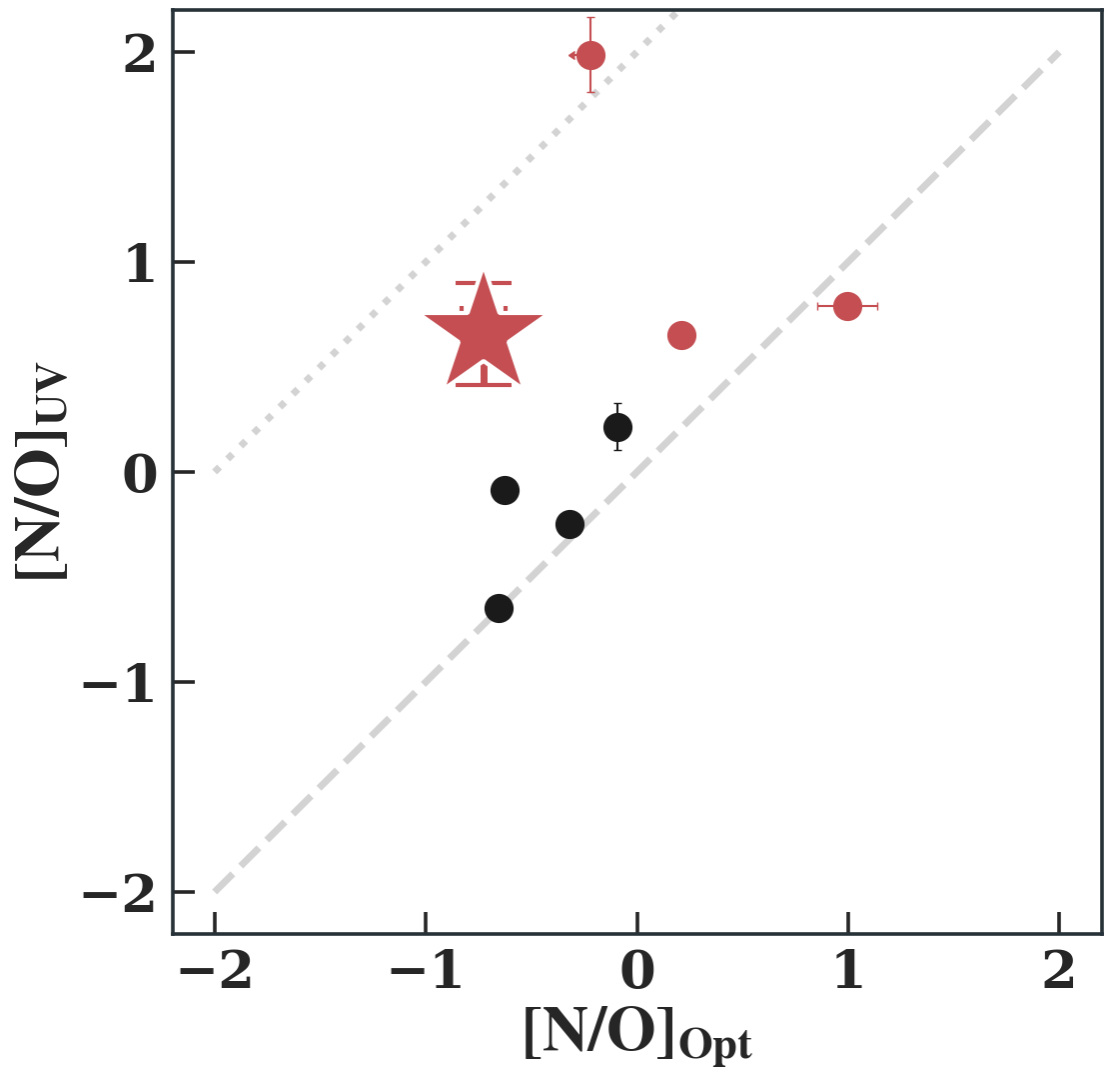}
    \caption{Comparison between the [N/O] value derived from the N\,{\sc iv}]\,\W\W1483,1486 ([N/O]$_{\rm UV}$) and [N\,{\sc ii}]\,\W6583 ([N/O]$_{\rm Opt}$). The red and black circles represent literature measurements of individual galaxies for which both UV- and optical-based nitrogen abundances are available, at $z\sim0$ \citep{martinez_under_2025} and $z>1$ \citep{martinez_under_2025,ji_ga-nifs_2024,pascale_nitrogen-enriched_2023}, respectively.
    The red star represents our measurement from the GAll stacked spectrum.
    The grey dashed and dotted lines represent where $\mathrm{[N/O]}_{\rm UV}=\mathrm{[N/O]}_{\rm Opt}$ and $\mathrm{[N/O]}_{\rm UV}=100\times\mathrm{[N/O]}_{\rm Opt}$, respectively.
    }
    \label{fig:no_compare}
  \end{figure}

\begin{deluxetable}{lcccc}
  \tablecaption{Best-fit coefficients for empirical relations\label{tab:fit}}
  \tablehead{
  \multicolumn{5}{c}{Empirical relation fits}
  }
  \startdata
  \hline\hline
  \multicolumn{5}{c}{Strong-Line Metallicity Calibration} \\
  \hline
  Quantity & $c_0$ & $c_1$ & $c_2$ & RMS \\
  \hline
  $\log R3$ & 0.0680 & 1.8282 & -1.0690 & 0.0669 \\
  $\log R2$ & -1.2276 & 1.7359 & -0.3747 & 0.0832 \\
  \hline\hline
  \multicolumn{5}{c}{MZR} \\
  \hline
  Quantity & $Z_8$ & $\alpha$ & $\chi^2/{\rm dof}$ & \multicolumn{1}{c}{} \\
  \hline
  Eq. \ref{eq:mzr} & $7.694\pm0.028$ & $0.318\pm0.037$ & $4.879/2$ & {}  \\
  \hline\hline
  \multicolumn{5}{c}{Stellar-Mass--Ionization-Sensitive Line-Ratio Relations} \\
  \hline
  Quantity & $b_0$ & $b_1$ & \multicolumn{2}{c}{} \\
  \hline
  $\log O_{32}$ & $2.89\pm0.27$ & $-0.243\pm0.004$ & \multicolumn{2}{c}{} \\
  $\log C_{43}$ & $4.33\pm8.26$ & $-0.60\pm0.15$ & \multicolumn{2}{c}{} \\
  \hline\hline
  \enddata
  \tablecomments{
  For the strong-line-ratio fits, the fitted variable is
  $\log R$ with $x=12+\log({\rm O/H})-7$.
  For $x>0$, $f(x)=\sum_{n=0}^{N}c_nx^n$;
  for $x\le0$, $f(x)=c_0+c_1x$.
  The RMS is computed for $\log R_{\rm obs}-f(x)$ using detected points only;
  upper-limit rows are excluded from the RMS calculation.
  For the MZR, Equation~\ref{eq:mzr} is a linear relation pivoted at $\log(M_\ast/M_\odot)=8$. Here, $Z_8$ is the metallicity at the pivot mass and $\alpha$ is the logarithmic stellar-mass slope. The fit uses the four grating-stack measurements based on the empirical metallicity calibration.
  The stellar-mass--line-ratio relations use the ionization-parameter-sensitive ratios $\log O_{32}$ and $\log C_{43}$; their fitting function has the form of $b_0+b_1x$ with $x=\log(M_\ast/M_\odot)$.
  }
\end{deluxetable}

\section{Discussion}\label{sec:disc}

\subsection{Interpretation of Mass-Metallicity Relation}\label{sec:disc:mzr}
\begin{figure}
  \centering
  \includegraphics[width=\linewidth]{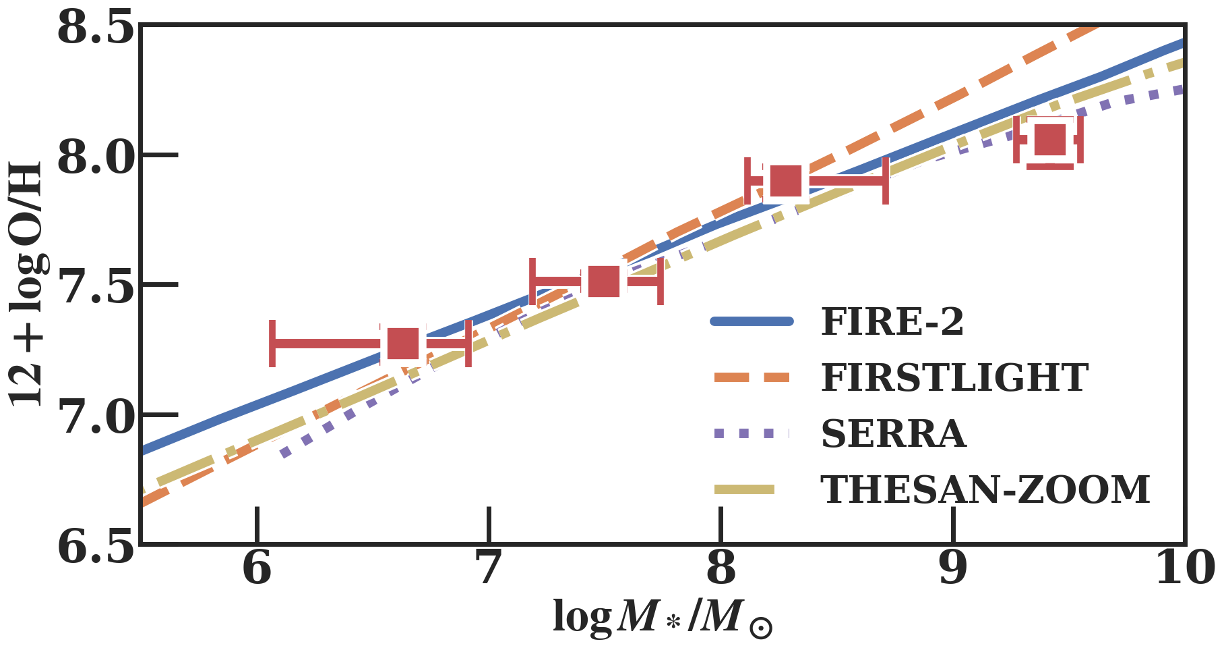} 
  \caption{The mass--metallicity relations predicted by simulations with different ISM assumptions. The blue solid, orange dashed, purple dotted, and yellow dash-dotted lines represent predictions from the \textsc{Fire-2} \citep{marszewski_high-redshift_2024}, \textsc{FirstLight} \citep{langan_weak_2020}, \textsc{Serra} \citep{pallottini_mass-metallicity_2025}, and \textsc{Thesan-zoom} \citep{mcclymont_span_2026} simulations, respectively. The black squares represent measurements from the grating stacks binned by stellar mass.}
  \label{fig:mzr_sims}
\end{figure}

Our fiducial linear fit gives an MZR slope of $\alpha=0.318\pm0.037$ over $M_\star\simeq10^{6.6}$--$10^{9.4}\,M_\odot$. The central value is only slightly steeper than the local low-mass slope of $\simeq0.28$ \citep{curti_massmetallicity_2020}. Given the statistical uncertainty and the different mass ranges probed, we do not interpret this small difference as evidence for strong evolution in the MZR slope.
To investigate the physical processes underlying the MZR over the low-mass range probed by our sample, we compare our measurements with predictions from several cosmological simulations, including \textsc{Fire-2} \citep{marszewski_high-redshift_2024}, \textsc{FirstLight} \citep{langan_weak_2020}, \textsc{Serra} \citep{pallottini_mass-metallicity_2025}, and \textsc{Thesan-zoom} \citep{mcclymont_span_2026}. These simulations incorporate different treatments of star formation, feedback, and chemical enrichment, which can lead to variations in the predicted MZR. As presented in Figure~\ref{fig:mzr_sims}, our stacking results based on lensed faint galaxies are broadly consistent with the range of cosmological simulation predictions down to the representative mass of GM1, $M_\star\simeq10^{6.6}\,M_\odot$ \citep[e.g.,][]{asada_glimpse-ddt_2026,chemerynska_extreme_2024}.
The continued decline in metallicity toward lower stellar masses is consistent with a framework in which metal enrichment is regulated by the baryon cycle. However, the comparison does not uniquely identify the dominant physical mechanism. For example, \textsc{Thesan-zoom} attributes the low-mass slope mainly to the larger gas fractions of low-mass galaxies, which dilute their ISM metallicities, while finding no clear evidence for a strongly mass-dependent metal retention efficiency over the mass range where the trend can be robustly tested. The high-redshift \textsc{FIRE-2} simulations predict a nearly redshift-invariant MZR at $z=5$--12 with a slope broadly consistent with our measurement, but do not explicitly decompose its origin. Instead, they highlight increasing scatter toward lower stellar masses, plausibly associated with bursty star formation and secondary dependences on SFR or gas fraction. As presented in Section~\ref{sec:res:mzr}, we find an intrinsic metallicity scatter of around 0.2 dex across $M_\star\sim10^6$--$10^9\,M_\odot$ and do not detect a clear increase toward lower stellar mass. The four stacked measurements do not provide sufficient leverage to constrain curvature or a high-mass turnover. Larger samples will be required to distinguish the roles of gas fractions, inflows, and metal retention and to test the MZR shape beyond a simple linear description.\par

\subsection{The Origin of Chemical Enrichment}\label{sec:disc:abuns}
To understand the physical origin of the chemical enrichment in the galaxies in our sample, we compare the observed abundance ratios with predictions from chemical evolution models and nucleosynthetic yield calculations. 
The measured C/O, N/O, and Ne/O abundance ratios, together with the upper limit on Ar/O, are sensitive to different aspects of stellar nucleosynthesis and chemical enrichment history. 
Because the UV and optical diagnostics may trace different ionized regions, we treat their combined comparison with the yield models as exploratory rather than as a single co-spatial abundance pattern.
We use the solar-neighborhood Galactic chemical-evolution (GCE) model of \cite{kobayashi_origin_2020} as a reference for enrichment by AGB stars, core-collapse supernovae (CCSNe), neutron-star mergers, and Type Ia SNe.

Motivated by the high $\mathrm{[N/O]}_{\rm UV}$, we also compare our measurements with the yield predictions for Wolf--Rayet (WR) stars \citep{berg_fleeting_2026,limongi_presupernova_2018,todt_potsdam_2015}, supermassive stars \citep[SMSs;][]{nandal_100010000_2025,ebihara_nitrogen_2026,watanabe_chemical_2026,nagele_multiple_2023}, and tidal disruption events \citep[TDEs;][]{isobe_jwst_2023,watanabe_chemical_2026}. 
WR stars enrich the ISM through their strong stellar winds, which can lead to enhanced nitrogen production.
\cite{watanabe_chemical_2026} calculate the chemical evolution by assuming a Kroupa IMF with an instantaneous burst of star formation. \cite{watanabe_chemical_2026} calculate chemical yield from rapidly rotating WR stars based on the model for $\rm [Fe/H]=-3$ with rotation speed of 300 km/s in \cite{limongi_presupernova_2018}. In this WR model, the stellar rotation enhances the chemical mixing. The carbon and oxygen are transported to the hydrogen burning layer, where they are converted to nitrogen via the CNO cycle. The strong stellar wind then ejects the nitrogen-rich and oxygen/carbon-poor material into the ISM.
For SMSs, \cite{watanabe_chemical_2026} assume the stellar evolution and chemical yield calculated by \cite{nagele_multiple_2023}. \cite{nagele_multiple_2023} calculate chemical yields of metal-enriched ($0.1~Z_\odot$) very/supermassive stars (V/SMS). Metal-rich V/SMS are theoretically predicted to form via runaway collisions of massive stars \citep[e.g.,][]{portegies_zwart_star_1999} and the collapse of the disk formed after metal-enriched galaxies \citep[e.g.,][]{mayer_direct_2010,mayer_direct_2015}. These V/SMS stars also undergo the CNO cycle in the hydrogen-burning layer, which leads to the nitrogen enhancement in the stellar wind. Nucleosynthesis continues in the helium burning layer, which produces carbon via triple-$\alpha$ reactions. The strong stellar wind then ejects the nitrogen-rich and carbon-rich material into the ISM. \cite{watanabe_chemical_2026} adopt the chemical yield of SMS with the mass of $10^5~M_\odot$ and the metallicity of $0.1~Z_\odot$.
For TDEs, \cite{watanabe_chemical_2026} assume that 50\% of the outer hydrogen layer is stripped from stars with 9-100 $M_\odot$ so N-rich material are ejected while the oxygen-rich core remains undisrupted. The TDE yield is calculated by \cite{watanabe_chemical_2026} based on the stellar evolution model by \cite{tominaga_supernova_2007} assuming the stellar metallicity at $Z=0.001$.
Besides the yields from WR stars, SMSs, and TDEs, \cite{watanabe_chemical_2026} also consider enrichment from ordinary CCSNe. 
They adopt the CCSN yields from \cite{nomoto_nucleosynthesis_2013}. 
To adjust the contribution from CCSNe, \cite{watanabe_chemical_2026} introduce the failed supernova fraction, which corresponds to the fraction of the massive stars above the transition mass that fail to explode as CCSNe.
In their fiducial prescription, the transition mass is set to $M_{\rm trans}=15\,M_\odot$. Stars with $9\,M_\odot \lesssim M_\star < M_{\rm trans}$ explode as ordinary CCSNe after lifetimes of $\gtrsim10$ Myr, whereas stars above $M_{\rm trans}$ undergo either ordinary CCSNe or failed SNe according to the adopted failed-SN fraction. Because stars undergoing failed SNe are assumed to collapse directly into black holes, increasing this fraction reduces the enrichment of the ISM by their inner, oxygen-rich ejecta.
Our comparison with the WR, SMS, and TDE yields of \cite{watanabe_chemical_2026} is intended to identify plausible prompt enrichment sources rather than to reconstruct a unique chemical-evolution history. The C/N/O composition of any pre-enriched ISM depends on its previous enrichment sources and star-formation history, and the final abundance pattern reflects its subsequent mixing with newly injected ejecta. For example, the self-consistent models of \cite{kobayashi_rapid_2024} show that an intermittent star-formation history can reproduce enhanced N/O through a short-lived WR-dominated enrichment phase.\par
\begin{figure*}[htbp] 
  \centering
  \includegraphics[width=0.75\linewidth]{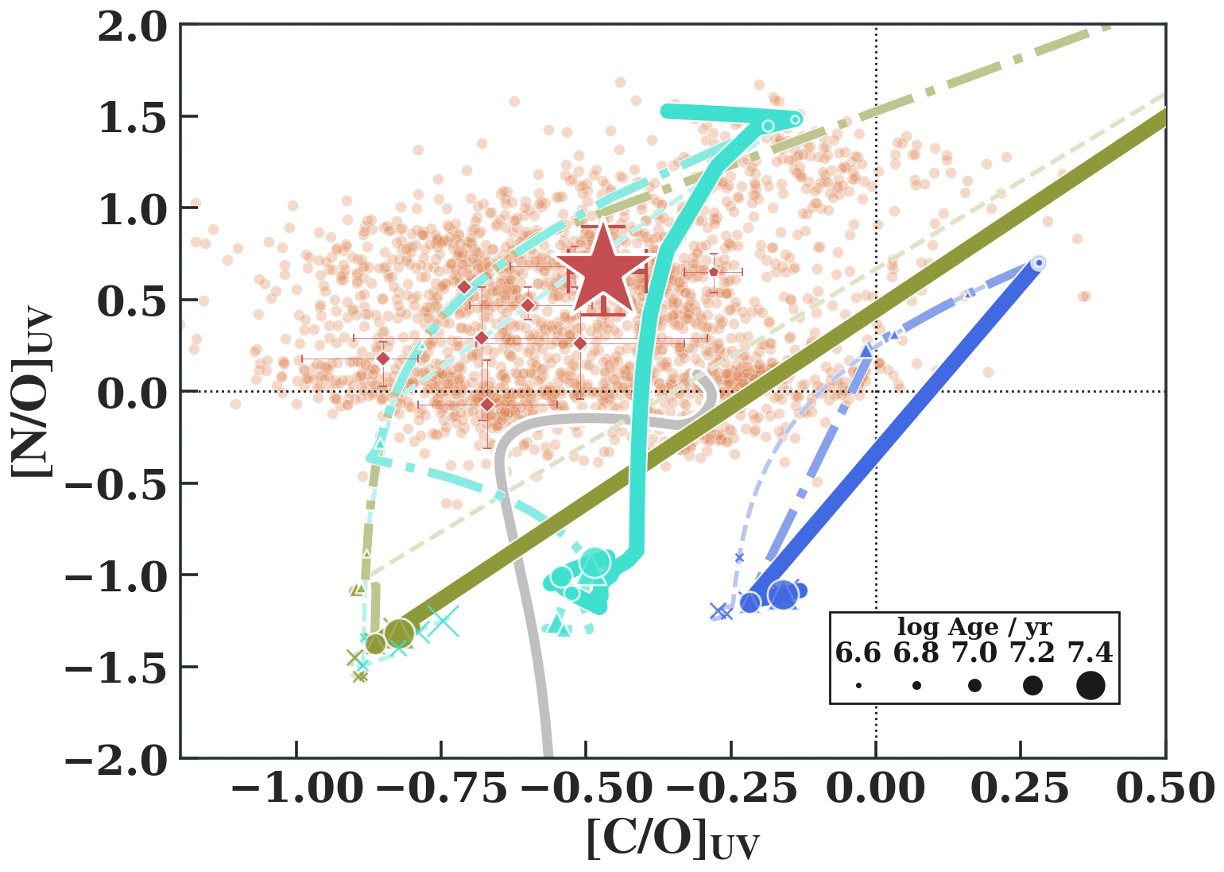} 
  \caption{The $\rm [C/O]_{UV}$ vs. $\rm [N/O]_{UV}$ abundance ratio. The red star represents the measurements for the stacked spectra from all galaxy grating spectra. The other red symbols represent the abundance measurements of N\,{\sc iv}] emitters from literature. The gray solid line represents the solar-neighborhood GCE model of \cite{kobayashi_origin_2020}. Cyan, olive, and blue solid lines represent the yield predictions by \cite{watanabe_chemical_2026} for the Wolf-Rayet stars, supermassive stars, and Tidal Disruption Event, respectively. Solid, dashed, and dotted lines correspond to the prediction with failed supernova fraction of 1, 0.01, and 0, respectively. The smallest through largest circles, triangles, and crosses on the thickest solid lines correspond to the predictions  with failed supernova fraction of 1, 0.01, and 0, respectively, at the $\log {\rm Age}=$6.6 to 7.4 with a 0.2 dex step size. The orange dots represent the measurements for the member stars in the globular cluster from APOGEE sample \citep{schiavon_apogee_2024}. The black dotted vertical and horizontal lines represent the solar abundances.}
  \label{fig:co_no} 
\end{figure*}

We first compare the observed C/O and N/O abundance ratios with the chemical evolution models, as shown in Figure \ref{fig:co_no}. We find that the observed C/O and N/O abundance ratios are offset from the GCE model, with higher N/O and lower C/O abundance ratios for the observations. We instead find that the observed C/O and N/O abundance ratios are more consistent with the yield predictions for the WR and SMS stars. The TDE model predicts higher C/O than the other models. While nitrogen enrichment process occurs both in WR and SMS stars, SMSs have higher N/O because SMS can sustain the CNO cycle for a long duration and convert oxygen to nitrogen efficiently. We note that N/O and C/O abundances are consistent with the models at the stellar age of 10 Myr or lower for both WR and SMS models, because after around 10 Myr, CCSNe eject oxygen into the ISM and N/O suddenly decreases.\par

\begin{figure*}[htbp] 
  \centering
  \begin{subfigure}{\linewidth}
    \centering
    \includegraphics[width=0.8\linewidth]{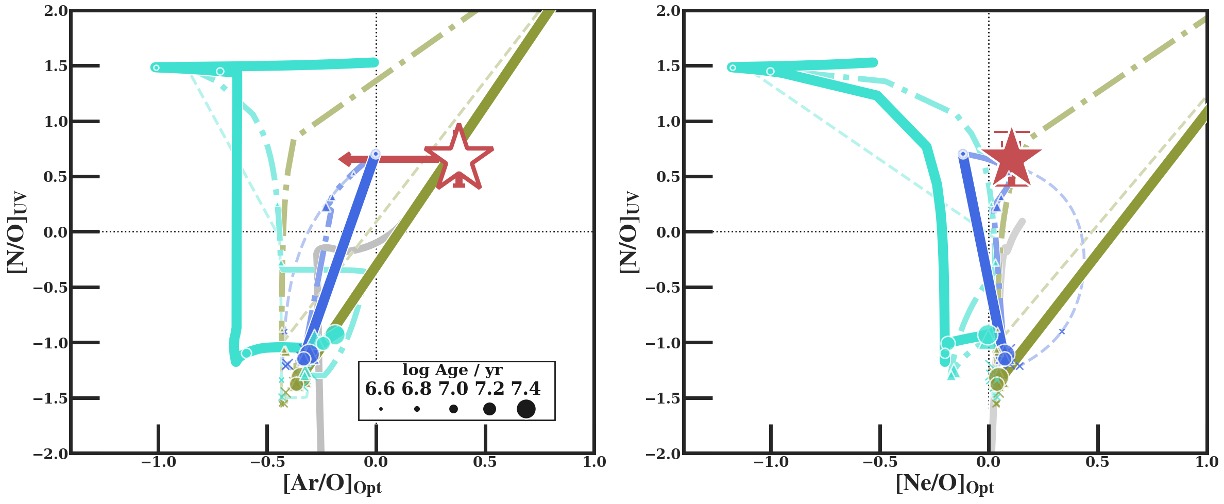}
    \label{fig:no_aro_neo_uv}
  \end{subfigure}
  \begin{subfigure}{\linewidth}
    \centering
    \includegraphics[width=0.8\linewidth]{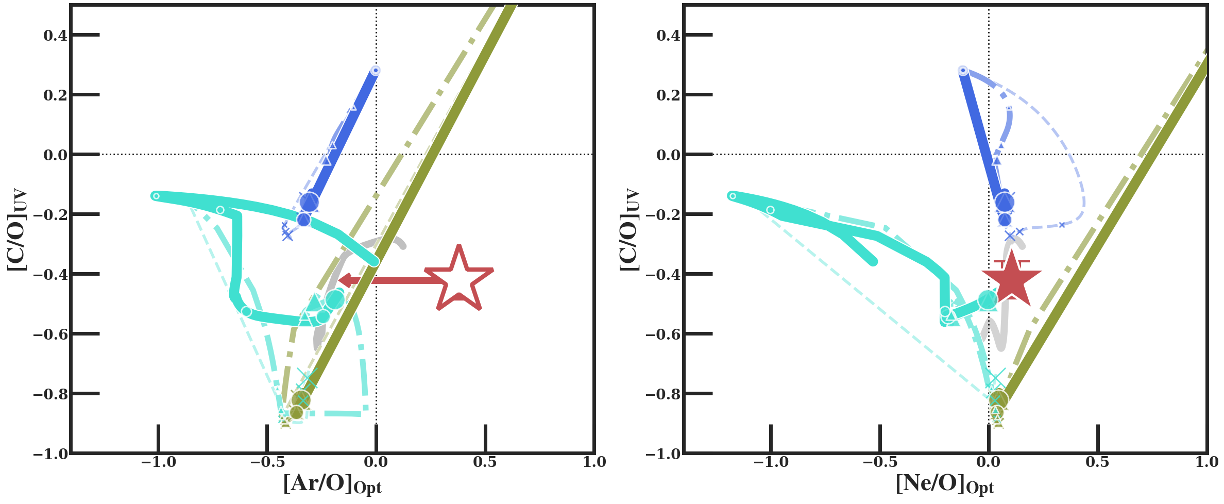}
  \end{subfigure}
  \caption{(Top): The [N/O] vs. [Ar/O] and [N/O] vs. [Ne/O] abundance ratios. (Bottom): The [C/O] vs. [Ar/O] and [C/O] vs. [Ne/O] abundance ratios. For GAll, Ar/O is a 1$\sigma$ upper limit, indicated by the leftward arrows, whereas Ne/O is a measurement. The symbols are otherwise the same as in Figure~\ref{fig:co_no}.}
  \label{fig:aro_neo} 
\end{figure*}

We consider similar comparisons for the Ar/O and Ne/O abundance ratios.
We show the [N/O] vs. [Ar/O], [N/O] vs. [Ne/O], [C/O] vs. [Ar/O], and [C/O] vs. [Ne/O] abundance ratios in Figure~\ref{fig:aro_neo}. Although neon, oxygen, and argon are all $\alpha$-elements, they are synthesized in different layers of massive stars. Their relative abundances, such as Ne/O and Ar/O, may also depend on the enrichment mechanism and on how deeply the stellar layers are ultimately ejected.
As shown in the bottom panel of Figure~\ref{fig:aro_neo}, the [Ne/O] abundance ratio is enhanced for SMS because oxygen is efficiently depleted via the CNO cycle. Our stack results suggest that the observed [Ne/O] abundance ratio is around a solar value, and lies around the N, Ne, O abundances in the values predicted between WR and SMS stars. TDE also predict the similar N/O and Ne/O abundance ratios, but the TDE scenario is disfavored by the observed C/O abundance ratio. Supermassive stars do overpredict [Ne/O] because of their efficient oxygen depletion via CNO cycle in the H-burning layer. The 1$\sigma$ upper limit on Ar/O is consistent with both WR and SMS yields and therefore does not discriminate between these enrichment scenarios.\par

As discussed above, the UV-based N/O and C/O ratios can be reproduced by enrichment from young WR or V/SMS populations. An alternative explanation is a selective-outflow scenario, in which oxygen-rich CCSN ejecta from short-lived massive stars ($\lesssim10$ Myr) are preferentially expelled, while nitrogen-rich ejecta from delayed channels, such as AGB stellar winds, are retained. \citet{mcclymont_span_2026} show that bursty star formation in low-mass galaxies can produce such differential metal retention on timescales of $\sim100$ Myr, followed by the recollapse of nitrogen-rich gas into GMCs and a subsequent burst. The decline in metallicity toward lower stellar mass is qualitatively compatible with efficient metal removal or dilution in low-mass systems, but the MZR slope alone does not distinguish selective outflows from other baryon-cycle processes. However, it remains to be tested whether the ionizing spectrum of the subsequent stellar population can reproduce the observed high-ionization UV lines and the discrepancy between the UV- and optical-based nitrogen abundances. Additional observables, including emission-line features unique to the WR and V/SMS populations, may help distinguish this scenario from prompt enrichment by WR or V/SMS populations.\par

\subsection{Discrepancies Between UV and Optical Nitrogen Abundance}\label{sec:disc:nuv_opt}
Before we further discuss on the inhomogeneous nitrogen abundance pattern in a single system, we need to evaluate alternative possibilities to explain the discrepancies between UV and optical nitrogen abundance measurements. In GAll, the UV--optical N/O discrepancy could partly reflect inter-galaxy mixing in the stack, because the UV and optical nitrogen lines may be dominated by different subsets of galaxies. 
The galaxies contributing at the wavelengths of the UV and optical nitrogen lines form partially overlapping subsets. In GAll, 45 galaxies contribute to N\,{\sc iv}]\,\W\W1483,1486, 74 to [N\,{\sc ii}]\,\W6583, and 19 to both, without requiring individual line detections. Among galaxies with measured photometric UV magnitudes, the median values are similar (i.e., $-18.56$ and $-18.47$). Differences between the contributing populations may therefore affect the inferred abundance offset, which alone cannot establish distinct ionization or abundance zones within the same galaxies. 
However, the literature measurements \citep[e.g.,][]{ji_ga-nifs_2024,pascale_nitrogen-enriched_2023,martinez_under_2025} shown in Figure~\ref{fig:no_compare} include individual galaxies with detections of both optical and UV nitrogen emission lines, and these galaxies also generally show higher UV-based N/O. Therefore, population mixing may contribute to the offset in GAll, but it cannot by itself explain the occurrence of the same trend in individual galaxies. Accordingly, the present stack cannot uniquely distinguish internal chemical stratification from population mixing, even though the same direction of offset in individual galaxies motivates exploring a multi-zone interpretation. The wavelength-by-wavelength median and bootstrap resampling reduce sensitivity to a small number of galaxies with exceptionally strong N\,{\sc iv}] emission, but do not establish that the UV and optical nitrogen lines originate from the same galaxies or spatial regions. Distinguishing these possibilities will require deeper observations of individual galaxies with both N\,{\sc iv}] and [N\,{\sc ii}] coverage, together with spatially resolved IFU spectroscopy capable of localizing the high- and low-ionization nitrogen-emitting regions.

Another possible way to enhance the N\,{\sc iv}] emission is to invoke a harder ionizing radiation field, which can produce high-ionization species such as $\mathrm{N}^{+3}$ more efficiently. We therefore test how the inferred N/O ratio changes when the underlying ionizing spectrum is switched from the stellar to AGN-like. Although we find that $\mathrm{[N/O]}_{\rm UV}$ decreases by 0.2 dex if an AGN radiation field is assumed, it still does not agree with $\mathrm{[N/O]}_{\rm Opt}$. Thus, the magnitude of the inferred nitrogen enhancement depends moderately on the assumed ionizing spectrum, but adopting an AGN-like radiation field does not remove the UV--optical N/O discrepancy. More generally, within the ranges of ionization parameter and electron density, and the stellar, AGN-like, and WNE radiation fields explored here, the observed N\,{\sc iv}]/O\,{\sc iii}] ratio continues to require elevated UV-based N/O. This is consistent with recent photoionization modeling showing that allowing for AGN excitation need not substantially change the inferred N/O \citep{zhu_nature_2026}. Median stacking also reduces the likelihood that the observed N\,{\sc iv}] emission is dominated solely by a few extreme contaminants, although it does not exclude a weak AGN contribution shared by a non-negligible fraction of the sample. We therefore cannot completely rule out an AGN contribution to the hard ionizing spectrum and regard the massive-star-wind scenario as a plausible interpretation rather than a unique identification of the ionizing source. Another possibility is a nitrogen-enhanced density-bounded ISM. Because [N\,{\sc ii}] emission lines are primarily emitted in the low-ionization outer region near the edge of an H\,{\sc ii} region, truncating this outer layer can suppress the [N\,{\sc ii}] emission. However, this scenario requires fine tuning, because [O\,{\sc ii}] emission line fluxes would also be reduced, given its only slightly higher ionization potential.\par

We also consider more exotic ionizing sources beyond typical stellar and AGN radiation fields. Although an AGN-like spectrum cannot simultaneously reproduce the UV and optical nitrogen line strengths, the GAll stack shows strong He\,{\sc ii} emission, with He\,{\sc ii}\,\W4686/H$\beta \simeq 0.03$. The UV He\,{\sc ii}\,\W1640 emission is likewise strong, with He\,{\sc ii}\,\W1640/(C\,{\sc iv}\,\W\W1548,1551+C\,{\sc iii}]\,\W\W1907,1909)$\simeq0.19$. At the observed C\,{\sc iv}/C\,{\sc iii}] ratio, this places GAll beyond the Pop~II model shown in \citet{nakajima_diagnostics_2022}. Local metal-poor galaxies are also known to exhibit He\,{\sc ii}\,\W4686 of comparable strength \citep{berg_window_2018,umeda_empress_2022,olivier_characterizing_2022}. Also, some nitrogen emitters also show strong He\,{\sc ii}\,\W4686/H$\beta\gtrsim0.01$ \citep[e.g.,][]{berg_fleeting_2026,thuan_spitzer_2008}. Such strong He\,{\sc ii}\,\W4686 is difficult to reproduce with standard stellar radiation fields alone \citep[e.g.,][]{berg_window_2018}. Several scenarios have been proposed to explain nebular He\,{\sc ii} emission, including radiative shocks propagating through highly ionized channels \citep{izotov_large_2021}, extremely density-bounded nebulae \citep{plat_constraints_2019}, and intrinsically hard ionizing spectra associated with exotic sources such as stripped stars, Wolf-Rayet stars, compact objects or intermediate-mass black holes \citep{umeda_empress_2022,olivier_characterizing_2022,hatano_empress_2024}. 
Motivated by the chemical abundance analysis in Section \ref{sec:disc:abuns}, we test whether a WR spectrum affects the inferred N/O measurements. For the WR spectrum, we use the nitrogen-rich phase of the WR stars modeled in the Potsdam Wolf-Rayet (PoWR) model \citep{grafener_line-blanketed_2002,hamann_temperature_2003,sander_consistent_2015}\footnote{\url{https://www.astro.physik.uni-potsdam.de/PoWR/powrgrid1.php}}. We adopt the Wolf-Rayet stars of the early nitrogen subclass (WNE) spectrum at the metallicity of Small Magellanic Cloud with the effective temperature of $10^{5.25}$ K with corresponding mass-loss rates (i.e., $\dot{M}$) ranging from $1.5\times10^{-6}$ to $3.5\times10^{-5}~M_\odot~{\rm yr}^{-1}$. As we have explained in the Appendix and shown in Figure \ref{fig:he2}, the WNE models can produce very high He\,{\sc ii}\,\W4686/H$\beta$ ratio of around 0.3 with $\dot{M}\geq1.2\times10^{-5}~M_\odot~{\rm yr}^{-1}$, while only raising the N\,{\sc iv}]\,\W\W1483,1486/O\,{\sc iii}]\,\W\W1661,1666 by 0.2 dex compared to the BPASS radiation at the fixed N/O. Therefore, while considering WNE star as an ionizing source could explain the observed strong He\,{\sc ii} recombination line strength, high nitrogen abundance is still required to explain the UV nitrogen emission line strength.\par 

\begin{figure}[htbp] 
  \centering
  \includegraphics[width=0.7\linewidth]{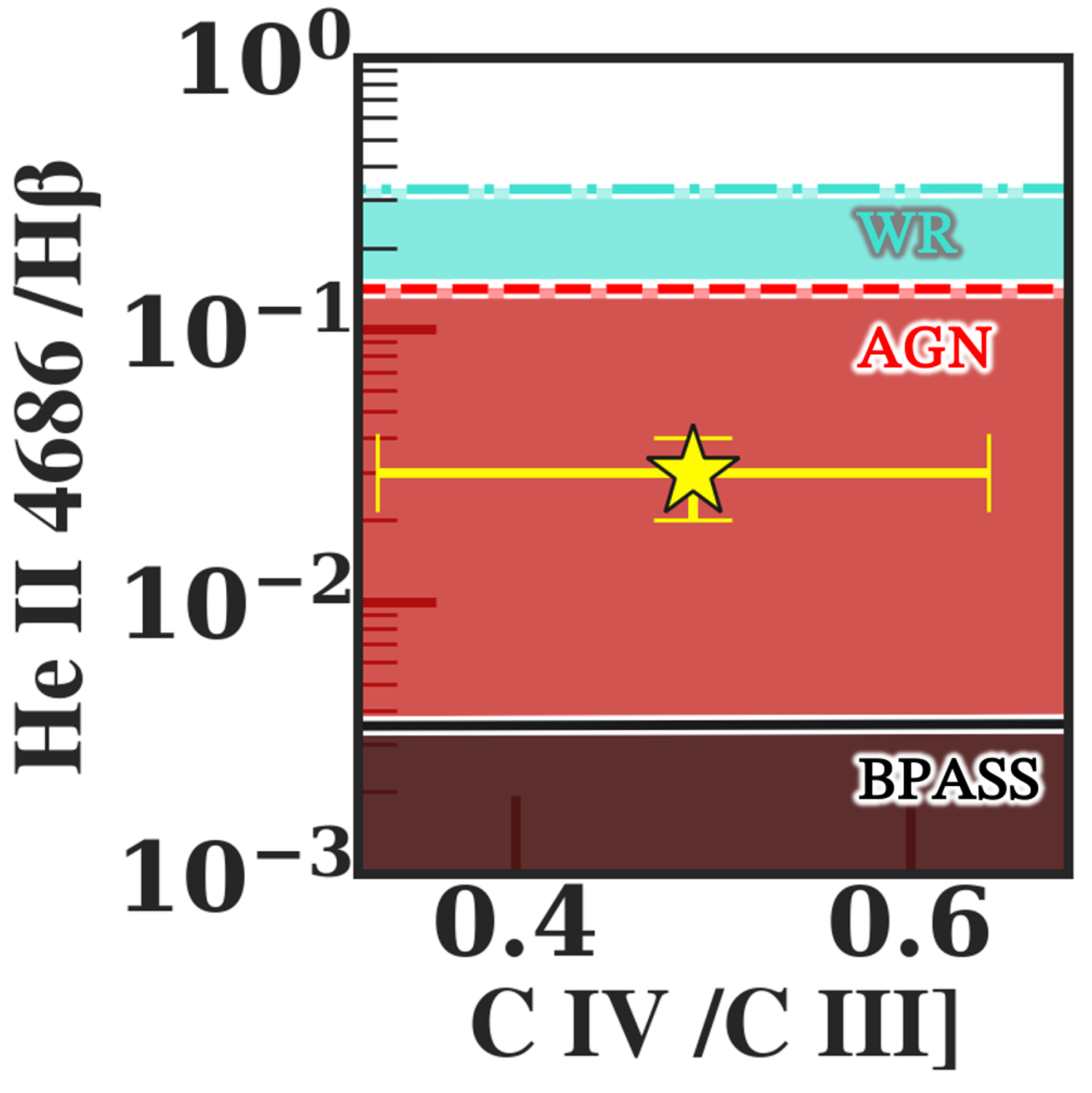}
  \caption{Comparison of the observed He\,{\sc ii}\,\W4686/H$\beta$ ratio with the maximum values predicted by the BPASS, AGN, and WNE models as a function of $C_{43}$. The observed value is shown by the yellow star, while the shaded regions extending up to the model maxima represent BPASS (black), AGN (red), and WNE (turquoise). For the model descriptions, see Appendix~\ref{app:icf}.}
  \label{fig:he2} 
\end{figure}

Although population mixing remains unresolved in GAll, the UV–optical offsets observed in individual galaxies motivate a multi-zone ISM interpretation, in which the UV and optical lines predominantly arise from physically distinct regions with different ionization conditions and possibly different N/O ratios. 
\cite{ji_ga-nifs_2024} proposes the stratified multi-zone ISM to explain the nitrogen abundance discrepancies within a single galaxy. \cite{pascale_nitrogen-enriched_2023} also introduces the high/low pressure ISM structure to explain the strong UV and weak optical nitrogen line within the Sunburst Arc. Moreover, the electron densities measured using $\rm C\,\textsc{iii}]\,\lambda\lambda1907,1909$  are systematically higher than those by $\rm [S\,\textsc{ii}]\,\lambda\lambda6716,6731$, indicating that highly ionized regions traced by $\rm C^{+2}$ also trace denser gas than low ionized region traced by $\rm S^+$. \cite{harikane_jwst_2025} show that, for three $z\sim6$ galaxies observed with ALMA and JWST, the optical oxygen lines trace denser regions than the far-infrared lines, requiring a multi-zone ISM to simultaneously reproduce [O\,{\sc iii}]\,\W5007 and [O\,{\sc iii}]\,\W$88\,\mu\mathrm{m}$ \citep[see also][]{topping_aurora_2025}. These observational evidence support the multi-zone ISM scenario, in which the UV and optical lines trace different physical regions with different ionization conditions and possibly different N/O ratios.\par 

Regarding the inhomogeneous nitrogen abundance in a single system, globular clusters are known to host distinct stellar populations with different N/O abundance ratios: P1 stars have N/O ratios similar to those of local H\,{\sc ii} regions, while P2 stars have ratios similar to the high-z $\rm [N/O]_{UV}$ measurements \citep[e.g.,][]{isobe_jwst_2023}. This resemblance in abundance pattern could suggest the possibility of a common enrichment environment between GC and high-z nitrogen-enhanced galaxies \citep[e.g.,][]{fukushima_impacts_2024,schaerer_nitrogen_2026}. Highly magnified galaxies demonstrate that high-redshift systems can contain stellar clumps with masses comparable to those of globular clusters. For example, the Firefly Sparkle at $z=8.304$ with a total stellar mass of $\simeq10^7 M_\odot$ shows at least 10 stellar clumps with masses of $M_\star\sim10^{5}-10^{6}~M_\odot$ in a compact region with a half-light radius of 100 pc \citep{mowla_formation_2024}. Although we do not resolve or dynamically characterize such clumps in the galaxies contributing to GAll, this result and other observations of clumpy high-redshift galaxies \citep[e.g.,][]{nakane_venus_2025} motivate a qualitative comparison with present-day globular clusters. 

\begin{figure*}[htbp]
  \centering
  \begin{minipage}{0.46\linewidth}
    \centering
    \includegraphics[width=\linewidth]{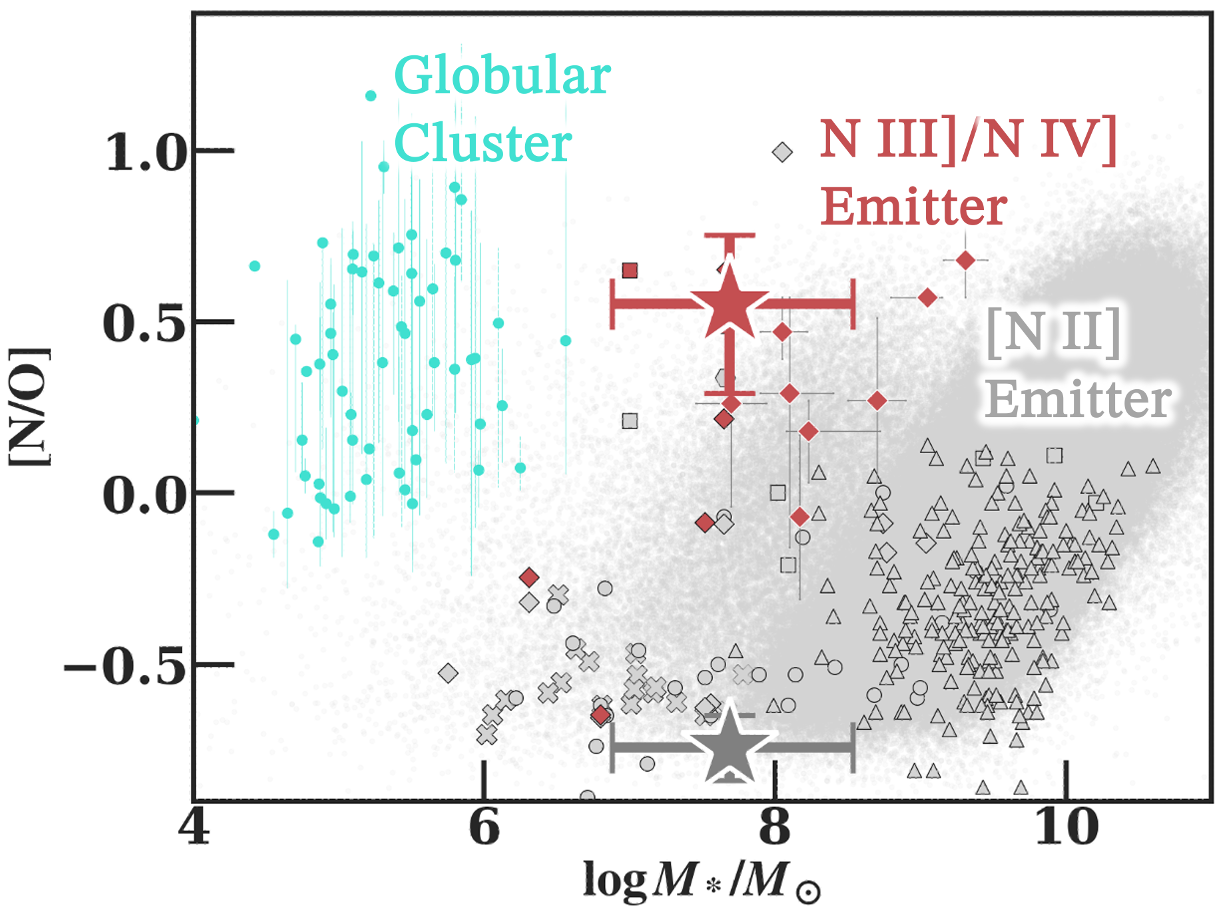}
    \par\centering (a)
  \end{minipage}\hfill
  \begin{minipage}{0.52\linewidth}
    \centering
    \includegraphics[width=\linewidth]{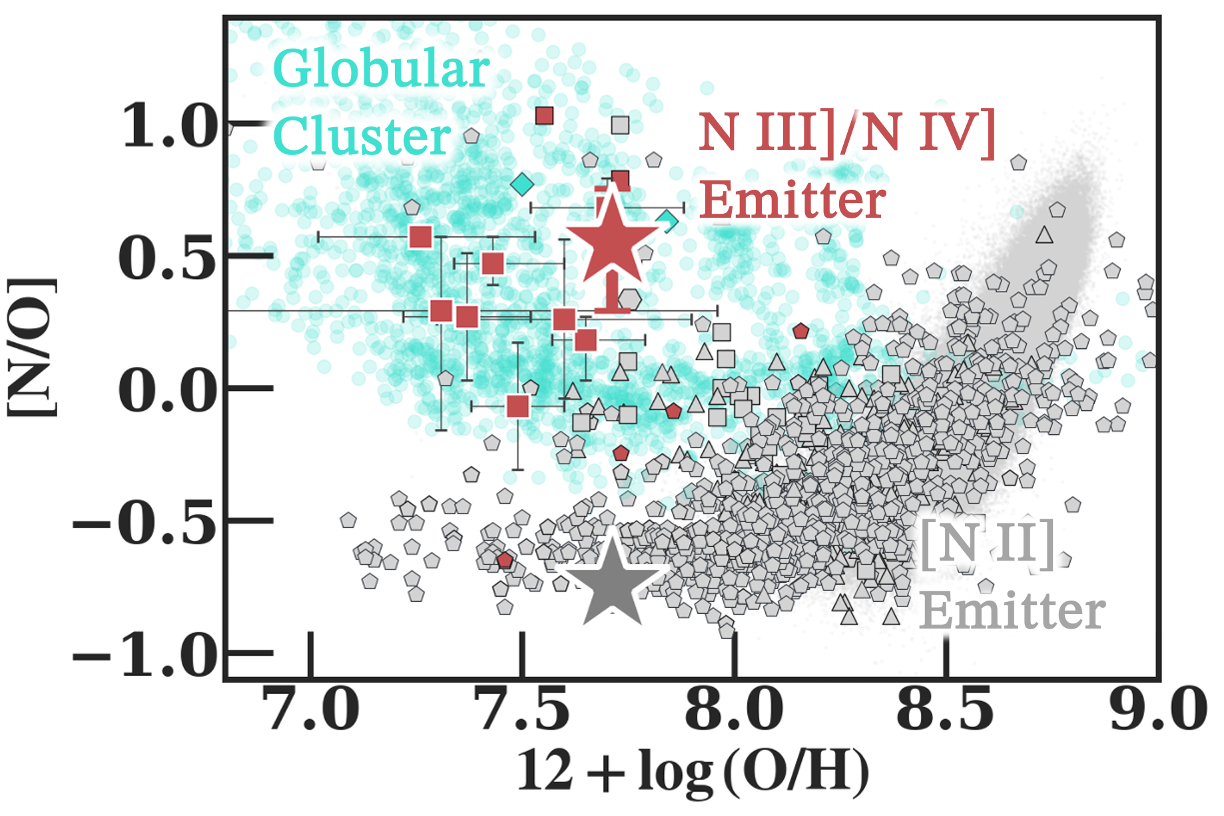}
    \par\centering (b)
  \end{minipage}

  \caption{The [N/O] abundance ratio as a function of (a) stellar mass and (b) gas-phase metallicity. The red and gray symbols represent UV- and optical-based N/O measurements, respectively, from this work and the literature samples compiled in Figures~\ref{fig:no_mass_all} and \ref{fig:no_oh_all}. The red and gray stars show the corresponding measurements for GAll. The turquoise symbols represent stellar-abundance measurements of globular-cluster members from the APOGEE survey \citep{schiavon_apogee_2024}.
  }
  \label{fig:no_mass_oh}
\end{figure*}

We show the [N/O] abundance ratios derived from N\,{\sc iv}]\,\W\W1483,1486 and [N\,{\sc ii}]\,\W6583 together in Figure \ref{fig:no_mass_oh} as a function of stellar mass and gas-phase metallicity. We also present the [N/O] measurements globular-cluster stars in the APOGEE sample \citep{schiavon_apogee_2024} in Figure \ref{fig:no_mass_oh}. The chemical abundances from the member stars of the clusters are measured from the stellar absorption features instead of ISM-origin emission lines. For the APOGEE sample, we use a value-added catalog from \citep{schiavon_apogee_2024}\footnote{\url{https://www.sdss.org/dr19/data_access/value-added-catalogs/?vac_id=105}} and present the abundance measurements derived from the high signal-to-noise ratio spectra \citep[i.e., $\rm SNR>150$;][]{schiavon_apogee_2024}. For the stellar mass of each globular cluster, we adopt another value listed in the value-added catalog of \cite{schiavon_apogee_2024}, and use the median abundance ratio of its member stars as the representative abundance measurement for the cluster. Most globular-cluster systems have the stellar mass of $M_\star\sim10^{4}-10^{6}~M_\odot$, and have  super-solar [N/O] abundance ratio. In the $M_\star$-[N/O] plane, we find that the $\rm [N/O]_{UV}$ measurements lie in between the globular cluster and the local H\,{\sc ii} regions, while the $\rm [N/O]_{\rm Opt}$ measurements are consistent with the local H\,{\sc ii} regions. In the $12+\log({\rm O/H})$-[N/O] plane, we find that the $\rm [N/O]_{UV}$ measurements are consistent with the globular cluster, while the $\rm [N/O]_{\rm Opt}$ measurements are consistent with the local H\,{\sc ii} regions. We also plot the [N/O] and [C/O] measurements of the APOGEE globular cluster sample in Figure \ref{fig:co_no}. We find that the $\mathrm{[N/O]}_{\rm UV}$ and $\mathrm{[C/O]}_{\rm UV}$ ratios of GAll and other high-redshift nitrogen emitters overlap with those of nitrogen-enhanced stars in APOGEE globular clusters. This resemblance suggests that globular-cluster-like enrichment environments may have been widespread in early galaxies and that the UV-bright, nitrogen-rich regions may trace sites or evolutionary phases associated with globular cluster formation.

A physical connection between the nitrogen enhancement in high-z galaxies and the abundance pattern in globular cluster would be more compelling if the same stellar population could account for both the hard ionizing spectrum and the localized nitrogen enrichment on globular-cluster scales. Massive-star winds provide a natural candidate for producing both observables \citep[see also][]{berg_fleeting_2026}. Here, we use a WNE atmosphere as one quantitative realization because the required stellar spectrum and wind properties are available from the PoWR models. We adopt the PoWR SMC WNE model with $\log T_\ast/~{\rm K}\simeq5.25$, $\dot{M}=1.2\times10^{-5}~{M_\odot}~{\rm yr}^{-1}$, and $M_\ast\simeq12\,M_\odot$. For the normal stellar component, we assume a BPASS binary single-burst population at $t\simeq10$ Myr and $Z\simeq0.1Z_\odot$, for which He~{\sc ii}\,\W4686/H$\beta\lesssim0.001$--0.003. We combine the WNE and BPASS components at the flux level. Setting the normalization of the WNE component to unity and that of the BPASS component to $b$, the mixed line ratio is
\begin{equation}
\left(\frac{{\rm He\,\textsc{ii}\,\lambda4686}}{{\rm H}\beta}\right)_{\rm mix}
=
\frac{F_{{\rm He\,\textsc{ii}\,\lambda4686},\rm WNE}+bF_{{\rm He\,\textsc{ii}\,\lambda4686},\rm BPASS}}
{F_{{\rm H}\beta,\rm WNE}+bF_{{\rm H}\beta,\rm BPASS}}.
\end{equation}
We determine $b$ by requiring the mixture to reproduce the observed He~{\sc ii}/H$\beta\simeq0.03$ for the GAll stack. For the adopted WNE and BPASS templates, the resulting WNE-like component contributes approximately 10\% of the total H$\beta$ luminosity. Combining this flux normalization with the H-ionizing photon production rates of the BPASS burst and the adopted PoWR WNE template, we estimate that a young stellar population with $M_\ast\simeq10^{7.7}\,M_\odot$ requires $N_{\rm WN}\sim1650$ WNE-like sources. For comparison, the BPASS broken-power-law IMF predicts $\simeq2\times10^5$ stars with $M_{\rm init}>15\,M_\odot$ for $M_\ast=10^{7.7}\,M_\odot$. Thus, the required $N_{\rm WN}\sim1650$ corresponds to only $\sim1\%$ of the available massive-star reservoir, although not all such stars are expected to become SMC-WNE-like sources.\par

The associated nitrogen-rich wind mass is estimated as
\begin{equation}
  M_{\rm N,wind}\simeq N_{\rm WNE} X_{\rm N}\dot{M}t_{\rm WNE},
\end{equation}
where $X_{\rm N}$ is the nitrogen mass fraction in the wind and $t_{\rm WNE}$ is the duration over which a star remains in a WN-like, nitrogen-rich, H-poor/H-free wind phase. We use $t_{\rm WNE}\sim0.5$ Myr as a representative effective WR duration \citep[e.g.,][]{rosslowe_spatial_2015}. With $X_{\rm N}=1.5\times10^{-3}$ assumed in the SMC WNE model, $N_{\rm WN}\sim1650$ gives $M_{\rm N,wind}\sim15\,M_\odot$.\par 

This nitrogen mass ejected via stellar winds is comparable to the nitrogen excess expected for a globular-cluster-scale enriched population. Galactic globular clusters have an observed mean enriched-star fraction of $f_{\rm enriched}\sim0.7$, based on number counts of long-lived low-mass stars \citep{bastian_globular_2015}, and Fornax globular clusters show internal nitrogen ranges up to $\Delta[{\rm N/Fe}]\simeq2$ dex between N-normal and N-enhanced stars \citep{larsen_nitrogen_2014}. For the mass budget, we adopt a representative enriched abundance of $[{\rm N/H}]_{\rm enriched}\simeq-0.7$, motivated by high-$[{\rm N/O}]$ globular-cluster-like stars selected from APOGEE, $[{\rm N/O}]>0.55$, together with a metal-poor GC abundance scale $[{\rm Fe/H}]\sim-1.5$ and $[{\rm O/Fe}]\sim+0.3$ \citep{belokurov_nitrogen_2023}. For a present-day cluster mass $M_{\rm GC}=2\times10^5\,M_\odot$, this corresponds to an enriched stellar mass $f_{\rm enriched}M_{\rm GC}\simeq1.4\times10^5\,M_\odot$ and a nitrogen mass
\begin{equation}
  M_{\rm N}\simeq f_{\rm enriched}M_{\rm GC}X_{{\rm N},\odot}10^{[{\rm N/H}]_{\rm enriched}}\sim20\,M_\odot,
\end{equation}
where we adopt $X_{{\rm N},\odot}\simeq7\times10^{-4}$. After subtracting a nitrogen-normal baseline with $[{\rm N/H}]_{\rm normal}\simeq-1.5$, the excess nitrogen mass is $\Delta M_{\rm N}\sim15\,M_\odot$. The estimated $\Delta M_{\rm N}\sim15\,M_\odot$ is around the same amount as the value predicted from the observed He\,{\sc ii}\,\W4686/H$\beta$ ratio assuming the contribution from WNE stars. Thus, at the order-of-magnitude level, the nitrogen mass supplied by $\sim10^3$ WNE-like sources is comparable to the nitrogen excess locked into a globular-cluster-scale enriched population.\par

The prompt enrichment need not arise exclusively from classical WNE stars. \citet{shi_very_2026-1} show that very massive stars ($M_\star\gtrsim100\,M_\odot$), which may appear as hydrogen-rich WNh stars while still undergoing core hydrogen burning, can rapidly enrich gas near a young compact cluster with nitrogen and produce a chemically stratified nebula. VMS/WNh stars therefore provide a related massive-star-wind channel, although applying their model quantitatively to our stack is beyond the scope of the present analysis.\par

Subsequent ultra-deep observations of GN-z11 have identified nitrogen enhancement together with massive-star/WR wind signatures, including a broad N\,{\sc iv}] component, and favor a picture in which dense nitrogen-emitting gas is locally enriched by stellar winds \citep{chen_spurs_2026-1,nakane_why_2026}. In extremely UV-luminous galaxies, broad wind-associated N\,{\sc iv}] has also been detected without accompanying nebular N\,{\sc iii}] emission. \citet{yang_quasar_2026} suggest that this may occur because nitrogen-rich wind ejecta have not yet mixed into the ionized ISM, while noting other possible explanations. Although these individual systems are substantially more UV-luminous than the low-mass galaxies emphasized here, they provide empirical examples of localized nitrogen enrichment and incomplete mixing between stellar-wind ejecta and the surrounding ISM. Whether the same process operates in the much fainter galaxies contributing to GAll remains to be established.\par

We caution, however, that WR stars are already included in BPASS, and the BPASS binary models also include envelope-stripped and spun-up stars. Nevertheless, these models still underpredict the strongest observed He~{\sc ii}\,\W4686/H$\beta$ ratios. Our WNE-like component should therefore be interpreted not as a simple addition of ordinary WR stars to BPASS, but as a phenomenological representation of a harder and/or more numerous WNE-like population than predicted by BPASS at the relevant age and metallicity. Indeed, \texttt{GALSEVN} photoionization models \citep{lecroq_nebular_2024} show that, after the first few Myr of a starburst, the He$^+$-ionizing photon budget can be dominated by WNE stars and can produce much stronger nebular He~{\sc ii} emission than BPASS binary models. Because both models use broadly similar PoWR spectra for WR-like stars, this discrepancy likely reflects differences in the number, timing, and duration of WNE-like stars rather than the adopted stellar atmosphere spectra alone. If such WNE-like stripped stars remain important over a longer or different post-burst interval than in the BPASS population used here, the interpretation of emission-line-based burstiness may also change. In standard BPASS-based models, strong nebular emission and high He~{\sc ii}\,\W4686/H$\beta$ ratios generally select a short-lived phase close to a recent burst. A more extended WNE-like contribution could broaden this observable time window, reducing the need to observe galaxies in an extremely short post-burst phase. This may help explain why strong He\,{\sc ii} emission and nitrogen enrichment are observed in some low-mass high-redshift galaxies. The origin of such binary star processes may be related to the environment of star formation. \cite{martinez_under_2025} show that N/O is positively correlated with electron density and star-formation surface density. The star-formation in the dense compact region may be related to the enhanced binary interactions. A detailed and self-consistent model would require coupling binary stellar evolution, WNE lifetimes, wind yields, stellar atmospheres, and photoionization calculations. We leave such modeling to future work. On the observational side, larger statistical samples, especially from lensed-field surveys such as VENUS \citep[e.g.,][]{nakane_venus_2025}, are needed to test whether current observations preferentially select bursty low-mass galaxies with strong nebular emission. Moreover, further spectroscopic follow-ups of individual stellar clumps in the highly magnified galaxies would help understand the multi-zone nature of the ISM and the connection to the globular cluster formation.

\section{Summary}
In this paper, we analyzed 405 deep JWST/NIRSpec spectra of gravitationally lensed galaxies at $z \simeq 4.5$--$9.8$ from multiple lensing-cluster surveys to investigate the chemical enrichment of low-mass galaxies in the early universe. The individual-galaxy sample reaches $M_\star \simeq 10^{5.7}\,M_\odot$. By combining individual spectra and stellar-mass-binned stacked spectra, we measured rest-frame UV and optical emission lines and derived gas-phase metallicities, ionization properties, and elemental abundance ratios. Our main results are summarized as follows.

\begin{enumerate}
    \item We derive new gas-phase metallicity calibrations using the stellar-mass-binned stacked spectra down to $\log(M_\star/M_\odot)=5.69$--$6.97$, with a representative stellar mass of $M_\star\simeq10^{6.6}\,M_\odot$. We find that these calibrations are consistent with those derived from more massive galaxies, even at the low-metallicity end, supporting the applicability of empirical strong-line calibrations to very low-mass galaxies at high redshift.
    
    \item Using the newly derived gas-phase metallicity calibration, we determine the mass--metallicity relation at $z\sim6$ over $M_\star\simeq10^{6.6}$--$10^{9.4}\,M_\odot$ based on the stellar-mass-binned stacked spectra. A two-parameter linear fit gives $Z_8=7.694\pm0.028$ at $\log(M_\star/M_\odot)=8$ and a slope of $\alpha=0.318\pm0.037$. The relation extends the observed decline in gas-phase metallicity to very low stellar masses; its slope is only slightly steeper than the local low-mass slope and lies within the range predicted by hydrodynamical simulations.
    
    \item We measure the N/O, C/O, and Ne/O abundance ratios and place a 1$\sigma$ upper limit on Ar/O using the stacked spectra with a representative stellar mass of $M_\star \simeq 10^{7.7}~\,M_\odot$. We find that the N/O abundance ratio inferred from the UV N\,{\sc iv}]\,$\lambda\lambda1483,1486$ doublet is 0.6 dex higher than the solar value, similar to values from the brightest galaxies at similar redshifts \citep[e.g.,][]{cameron_nitrogen_2023,schaerer_nitrogen_2026} and globular clusters with stellar masses around $10^4$--$10^6\,M_\odot$ \citep[e.g.,][]{schiavon_apogee_2024}. 

    \item We find differing N/O abundance ratios from the high-ionization UV and low-ionization optical diagnostics. This difference is consistent with, but does not uniquely require, a multi-zone ISM in which the highly ionized regions traced by the UV lines are preferentially nitrogen-enhanced; population mixing within the stack remains possible. Using the number of WNE stars required to reproduce the observed He\,{\sc ii}/H$\beta$ strength, we estimate that their stellar winds can supply a nitrogen mass comparable to that required to produce the nitrogen-enhanced population in a $\sim10^5\,M_\odot$ globular cluster. This result suggests that localized enrichment by clustered WR stars may be connected to the formation of nitrogen-enhanced stellar populations in globular-cluster-like environments.
\end{enumerate}

These results suggest that the earliest low-mass galaxies were already undergoing chemically diverse enrichment processes while still following a global metallicity sequence. Larger spectroscopic samples, improved photoionization-model abundance calibrations, and deeper rest-frame UV and optical observations with JWST will be essential for establishing how widespread such chemically inhomogeneous regions are, and for clarifying the role of massive stars in shaping the earliest phases of galaxy evolution. Future observations of highly magnified clumpy galaxies could also clarify the connection between inhomogeneous chemical properties in low-mass galaxies and clumpy star formation.

\begin{acknowledgments}
We thank Charlotte Mason, Dan Stark, Seiji Fujimoto, Chris Willott, Rachel Bezanson, Ivo Labbe, Tomasso Treu, and Klaus Pontoppidan and all the SPURS, GLIMPSE(-D), CANUCS, UNCOVER, GLASS, and ERO teams for designing the observations and obtaining the JWST datasets we used in this paper. We thank Tomoya Suzuguchi, Hajime Fukushima, Xihan Ji, Daniela Calzetti, Minami Nakane, James Trussler, Mana Ito, Reishi Ishida, Sho Ebihara, Shunta Shimizu, Ken Ohsuga, Masao Mori, Kohji Yoshikawa, Liang Dai, Chiaki Koabayshi, Stéphane Charlot, Seiji Fujimoto, Roberta Tripodi, Antonio Giménez Alcázar, and the anonymous referee for useful comments and discussions. This work is based on observations with the NASA/ESA/CSA James Webb Space Telescope at the Space Telescope Science Institute, which is operated by the Association of Universities for Research in Astronomy, Inc., under NASA contract NAS 5-03127 for JWST. Some of the data products presented herein were retrieved from the Dawn JWST Archive (DJA). DJA is an initiative of the Cosmic Dawn Center (DAWN), which is funded by the Danish National Research Foundation under grant DNRF140. We thank Chiaki Kobayashi and the collaboration of the Galactic Chemical Evolution (GCE) model for providing the GCE model data. This work utilizes gravitational lensing models produced by PIs Bradač, Natarajan \& Kneib (CATS), Merten \& Zitrin, Sharon, Williams, Keeton, Bernstein and Diego, and the GLAFIC group. This lens modeling was partially funded by the HST Frontier Fields program conducted by STScI. STScI is operated by the Association of Universities for Research in Astronomy, Inc. under NASA contract NAS 5-26555. The lens models were obtained from the Mikulski Archive for Space Telescopes (MAST).

Funding for the Sloan Digital Sky Survey V has been provided by the Alfred P. Sloan Foundation, the Heising-Simons Foundation, the National Science Foundation, and the Participating Institutions. SDSS acknowledges support and resources from the Center for High-Performance Computing at the University of Utah. SDSS telescopes are located at Apache Point Observatory, funded by the Astrophysical Research Consortium and operated by New Mexico State University, and at Las Campanas Observatory, operated by the Carnegie Institution for Science. The SDSS web site is www.sdss.org. This research used data obtained with the Dark Energy Spectroscopic Instrument (DESI). DESI construction and operations is managed by the Lawrence Berkeley National Laboratory. This material is based upon work supported by the U.S. Department of Energy, Office of Science, Office of High-Energy Physics, under Contract No. DE-AC02-05CH11231, and by the National Energy Research Scientific Computing Center, a DOE Office of Science User Facility under the same contract. Additional support for DESI was provided by the U.S. National Science Foundation (NSF), Division of Astronomical Sciences under Contract No. AST-0950945 to the NSF's National Optical-Infrared Astronomy Research Laboratory; the Science and Technology Facilities Council of the United Kingdom; the Gordon and Betty Moore Foundation; the Heising-Simons Foundation; the French Alternative Energies and Atomic Energy Commission (CEA); the National Council of Humanities, Science and Technology of Mexico (CONAHCYT); the Ministry of Science and Innovation of Spain (MICINN), and by the DESI Member Institutions: www.desi.lbl.gov/collaborating-institutions. The DESI collaboration is honored to be permitted to conduct scientific research on I'oligam Du'ag (Kitt Peak), a mountain with particular significance to the Tohono O'odham Nation. Any opinions, findings, and conclusions or recommendations expressed in this material are those of the author(s) and do not necessarily reflect the views of the U.S. National Science Foundation, the U.S. Department of Energy, or any of the listed funding agencies.

We acknowledge support from the World Premier International Research Center Initiative (WPI Initiative), MEXT, Japan, and KAKENHI (20H00180, 24KJ0202, 26H02061, 21H04467, 25H00674, 24H00245, 24KJ1160) through the Japan Society for the Promotion of Science, JST FOREST Program (JP-MJFR202Z), the JSPS International Leading Research (22K21349), the Sumitomo Foundation, the Ito Science Promotion Society, and the Yamaguchi Scholarship Foundation, the joint research program of the Institute for Cosmic Ray Research (ICRR), University of Tokyo, and FoPM WINGS Program in the University of Tokyo. This research was partially supported by the Hayakawa Yukio Foundation.

The authors acknowledge the use of ChatGPT (OpenAI, GPT5.5) and Codex (OpenAI, Codex 5.4) to assist with language editing, code development, and figure generation.
All AI-assisted outputs were carefully reviewed and validated by the authors. The authors take full responsibility for all analyses, interpretations, and conclusions
presented in this work.
\end{acknowledgments}

\appendix
\section{Photoionization-model Abundance Calibrations}\label{app:icf}
\begin{figure*}[htbp]
  \centering
  \begin{subfigure}{0.99\linewidth}
    \centering
    \includegraphics[width=\linewidth]{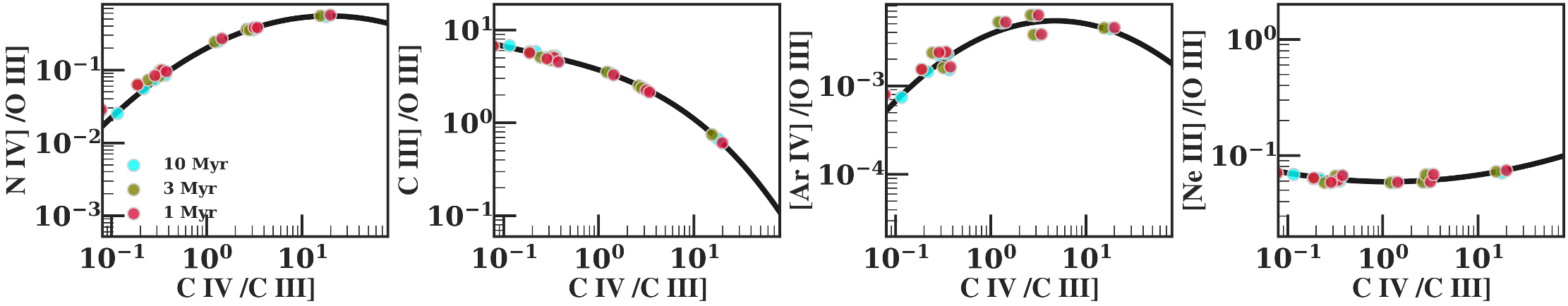}
    \subcaption{BPASS binary stellar-population radiation: $Z_\ast=Z_{\rm gas}=0.1\,Z_\odot$ and ages of 1--10 Myr.}
    \label{fig:icf_bpass}
  \end{subfigure}

  \vspace{0.5em}

  \begin{subfigure}{0.99\linewidth}
    \centering
    \includegraphics[width=\linewidth]{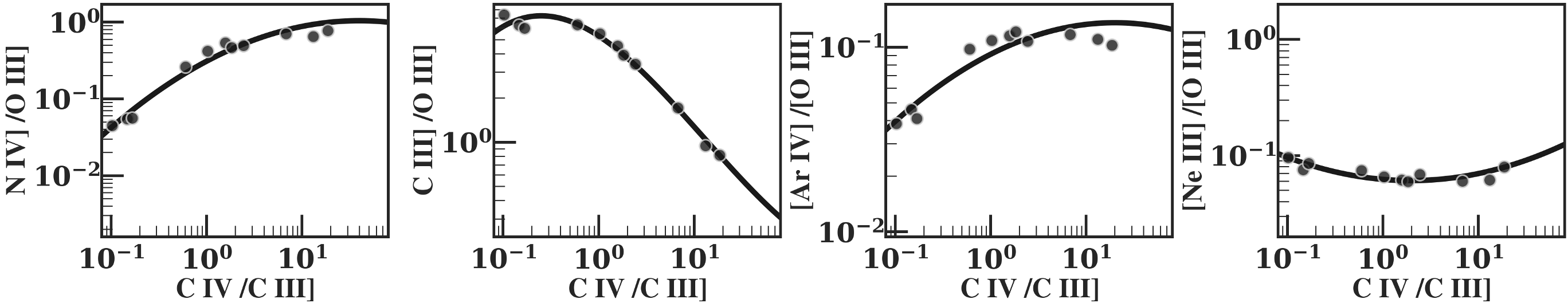}
    \subcaption{\textsc{Cloudy} default AGN radiation field, comprising a power law with index $-1.5$ and a $1.5\times10^5$ K blackbody; $Z_{\rm gas}=0.1\,Z_\odot$.}
    \label{fig:icf_agn}
  \end{subfigure}

  \vspace{0.5em}

  \begin{subfigure}{0.99\linewidth}
    \centering
    \includegraphics[width=\linewidth]{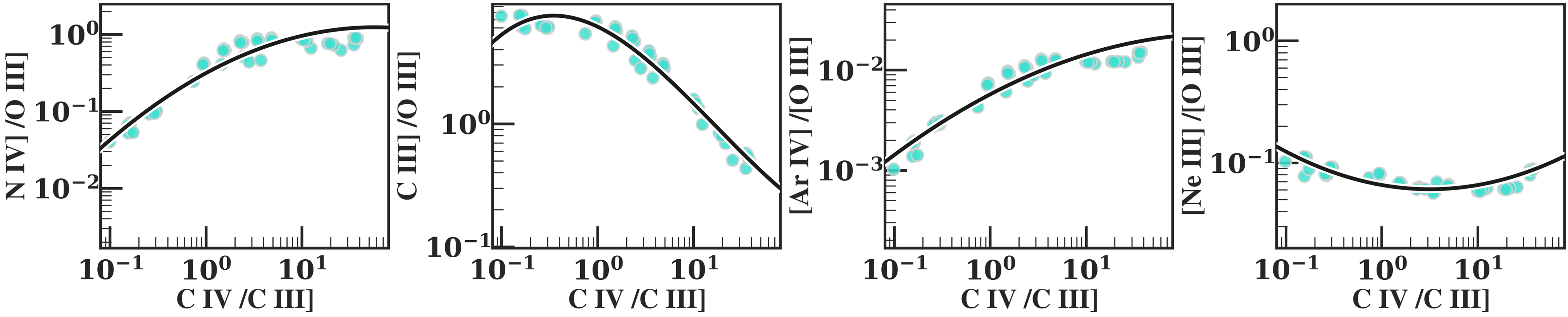}
    \subcaption{PoWR SMC WNE stellar-atmosphere radiation, with $\log(T_\ast/{\rm K})=5.25$, $M_\ast\simeq12\,M_\odot$, and $1.5\times10^{-6}\leq\dot{M}/(M_\odot\,{\rm yr}^{-1})\leq3.5\times10^{-5}$.}
    \label{fig:icf_wr}
  \end{subfigure}

  \vspace{0.5em}

  \begin{subfigure}{0.99\linewidth}
    \centering
    \includegraphics[width=\linewidth]{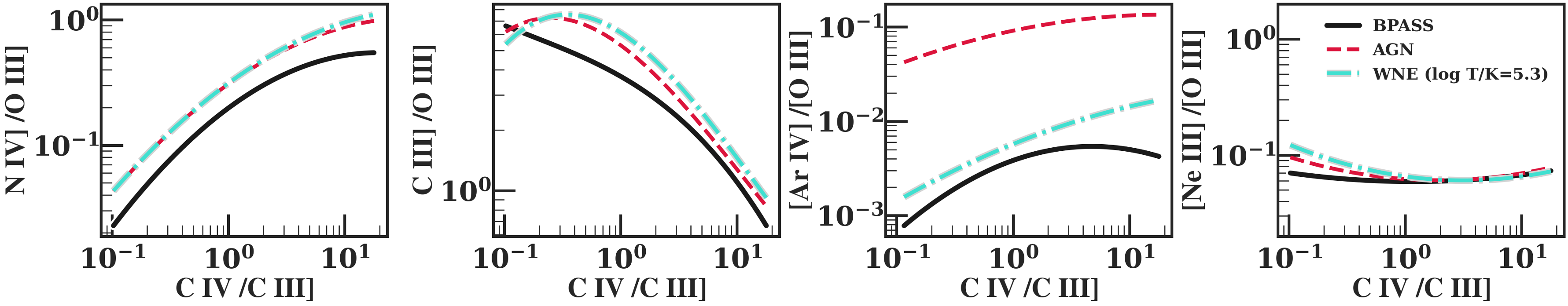}
    \subcaption{Comparison of the stellar and AGN best-fit relations.}
    \label{fig:icf_compare}
  \end{subfigure}
  \caption{Photoionization-model relations between C\,{\sc iv}/C\,{\sc iii}] and the abundance-sensitive N\,{\sc iv}]/O\,{\sc iii}], C\,{\sc iii}]/O\,{\sc iii}], [Ar\,{\sc iv}]/O\,{\sc iii}], and [Ne\,{\sc iii}]/O\,{\sc iii}] emission-line ratios. All models use a plane-parallel, constant-density cloud with solar relative abundances and span $-3\leq\log U\leq0$ and $\log(n_e/{\rm cm}^{-3})=1$, 3, and 5. Colors show the ionization parameter, electron density, and input radiation field. The radiation-field assumptions for panels (a)--(c) are given in the subcaptions. Panel (d) compares the best-fit relations for the BPASS stellar (black solid) and AGN (red dashed) radiation fields.}
  \label{fig:icf}
\end{figure*}

We construct photoionization-model relations for inferring elemental abundance ratios from the observed emission-line ratios using the \textsc{Cloudy} code \citep{ferland_2017_2017}. Based on our findings in Section~\ref{sec:res:sp}, \ref{sec:res:ion}, and \ref{sec:res:ne}, we calculate the emission-line strengths for the different ionization parameters, electron densities, and stellar radiation fields. For the ionization parameter, we vary $\log~U$ from -3 to 0. For the electron density, we vary the $\log~n_e$ from 1 to 5 in steps of 2. For the stellar radiation, we assume the BPASS stellar population synthesis models \citep{eldridge_binary_2017} with a metallicity of 0.1~$Z_\odot$ and ages between 1 and 10 Myr. We assume the gas-phase metallicity of 0.1~$Z_\odot$ and the solar abundance pattern \cite{asplund_chemical_2009} for the relative abundance of different elements. We assume a plane-parallel geometry for the gas cloud and the constan- density profile.
We calculate the emission-line ratios of C\,{\sc iv}\,\W\W1548,1551/C\,{\sc iii}]\,\W\W1907,1909, N\,{\sc iv}]\,\W\W1483,1486/O\,{\sc iii}]\,\W\W1661,1666, C\,{\sc iii}]\,\W\W1907,1909/O\,{\sc iii}]\,\W\W1661,1666, [Ar\,{\sc iv}]\,\W4711/[O\,{\sc iii}]\,\W\W4959,5007, and [Ne\,{\sc iii}]\,\W3869/[O\,{\sc iii}]\,\W\W4959,5007. We show our result in Figure~\ref{fig:icf}. We find that the N\,{\sc iv}]\,\W\W1483,1486/O\,{\sc iii}]\,\W\W1661,1666, C\,{\sc iii}]\,\W\W1907,1909/O\,{\sc iii}]\,\W\W1661,1666, [Ar\,{\sc iv}]\,\W4711/[O\,{\sc iii}]\,\W\W4959,5007, and [Ne\,{\sc iii}]\,\W3869/[O\,{\sc iii}]\,\W\W4959,5007 lie on sequences when plotted against C\,{\sc iv}\,\W\W1548,1551/C\,{\sc iii}]\,\W\W1907,1909. At fixed $C_{43}$, these relations give the emission-line ratios expected for the adopted solar relative abundance pattern. We infer each abundance ratio by comparing the observed line ratio with the corresponding model relation and interpreting the logarithmic offset as the abundance offset from solar. We fit the model line-ratio relations with third- or second-order polynomial functions and present the coefficients in Table~\ref{tab:icf}. We then use the best-fit relations as photoionization-model abundance calibrations for N/O, C/O, Ar/O, and Ne/O.

We additionally calculate photoionization models using PoWR stellar-atmosphere spectra for SMC-metallicity WNE stars \citep{grafener_line-blanketed_2002,hamann_temperature_2003,sander_consistent_2015}. We adopt $\log(T_\ast/{\rm K})=5.25$, $M_\ast\simeq12\,M_\odot$, and mass-loss rates from $1.5\times10^{-6}$ to $3.5\times10^{-5}\,M_\odot\,{\rm yr}^{-1}$. The gas metallicity, geometry, density grid, and ionization-parameter grid are the same as in the BPASS calculations described above. The models with $\dot{M}\geq1.2\times10^{-5}\,M_\odot\,{\rm yr}^{-1}$ reach He\,{\sc ii}\,$\lambda4686$/H$\beta\simeq0.3$, which we adopt as the WNE reference value in Section~\ref{sec:disc:nuv_opt}.

For comparison, we also calculate the corresponding abundance-calibration relations for N/O, C/O, Ar/O, and Ne/O using photoionization models from the \textsc{Cloudy} code with an AGN radiation field. 
Following the methodology of \citet{isobe_jades_2025}, we adopt the default AGN radiation field in the \textsc{Cloudy} code, which is a combination of a power-law component with an index of -1.5 and a blackbody component with a temperature of $1.5\times10^5$ K. We vary the ionization parameter and electron density in the same way as we do for the stellar radiation field.
We fix the gas-phase metallicity at 0.1\,$Z_\odot$ and the solar abundance pattern \citep{asplund_chemical_2009} for the relative abundance of different elements. We assume the plane-parallel geometry for the gas cloud and the constant density profile. We calculate the emission-line ratio of C\,{\sc iv}/C\,{\sc iii}], N\,{\sc iv}]/O\,{\sc iii}], C\,{\sc iii}]/O\,{\sc iii}], [Ar\,{\sc iv}]/O\,{\sc iii}], and [Ne\,{\sc iii}]/O\,{\sc iii}]. We show our results in Figure~\ref{fig:icf}. We find that the N\,{\sc iv}]/O\,{\sc iii}], C\,{\sc iii}]/O\,{\sc iii}], [Ar\,{\sc iv}]/O\,{\sc iii}], and [Ne\,{\sc iii}]/O\,{\sc iii}] are also on sequences when plotted against C\,{\sc iv}/C\,{\sc iii}] for the AGN radiation field. We find that the relation between the C\,{\sc iv}/C\,{\sc iii}] ratio and the N\,{\sc iv}]/O\,{\sc iii}], C\,{\sc iii}]/O\,{\sc iii}], [Ar\,{\sc iv}]/O\,{\sc iii}], [Ne\,{\sc iii}]/O\,{\sc iii}] ratios for the AGN radiation field are higher than the relation for the stellar radiation field. This is because the AGN radiation field is harder than the stellar radiation field, which leads to a higher ionization state of the gas. We fit the relation between the C\,{\sc iv}/C\,{\sc iii}] ratio and the N\,{\sc iv}]/O\,{\sc iii}], C\,{\sc iii}]/O\,{\sc iii}], [Ar\,{\sc iv}]/O\,{\sc iii}], and [Ne\,{\sc iii}]/O\,{\sc iii}] ratios for the AGN radiation field with a third-order polynomial function. We present the coefficients of the polynomial function in Table~\ref{tab:icf}. We then use the best-fit AGN relations to infer N/O, C/O, Ar/O, and Ne/O for galaxies with AGN signatures.

\begin{deluxetable}{lcccc}
  \tablecaption{Coefficients of the third-order polynomial fits to the relations between $\log C_{43}$ and the UV/optical ionic line ratios.\label{tab:icf}}
  \tablewidth{0pt}
  \tablehead{
  \colhead{Line ratio} & \colhead{$a_0$} & \colhead{$a_1$} & \colhead{$a_2$} & \colhead{$a_3$}
  }
  \startdata
  $\log(\mathrm{N\,\textsc{iv}]/O\,\textsc{iii}]})~(\rm stellar)$      & -0.70 & 0.69 & -0.27 & \nodata \\
  $\log(\mathrm{C\,\textsc{iii}]/O\,\textsc{iii}]})~(\rm stellar)$     & 0.57 & -0.33 & -0.14 & -0.06 \\
  $\log(\mathrm{[Ar\,\textsc{iv}]/[O\,\textsc{iii}]})~(\rm stellar)$    & -2.42 & 0.44 & -0.32 & \nodata \\
  $\log(\mathrm{[Ne\,\textsc{iii}]/[O\,\textsc{iii}]})~(\rm stellar)$    & -1.23 & -0.01 & 0.07 & \nodata \\
  $\log(\mathrm{N\,\textsc{iv}]/O\,\textsc{iii}]})~(\rm AGN)$      & -0.51 & 0.66 & -0.21 & \nodata \\
  $\log(\mathrm{C\,\textsc{iii}]/O\,\textsc{iii}]})~(\rm AGN)$     & 0.72 & -0.42 & -0.28 & 0.08 \\
  $\log(\mathrm{[Ar\,\textsc{iv}]/[O\,\textsc{iii}]})~(\rm AGN)$    & -1.04 & 0.27 & -0.10 & \nodata \\
  $\log(\mathrm{[Ne\,\textsc{iii}]/[O\,\textsc{iii}]})~(\rm AGN)$    & -1.21 & -0.07 & -1.21 & \nodata \\
  $\log(\mathrm{N\,\textsc{iv}]/O\,\textsc{iii}]})~(\rm WNE)$      & -0.51 & 0.68 & -0.19 & \nodata \\
  $\log(\mathrm{C\,\textsc{iii}]/O\,\textsc{iii}]})~(\rm WNE)$     & 0.79 & -0.37 & -0.34 & 0.09 \\
  $\log(\mathrm{[Ar\,\textsc{iv}]/[O\,\textsc{iii}]})~(\rm WNE)$    & -2.24 & 0.50 & -0.10 & \nodata \\
  $\log(\mathrm{[Ne\,\textsc{iii}]/[O\,\textsc{iii}]})~(\rm WNE)$    & -1.18 & -0.14 & 0.14 & \nodata \\
  \enddata
  \tablecomments{The fitted relation is given by $y = a_0 + a_1 x + a_2 x^2 + a_3 x^3$, where $x=\log C_{43}=\log(\mathrm{C\,\textsc{iv}}/\mathrm{C\,\textsc{iii}]})$. The stellar model corresponds to the BPASS spectrum with stellar metallicity of 0.1\,$Z_\odot$. The AGN and WNE models are described in Appendix~\ref{app:icf} and Section~\ref{sec:disc}.}
  \end{deluxetable}


\bibliography{main}{}
\bibliographystyle{aasjournalv7}



\end{document}